# Systematic Variation of Diffusion Rates of Components in Silicides Depending on Atomic Number of Refractory Metal Component

Aloke Paul

Department of Materials Engineering, Indian Institute of Science, C.V. Raman Road, Bengaluru, India 560012

aloke@iisc.ac.in; aloke.paul@gmail.com

**Keywords:** Diffusion coefficients; Diffusion-controlled growth; metal-silicon systems.

**Abstract.** Interdiffusion studies conducted in group IVB, VB and VIB metal-silicon systems are discussed in detail to show a pattern in the change of diffusion coefficients with the change in atomic number of the refractory metal (M) component. $MSi_2$ and $M_5Si_3$ phases are considered for these discussions. It is shown that integrated diffusion coefficients increase with the increase in atomic number of the refractory component when the data are plotted with respect to the melting point normalized annealing temperature. This indicates the increase in overall defect concentration facilitating the diffusion of components. This is found to be true in both the phases. Additionally, the estimated ratios of tracer diffusion coefficients indicate the change in concentration of antisite defects in certain manner with the change in atomic number of the refractory components.

## 1. Introduction

Silicide intermetallic compounds are used in wide range of applications for many decades. At present, different refractory metal silicides are used as bulk structural materials or as coatings. These are also used extensively as a thin film in electronic industries. Molybdenum disilicide is used as heating elements for more than a century. This is also used extensively as oxidation resistant coatings on many structural applications. Mo is used as an interlayer to the diffusion bond of $Si_3N_4$ and the growth of the silicide phases control the performance of the structure. Not only as the structural materials, but the molybdenum-silicon system draws special attention because of use in various applications such as Schottky contacts, VLSI (very large scale integration) interconnects and photomasks for VLSI fabrication, soft X-ray mirror etc. Extensive efforts are being made to develop Mo- and Nb-based silicides as structural materials in various applications. Vanadium disilicide alone is not so useful alone but it is developed as a protective coating in combination with successfully used $MoSi_2$. This is also considered for Ohmic contact in Schottky diodes**.** Initially, pure Ta and W were tested as the diffusion barrier layer in Cu interconnects. Cu does not dissolve in these components and therefore it cannot diffuse through this barrier layer to contaminate Si. At present various compounds based on these components are used with even higher performance. The reactive diffusion process is followed to grow tungsten disilicides for the use as integrated circuits due to their beneficial properties of low electrical resistivity and good thermal stability.

Other than the importance of understanding the diffusion-controlled growth process, the estimation of the diffusion coefficients is important to understanding many physical and mechanical properties of materials. Because of relevance, most of the studies concentrate on the diffusion controlled growth in thin film condition. In such a situation, mainly the disilicide phases are found to grow although there are many other phases present in the phase diagram. It is already known that the growth mechanism of the phases in the thin film is complicated. A few equilibrium phases do not grow because of stresses and nucleation problems. Sometimes, even the metastable phases are be found in the interdiffusion zone. Even, a sequential instead of a simultaneous growth of the phases is also common [1]. Very frequently, the Kirkendall marker experiments are required to conduct for the estimation of the intrinsic or tracer diffusion coefficients, which are important for the understanding of the atomic mechanism of diffusion. Since the size of inert particles used in bulk diffusion couple is relatively bigger in size compared to the thin film diffusion couple, these experiments were conducted by entrapping a gas bubble at the interface. However, these small bubbles were dragged by the grain boundaries and could not indicate the relative mobilities of the components successfully. Based on the various analysis in different systems, a higher diffusion rate of Si compared to the refractory metal component especially in the disilicide phases were considered to be true.

Bulk diffusion couple experiments are important for an extensive analysis of the diffusion process without the influence of other factors. One can study the Kirkendall marker experiments for the sake of determination of the relative mobilities of the components. It should be noted here that many publications concentrated mainly on the determination of the parabolic growth constants. However, these are not a material constant and depend on the composition of the end members. The values will change if the end member composition is changed. There is an only a limited number of studies in which the diffusion coefficients were determined There are also few studies in which the inert markers were used to detect the position of the Kirkendall plane. However, the diffusion rates of components were not estimated in these studies. Diffusion couple technique is already established as the reliable technique for the estimation of the relative mobilities of the components. Recently, we have conducted extensive analysis in the group IVB (Ti, Zr, Hf), VB (V, Nb, Ta) and VIB (Mo, W) refractory metal (M)-silicon (Si) systems. The integrated diffusion coefficients and the relative mobilities of the components are estimated in $MSi_2$ phase. The integrated diffusion coefficients are also estimated in another phase $M_5Si_3$. It should be noted here that the crystal structure of a particular phase in a particular group is similar.

It should be noted here that the general atomic mechanism of diffusion is known in the intermetallic compounds [1-3]. In most of the compounds, sublattice diffusion mechanism is operative, which is controlled by defects such as vacancies and antisites present on different sublattices. The diffusion coefficients are rather straightforward to estimate. However, there are no reliable experimental techniques available, which can be used to estimate these defects since different sublattices can have different concentration of these defects. Theoretical estimation of these defects is also not easy because of unavailability of basic parameters required for such estimation. In this situation, one can rather understand the relative concentration of the defects in a particular compound based on estimated diffusion coefficients. In this review, I shall discuss first the growth mechanism of the phases and estimation of the diffusion coefficients in different systems. Following, these systems will be compared to show that diffusion coefficients and therefore defect concentration changes following a certain pattern with the change in the atomic number of the refractory components. This discussion is developed based on the results published in Refs. [4-13].

## 2. Diffusion study in Group IVB (Ti, Zr and Hf) Metal (M)-Silicon (Si) systems

### 2.1 Interdiffusion study in the Ti-Si system

All the equilibrium phases are expected to grow in a bulk diffusion couple unless there is a significant difference in growth rate of the product phases such that one or more phases are difficult to detect in the optical and scanning electron microscopes. Different phases grow with different thicknesses because of the difference in diffusion rates of the components through the phase of interest. The Ti-Si phase diagram is shown in Figure 1 [14]. It can be seen that four intermetallic compound $TiSi_2$, TiSi, $Ti_5Si_4$ and $Ti_5Si_3$ are present with a narrow homogeneity range in the range of annealing temperature of this study *i.e.* (1150–1250 °C). Therefore, all thee phases are expected to grow in the interdiffusion zone in this temperature range. It should be noted that a phase mixture cannot grow in an interdiffusion zone of a binary system [1-3]. If experiments are conducted at a temperature below 1170 °C, $Ti_3Si$ is also expected to grow. It can be seen in Figure 2 that three phases $TiSi_2$, TiSi, $Ti_5Si_4$ grow in the interdiffusion zone at 1200 °C. A close examination near the Ti end member of the interdiffusion zone revealed the presence of two other phases, $Ti_5Si_3$ and $Ti_3Si$. This is further confirmed by the composition profile measurement as shown in Figure 3.

Therefore, the $Ti_3Si$ phase is found to be present at the temperature above 1170 °C and this indicates the problem with the published available phase diagram. To investigate it further, the interdiffusion zones were examined at 1225 and 1250 °C, as shown in Figure 4. It can be seen that this phase is present at 1225 °C. However, this is not present at 1250 °C. Therefore, this study indicates that this phase is stable up to a temperature between 1225-1250 °C instead of 1170 °C.

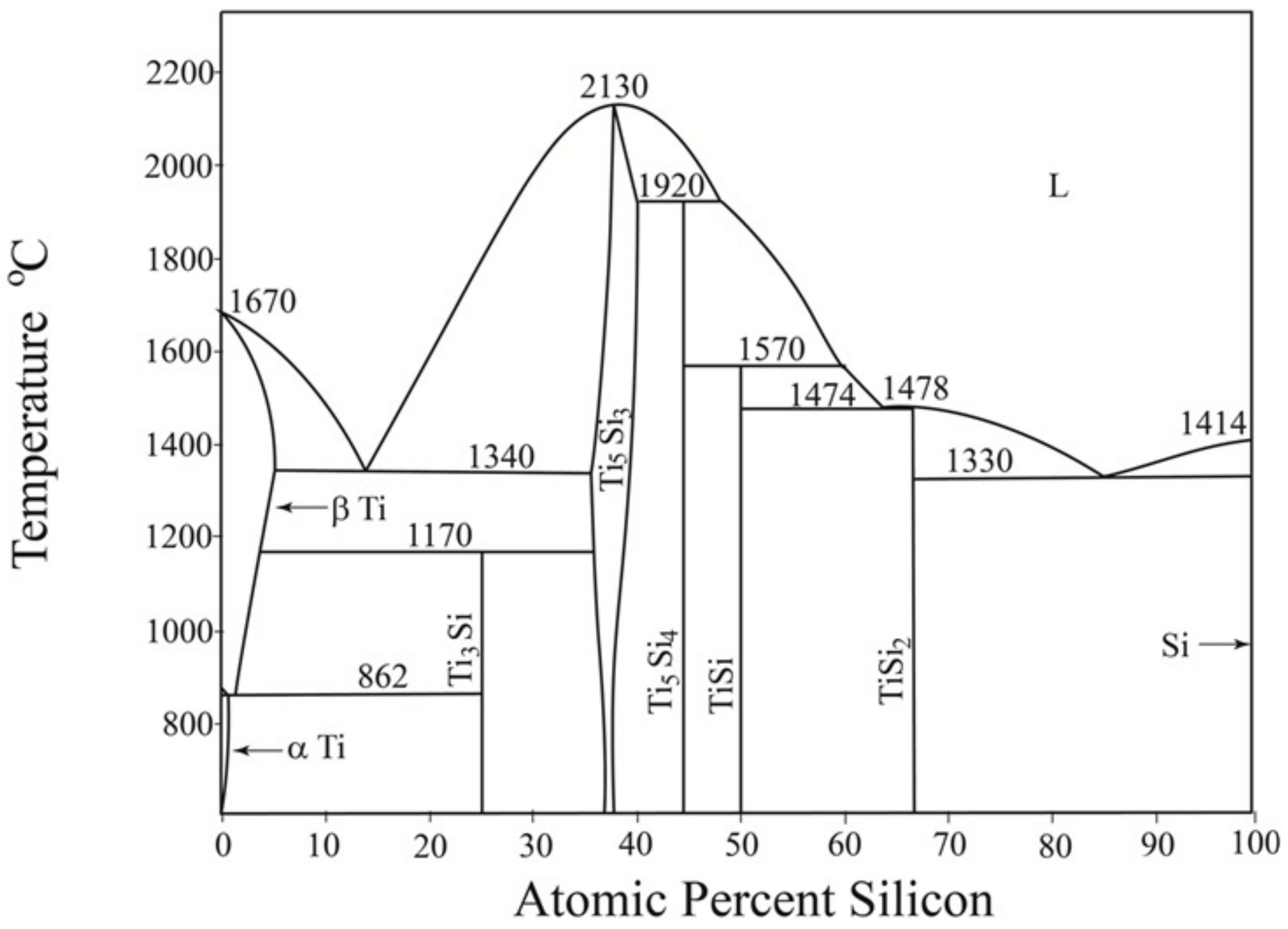

Figure 1 Ti-Si phase diagram, which is redrawn from reference [14].

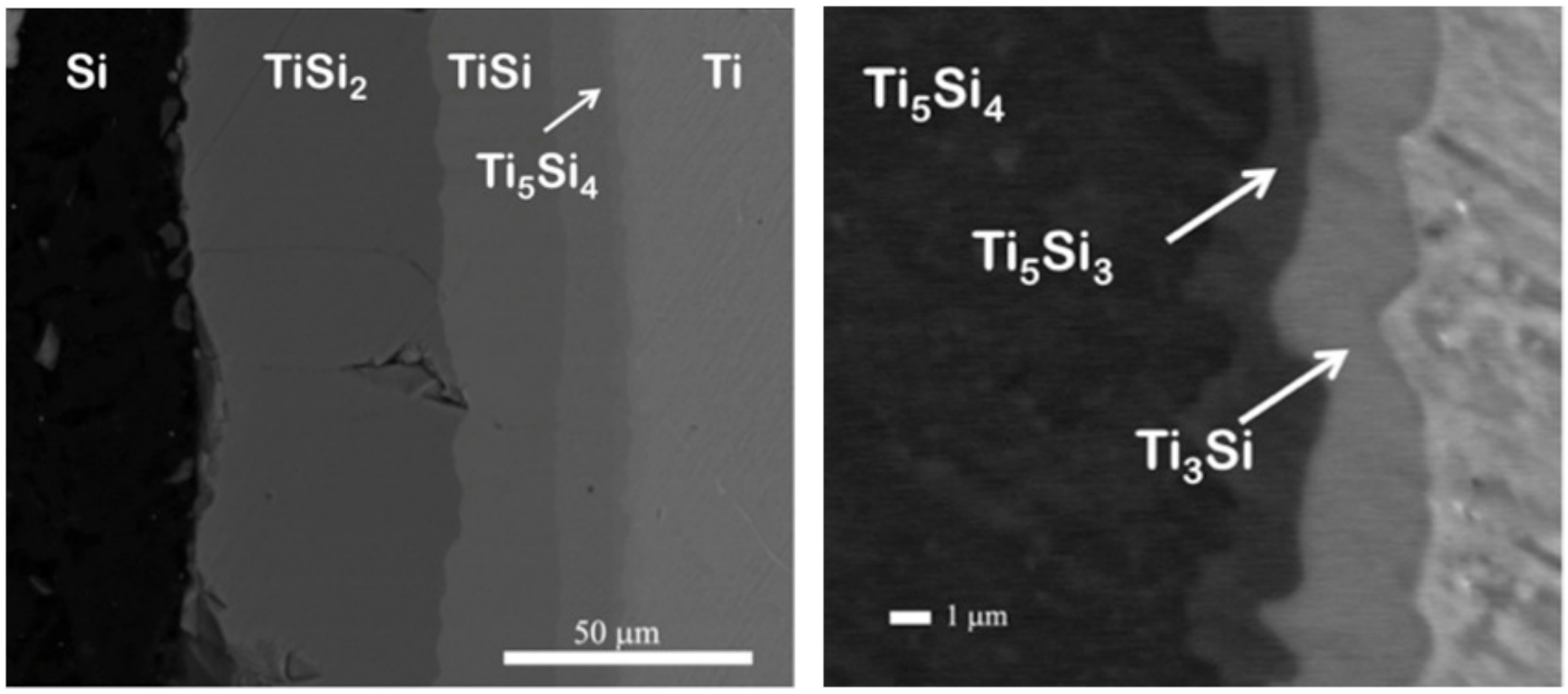

Figure 2 Scanning electron micrograph of Ti/Si diffusion couple annealed at 1200 °C for 16 hrs and the magnified image of the Ti-rich part showing the presence of $Ti_5Si_3$ and $Ti_3Si$ phases [4].

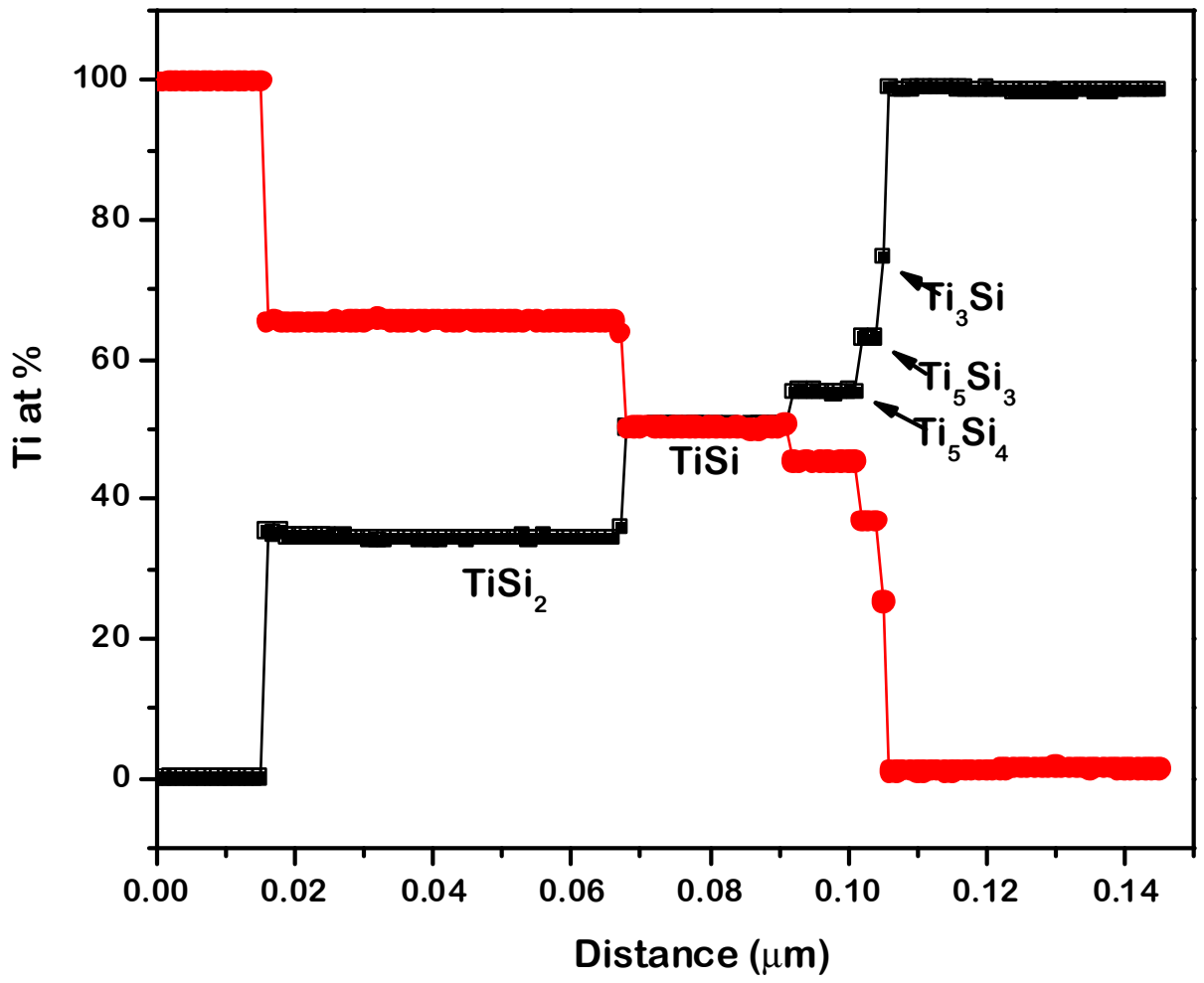

Figure 3 The composition profile of the interdiffusion zone developed at 1200 °C [4].

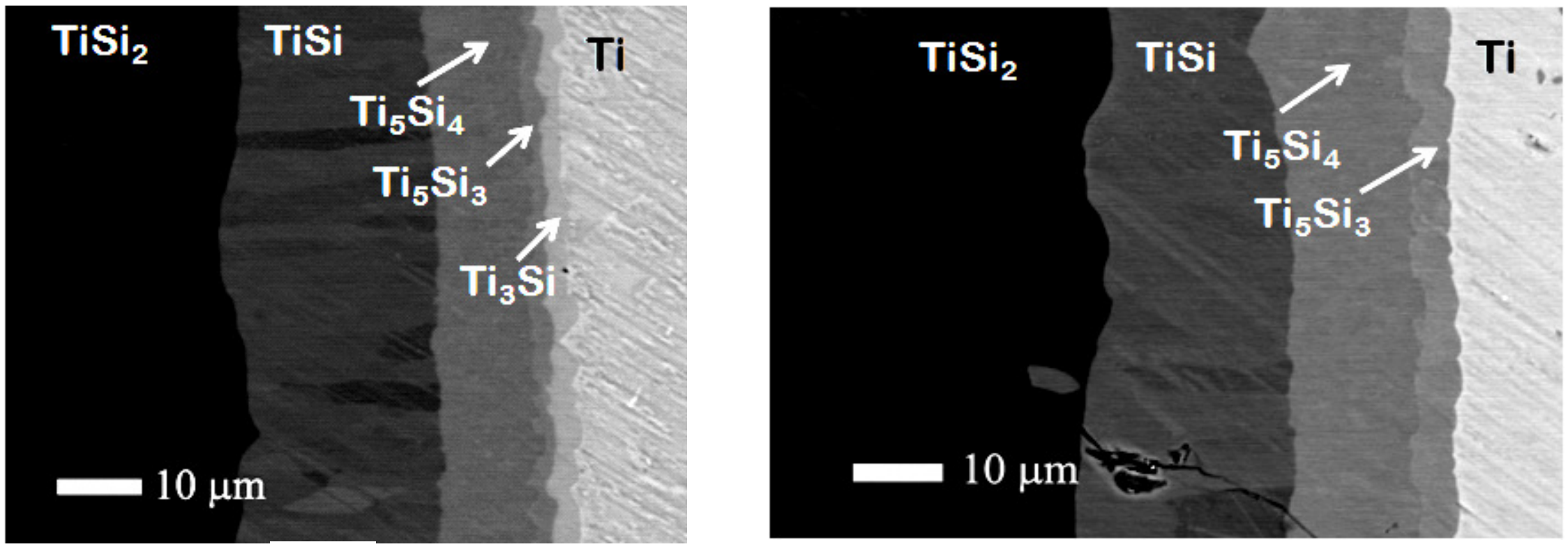

(a) (b)

Figure 4 (a) Magnified image of the Ti/Si diffusion couples annealed at (a) 1225 °C for 16 hrs and (b) 1250 °C for 16 hrs showing the presence of Ti-rich phases [4].

The location of the Kirkendall marker plane is detected by the microstructural evolution at the Si/$TiSi_2$ interface [4]. It should be noted here that microstructure analysis is very handy to locate the position of this plane in many systems as discussed in Reference [15]. As a mandatory step, the time-dependent experiments should be conducted before the quantitative diffusion analysis. In fact, there are many studies available reporting the time-dependent parabolic growth of the product phase $TiSi_2$ and therefore, there was no need to repeat the similar experiments. For the estimation of the activation energy for growth, the temperature dependent experiments were conducted. Table 1 shows the thickness of the phase layers *i.e.* $TiSi_2$, TiSi and $Ti_5Si_4$, which grow with measurable thickness. Surprisingly, two phases $TiSi_2$ and $Ti_5Si_4$ show a systematic increase in layer thickness with the increase in temperature; however, this change is negligible for the phase TiSi.

It is well understood that when only one phase layer grows in the interdiffusion zone, it is not affected by the growth of other phases. On the other hand, the growth kinetics is affected when it grows along with other phases. In fact, the diffusion coefficients are the material constants and the thickness of a phase is grown according to end member compositions leading to single phase or multiphase growth. Therefore, it is possible that the thickness of TiSi does not change significantly because of a certain way of change in diffusion parameters of the phase of interest and the other phases. This can be checked by the parabolic growth of the phase layer. A similar behaviour was found in the Ti-Al system [16]. To investigate this further, the time-dependent experiments were conducted at 1200 °C. It is evident from $\Delta x^2$ vs. time $t$ in Figure 5 that all the phases grow parabolically with time. These are also listed in Table 2.

Table 1 Thicknesses of phases grown at different temperatures after 16 hrs of annealing [4]

| Temperature (°C) | $\Delta$x $TiSi_2$ (μm) | $\Delta$x TiSi (μm) | $\Delta$x $Ti_5Si_4$ (μm) |
|---|---|---|---|
| 1150 | 39 ± 1.1 | 23 ± 0.5 | 9 ± 0.3 |
| 1175 | 43 ± 0.6 | 24 ± 0.4 | 12 ± 0.3 |
| 1200 | 52 ± 0.8 | 24 ± 0.6 | 14 ± 0.4 |
| 1225 | 60 ± 0.5 | 25 ± 0.6 | 15 ± 0.7 |
| 1250 | 67 ±0.9 | 25 ± 0.7 | 18 ± 0.4 |

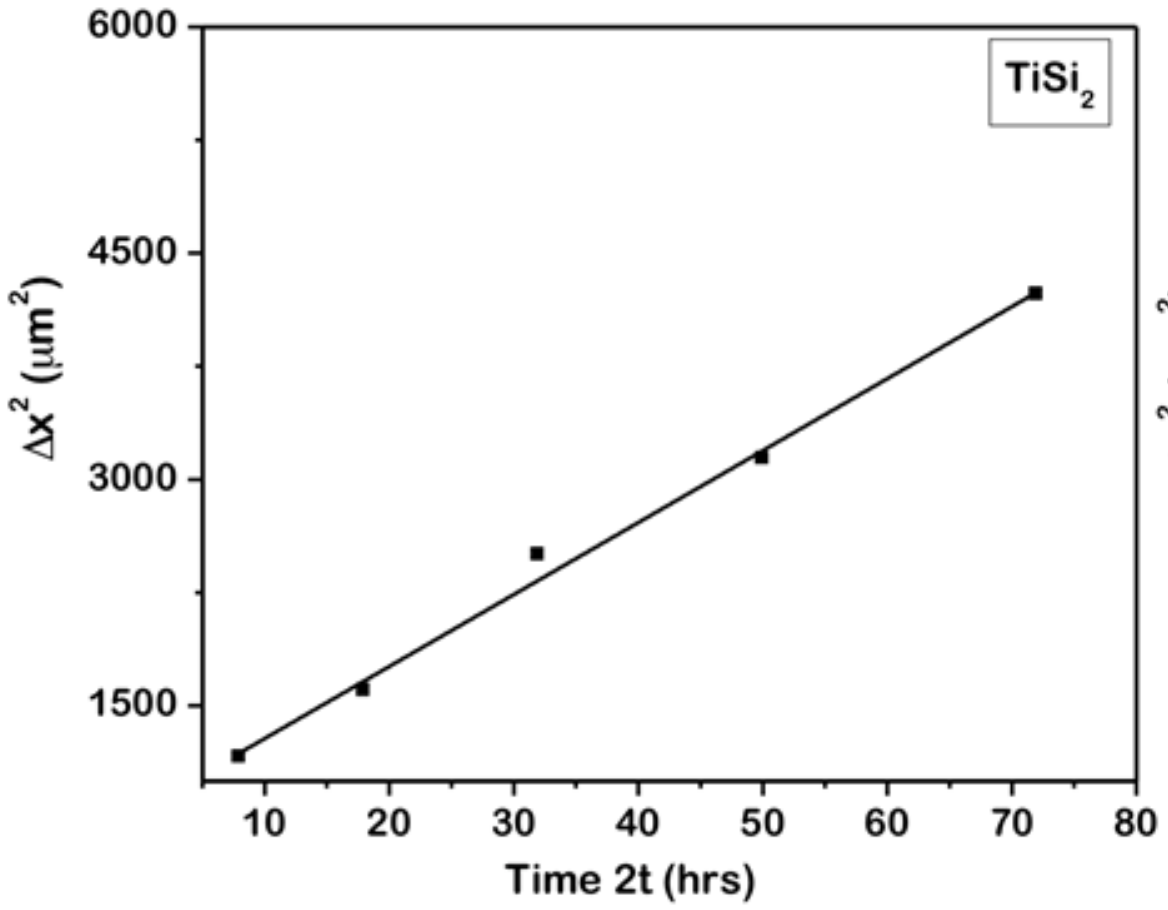

(a)

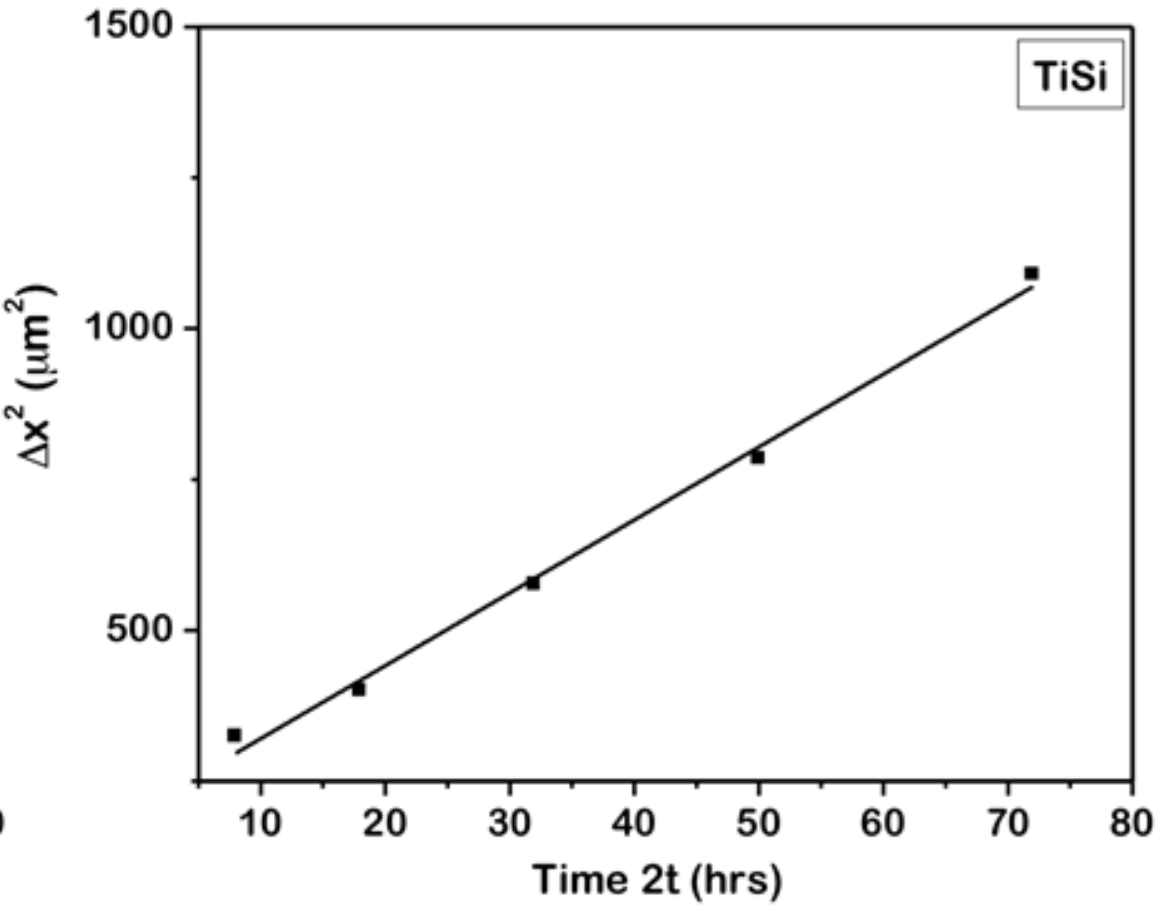

(b)

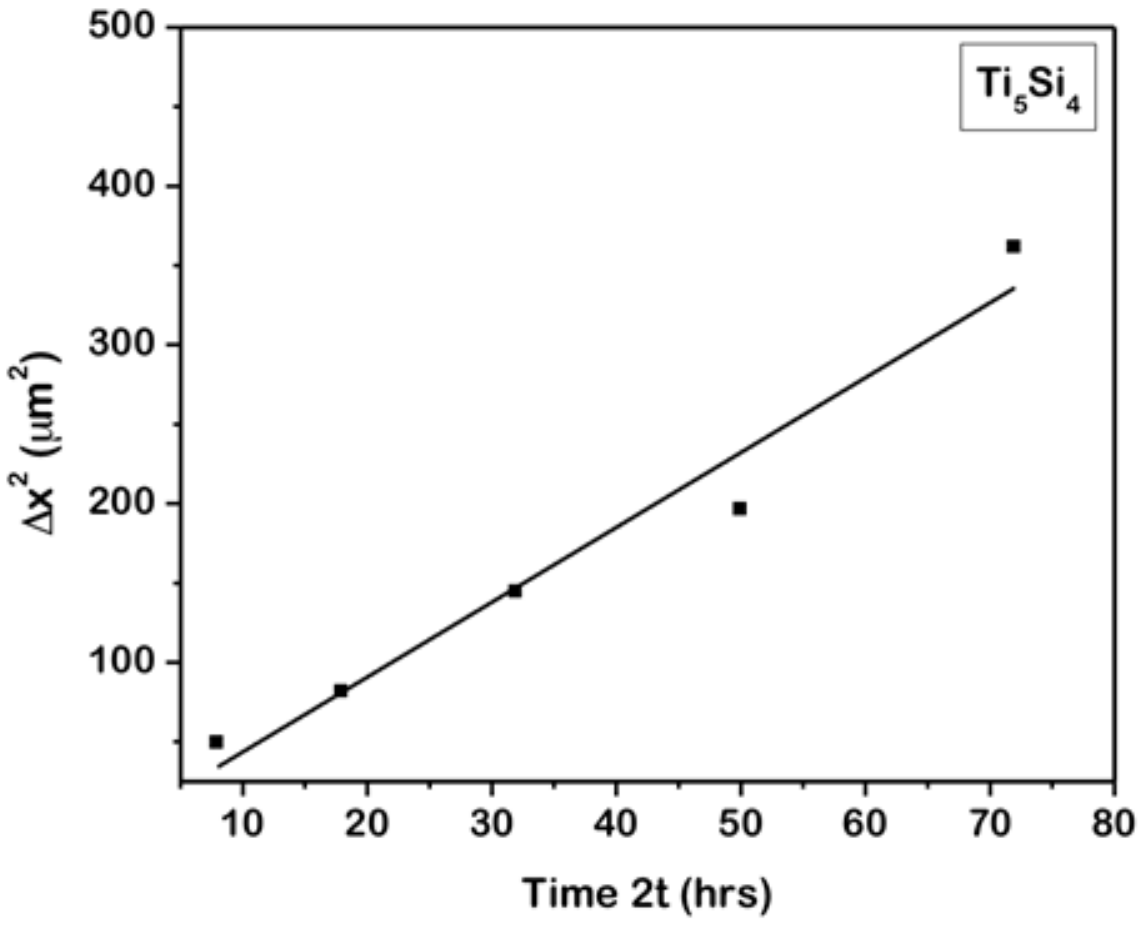

(c)

Figure 5 Time dependant experiments at 1200 °C for 4, 9, 16, 25 and 36 hrs are shown for (a) $TiSi_2$ (b) TiSi and (c) $Ti_5Si_4$ [4].

Table 2 Thickness of phases annealed for different times at 1200 °C [4].

| Time (hrs) | Δx $TiSi_2$ (μm) | Δx TiSi (μm) | Δx $Ti_5Si_4$ (μm) |
|---|---|---|---|
| 4 | 34 ± 1.1 | 18 ± 0.7 | 7 ± 0.4 |
| 9 | 40 ± 0.6 | 20 ± 0.4 | 9 ± 0.3 |
| 16 | 50 ± 0.8 | 24 ± 0.6 | 12 ± 0.4 |
| 25 | 56 ± 1 | 28 ± 0.8 | 14 ± 0.3 |
| 36 | 65 ± 0.5 | 33 ± 0.9 | 19 ± 0.3 |

The integrated diffusion coefficient $\tilde{D}_{\text{int}}$ of a phase β in a multiphase interdiffusion zone can be expressed as [1]

$$\tilde{D}_{\text{int}}^{\beta} = \int_{N_i'}^{N_i''} \tilde{D}^{\beta} dN_i , \quad (1)$$

where $\tilde{D}^{\beta}$ ($m^2/s$) is the interdiffusion coefficient of the phase β, $N_i$ is the mole fraction of component $i$ and $N_i'$ and $N_i''$ are the mol fractions of the phase boundaries. Since the composition profiles developed in the solid solution phases are negligible, $\tilde{D}_{\text{int}}$ can be calculated directly from the composition profile utilizing [1]

$$\tilde{D}_{\text{int}}^{\beta} = \frac{\left(N_i^{\beta} - N_i^{-}\right)\left(N_i^{+} - N_i^{\beta}\right)}{N_i^{+} - N_i^{-}} \frac{\Delta x_{\beta}^{2}}{2t} + \frac{\Delta x_{\beta}}{2t}\left[\frac{\left(N_i^{+} - N_i^{\beta}\right)\sum_{\nu=2}^{\nu=\beta-1} \frac{V_m^{\beta}}{V_m^{\nu}}\left(N_i^{\nu} - N_i^{-}\right)\Delta x_{\nu} + \left(N_i^{\beta} - N_i^{-}\right)\sum_{\nu=\beta+1}^{\nu=n-1} \frac{V_m^{\beta}}{V_m^{\nu}}\left(N_i^{+} - N_i^{\nu}\right)\Delta x_{\nu}}{N_i^{+} - N_i^{-}}\right] \quad (2)$$

where $N_i^{-}$ and $N_i^{+}$ are the mole fractions of component $i$ in the unreacted left and right hand side of the ends of the couple, respectively, with respect to element $i$, $N_i^{\beta}$ is the mol fraction of component $i$ in the phase of interest $v$, $V_m^{\nu}$ and $\Delta x_{\nu}$ are the molar volume and the layer thickness of the $\nu$th phase and $t$ is the annealing time.

It should be noted here that the first part of Equation 2 is related to the phase of interest and the second part is related to the growth of other phases. This is the reason that there is a reasonable variation of the integrated diffusion coefficient irrespective of the fact there is not much variation in the thickness of TiSi with temperature. The molar volumes of the phases $TiSi_2$, TiSi, Ti5Si4 are 8.41, 8.92 and 9.18 $cm^3$/mol. The calculated data are plotted in Figure 6. These are not calculated for $Ti_5Si_3$ and $Ti_3Si$ because of very small thickness. The activation energy for $TiSi_2$ is estimated as 190±9 kJ/mol, which is similar to the values reported earlier. It should note here that the activation energy of parabolic growth constant is the same as the activation energy of the integrated diffusion coefficient when only a single phase layer grows in an interdiffusion zone [1].

It is already mentioned earlier that the location of the Kirkendall marker plane is found to be at Si/$TiSi_2$ interface. Therefore, it is evident that this phase grows through the diffusion of Si. However, we can estimate the ratio of the intrinsic diffusion coefficients of the components following [1, 2]

$$\frac{V_M D_{Si}}{V_{Si} D_M} = \frac{D^*_{Si}}{D^*_M} = \frac{\left[ N^+_{Si} \int_{x^{-\infty}}^{x_K} \left(N_{Si} - N^-_{Si}\right) dx - N^-_{Si} \int_{x_K}^{x^{+\infty}} \left(N^+_{Si} - N_{Si}\right) dx \right]}{\left[ -N^+_M \int_{x^{-\infty}}^{x_K} \left(N_{Si} - N^-_{Si}\right) dx + N^-_M \int_{x_K}^{x^{+\infty}} \left(N^+_{Si} - N_{Si}\right) dx \right]} \tag{3}$$

where $D_i$ and $D_i^*$ are the intrinsic and the tracer diffusion coefficients of element *i*. $x_K$ is the Kirkendall marker plane location. $x^{-\infty}$ and $x^{+\infty}$ correspond to the unaffected ends of the diffusion couple. M stands for the metal component. This relation does not consider the vacancy-wind effect [35]. It is generally assumed as unity when we cannot estimate it. Additionally, we cannot estimate the partial molar volumes and therefore, we actually estimate the ratio of the tracer diffusion coefficients. Since the marker plane is found at the Si/$TiSi_2$ interface, we have this value as $\frac{D^*_{Si}}{D^*_{Ti}} = \infty$. It further means that the diffusion rate of the metal component is not zero but negligible compared to the diffusion rate of Si. Such an analysis is already established by diffusion estimation in the Mo-Si system [10].

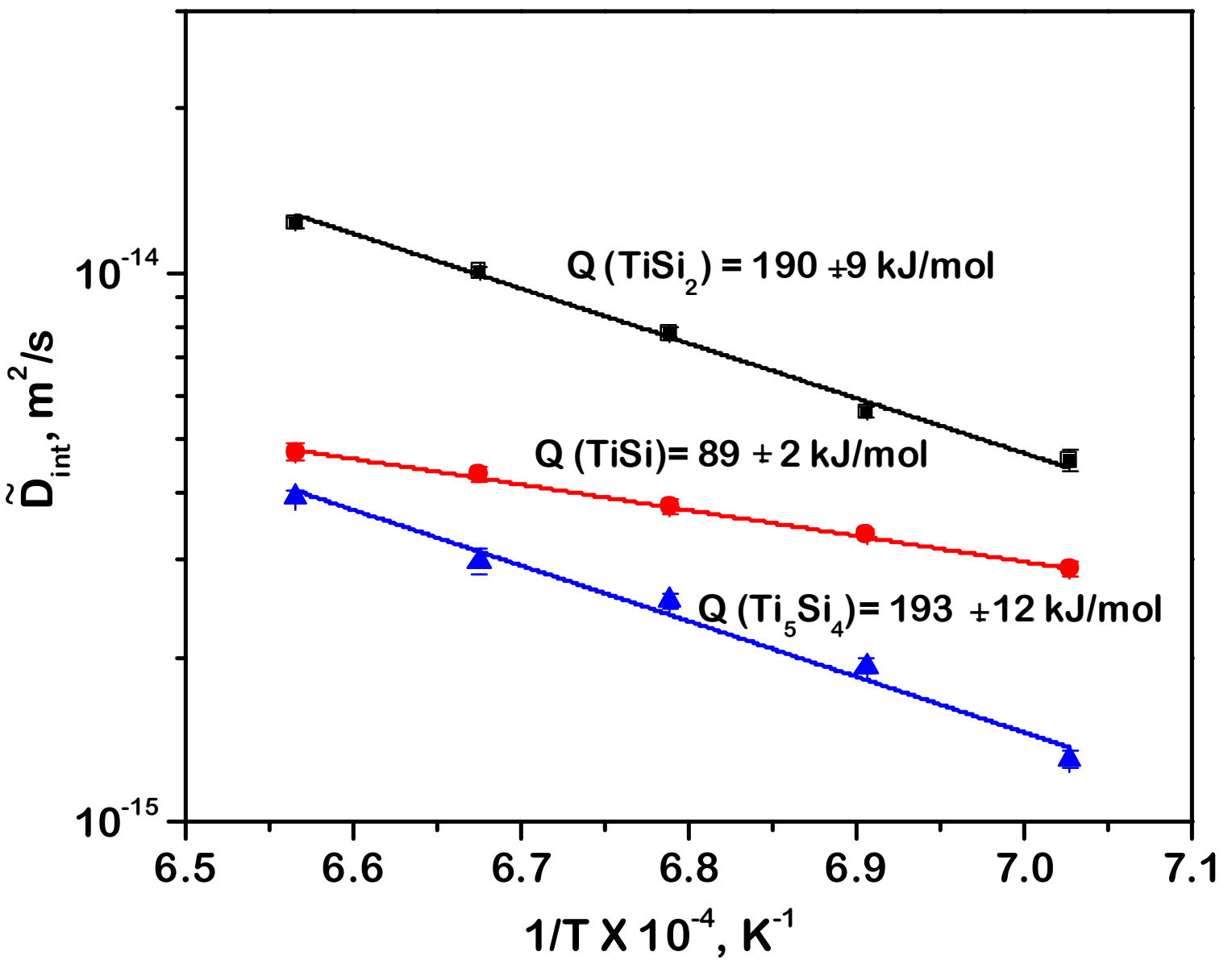

Figure 6 The integrated diffusion coefficients for $TiSi_2$, TiSi and $Ti_5Si_4$ plotted with respect to the Arrhenius equation [4].

Utilizing the knowledge of the ratio of diffusivities, one can even estimate the absolute values of the tracer diffusion coefficients if the necessary thermodynamic parameters are known [1] following

$$\begin{aligned} \tilde{D}^{\beta}_{int} &= -\left(N_M D^*_{Si} + N_{Si} D^*_M\right) \frac{N_M\left(\mu^I_M - \mu^{II}_M\right)}{RT} = -\left(N_M D^*_{Si} + N_{Si} D^*_M\right) \frac{N_{Si}\left(\mu^{II}_{Si} - \mu^I_{Si}\right)}{RT} \\ &= -\left(N_M D^*_{Si} + N_{Si} D^*_M\right) \frac{N_M \Delta_d G_M}{RT} = -\left(N_M D^*_{Si} + N_{Si} D^*_M\right) \frac{N_{Si} \Delta_d G_{Si}}{RT} \end{aligned} \tag{4}$$

$\mu_i$ is the chemical potential of element *i*, and $\Delta_d G_{Si}$ is the driving force for the diffusion of Si in the $TiSi_2$ phase. It should be noted that $N_M \Delta_d G_M = N_{Si} \Delta_d G_{Si}$ (the Gibbs-Duhem relation). Since the diffusion rate of Ti is negligible, the tracer diffusion coefficient of Si can be estimated by

$$\tilde{D}_{int}^{TiSi_2} = -N_M D_{Si}^* \frac{N_{Si}\Delta_d G_{Si}}{RT} \quad (5)$$

Note, here, that the driving forces should be estimated considering an incremental diffusion couple in which the product phase grows from the next neighbouring phases in the phase diagram [1], which is explained in Figure 7. These are estimated utilizing the data available in Refs. [17, 18], which are listed in Table 3. The temperature dependent Arrhenius plot of the tracer diffusion coefficient of Si $D_{Si}^*$ is shown in Figure 8. The activation energy is estimated at 197±8 kJ/mol.

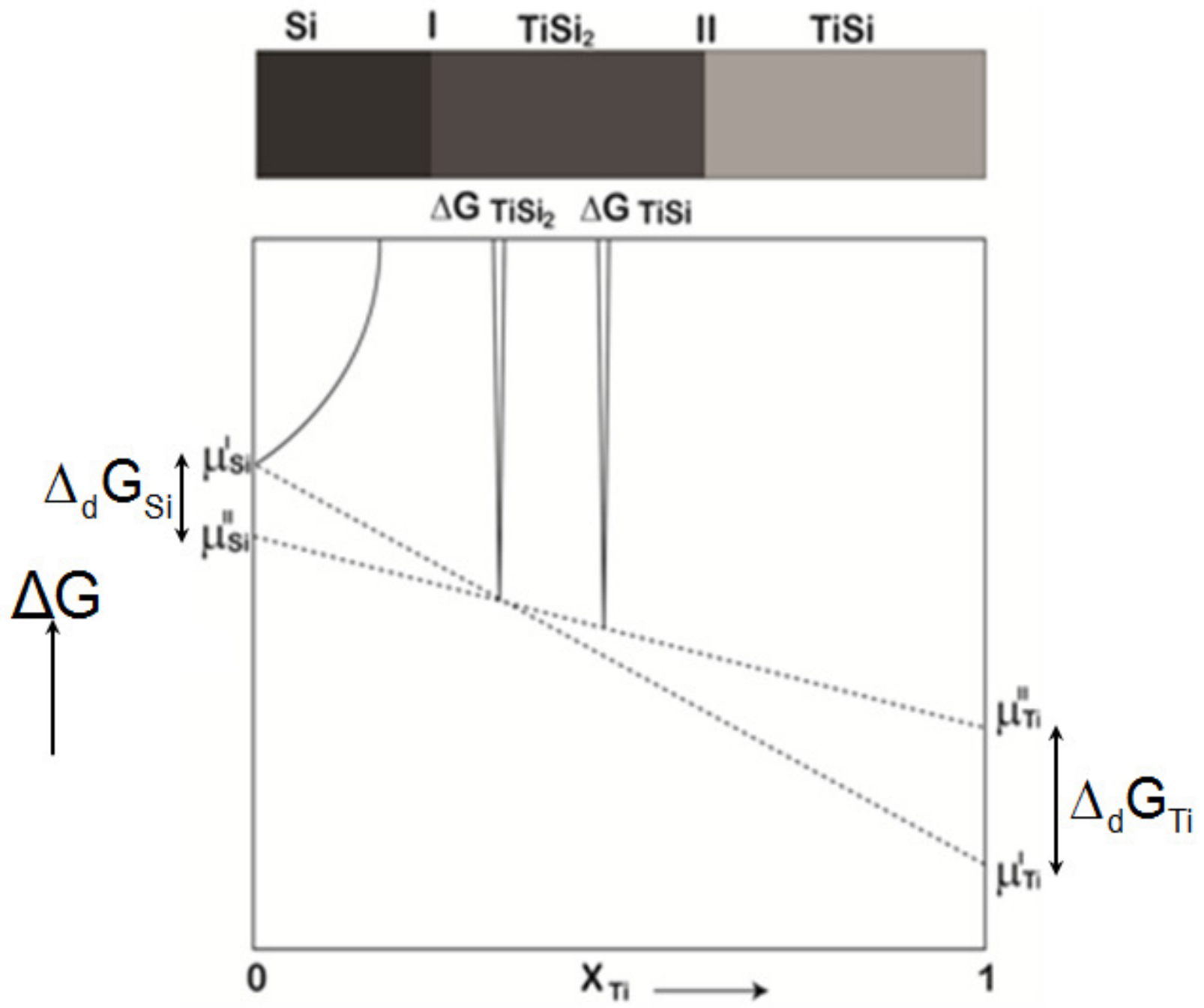

Figure 7 Schematic representation of the estimation of the driving forces for the diffusion-controlled growth of the $TiSi_2$ phase [4].

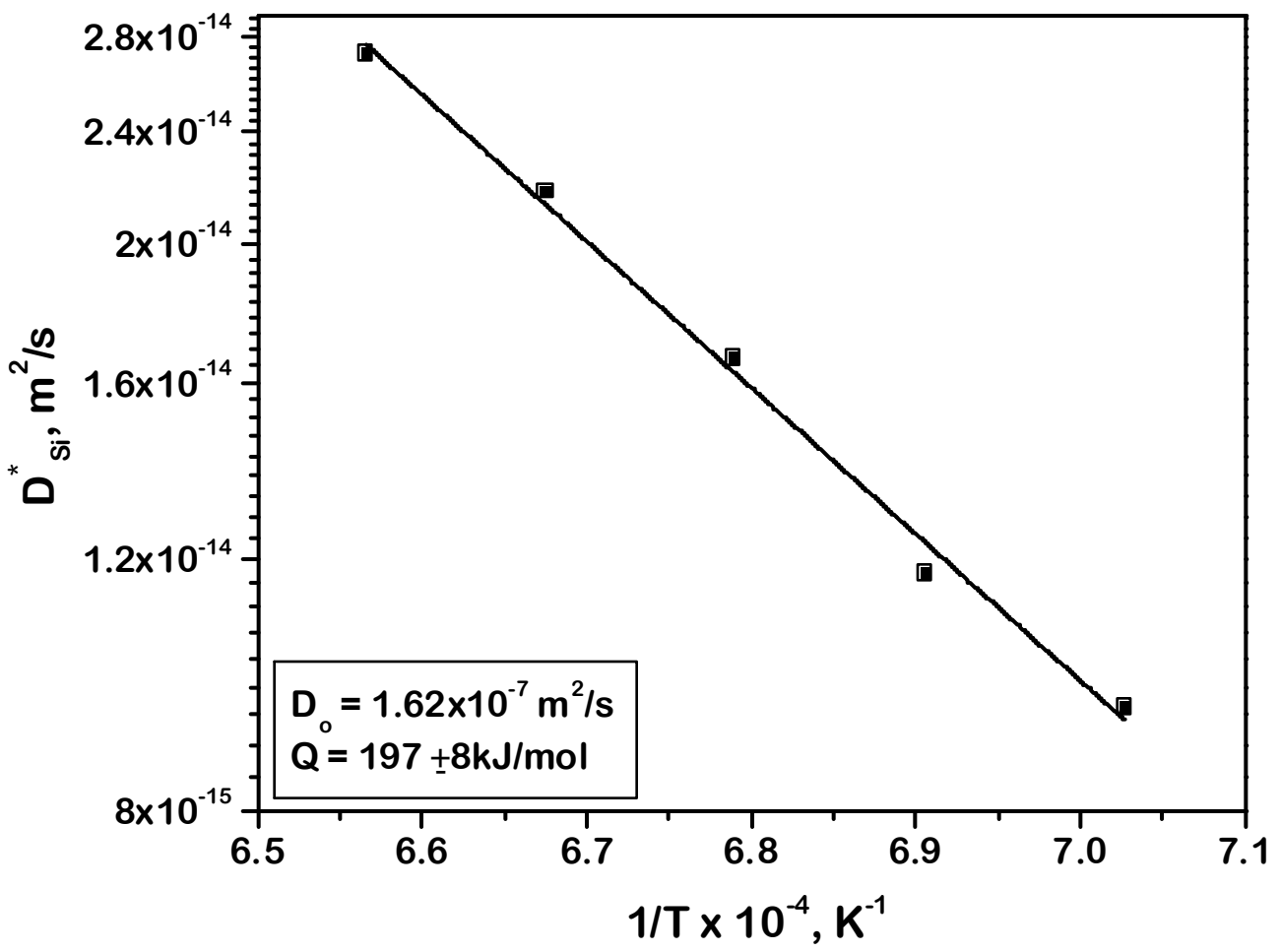

Figure 8 Temperature dependent Arrhenius plot of the Si tracer diffusion coefficients in the $TiSi_2$ phase [4].

Table 3 Free energy of the $\Delta G_{si}$ (diamond), $\Delta G$ ($TiSi_2$), $\Delta G$ (TiSi) and $\Delta_d G_{Si}$ at the temperature of our interest [4].

| T °C | $\Delta G_{si}$ J/mol | $\Delta G$ ($TiSi_2$) J/mol | $\Delta G$ (TiSi) J/mol | $\Delta_d G_{Si}$ J/mol |
|---|---|---|---|---|
| 1250 | -58329.1 | -119422.6 | -137100.6 | -25900 |
| 1225 | -56863.6 | -117889.8 | -135536.6 | -25900 |
| 1200 | -55410.1 | -116370.2 | -133986.4 | -25800 |
| 1175 | -53968.5 | -114863.9 | -132450.2 | -25800 |
| 1150 | -52539.2 | -113371.2 | -130928.1 | -25800 |

## 2.2 Interdiffusion study in the Zr-Si system

The diffusion-controlled growth studies in the Zr/Si system are very limited [5, 17, 18]. In one of the studies [19], only the growth process was discussed without any analysis of the diffusion coefficients. As can be seen in the Zr-Si phase diagram, there are six intermetallic compounds, which are expected to grow in the interdiffusion zone. However, only two phase ($ZrSi_2$ and ZrSi) were reported to grow in Ref. [19]. Other phases were found to grow only after the consumption of Si indicating a much lower growth rate of this phases. The same was found by Roy and Paul [5], which is described in detail here.

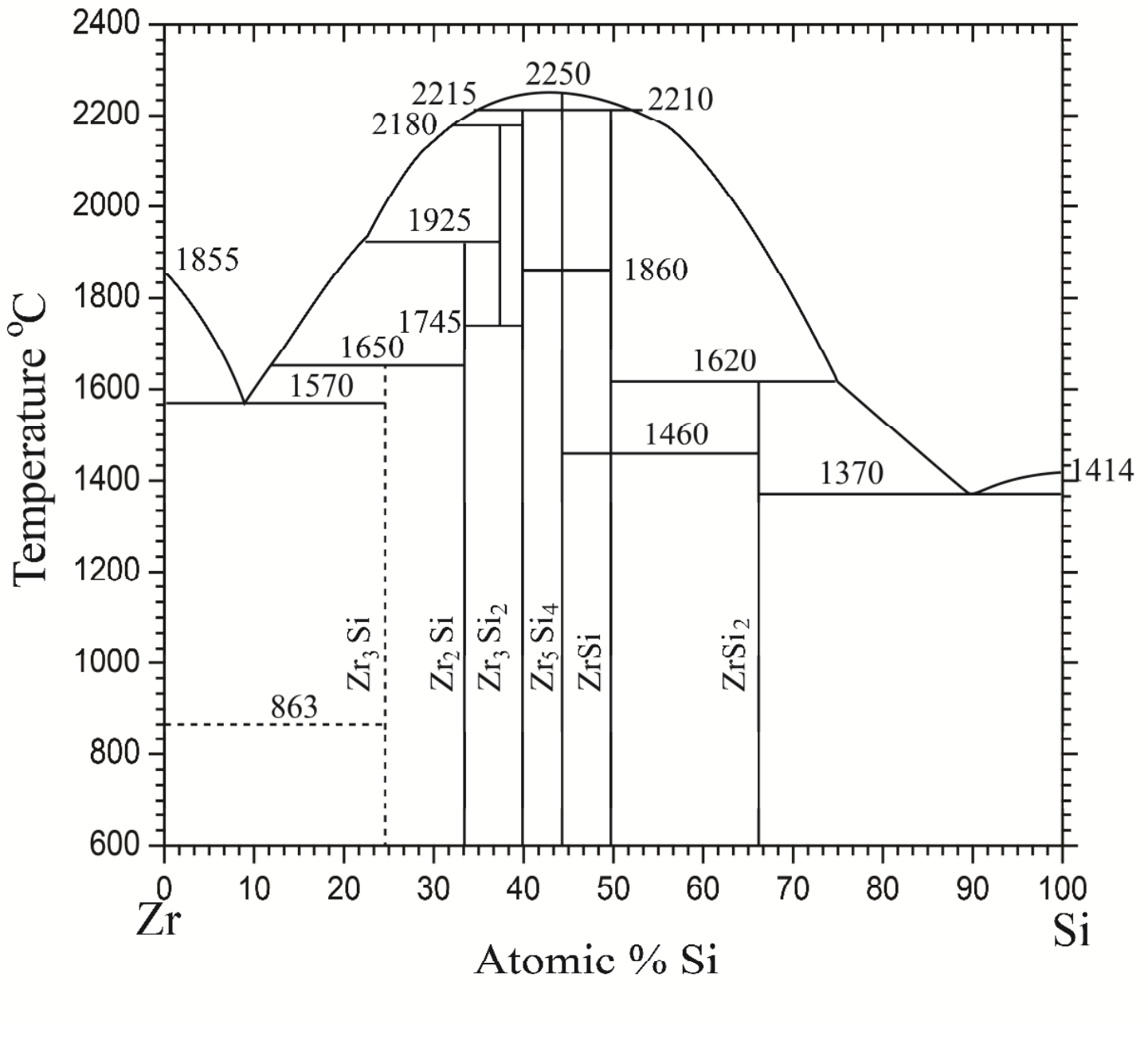

Figure 9 Zr-Si phase diagram [5].

The Backscattered electron image (BSE) of Zr/Si diffusion couple annealed at 1200 °C annealed for 16 hrs is shown in Figure 10. Indeed only two phases, the disilicide and monosilicide, are found in the interdiffusion zone. Since the thickness of ZrSi is very small, the presence of this phase is confirmed by X-ray diffractions, as shown in Figure 11. The thickness of the disilicide phase is found to be much higher compared to the monosilicide phase. The time-dependent experiments are conducted to find the parabolic nature of the growth, as shown in Figure 12.

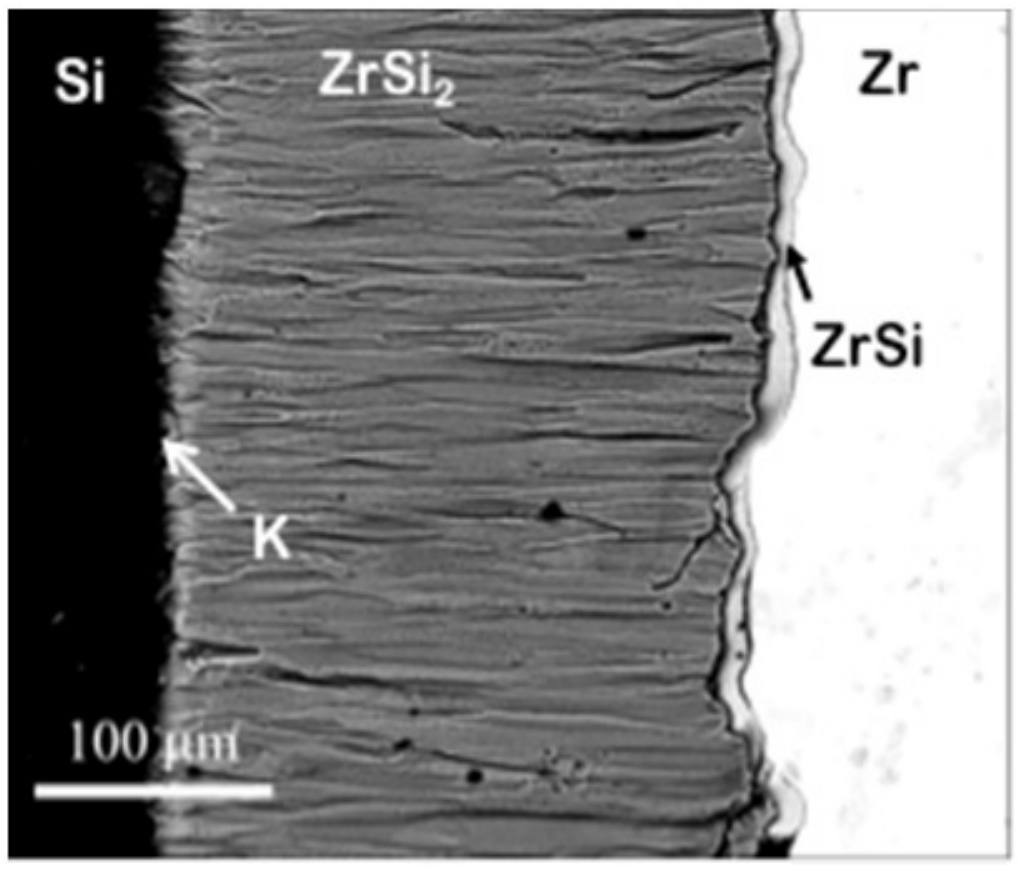

Figure 10 Zr-Si interdiffusion zone annealed 1200 °C for 16 hrs [5].

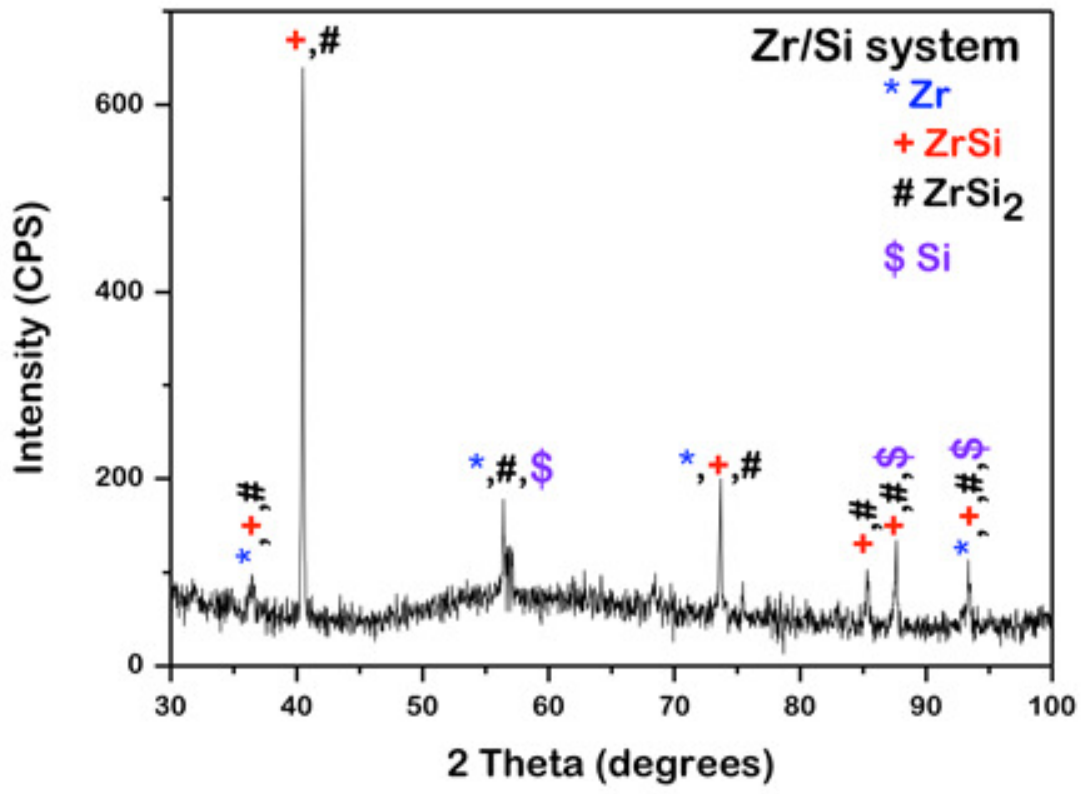

Figure 11 X-ray diffraction pattern of the Zr/Si diffusion couple identifying the phase layers [5].

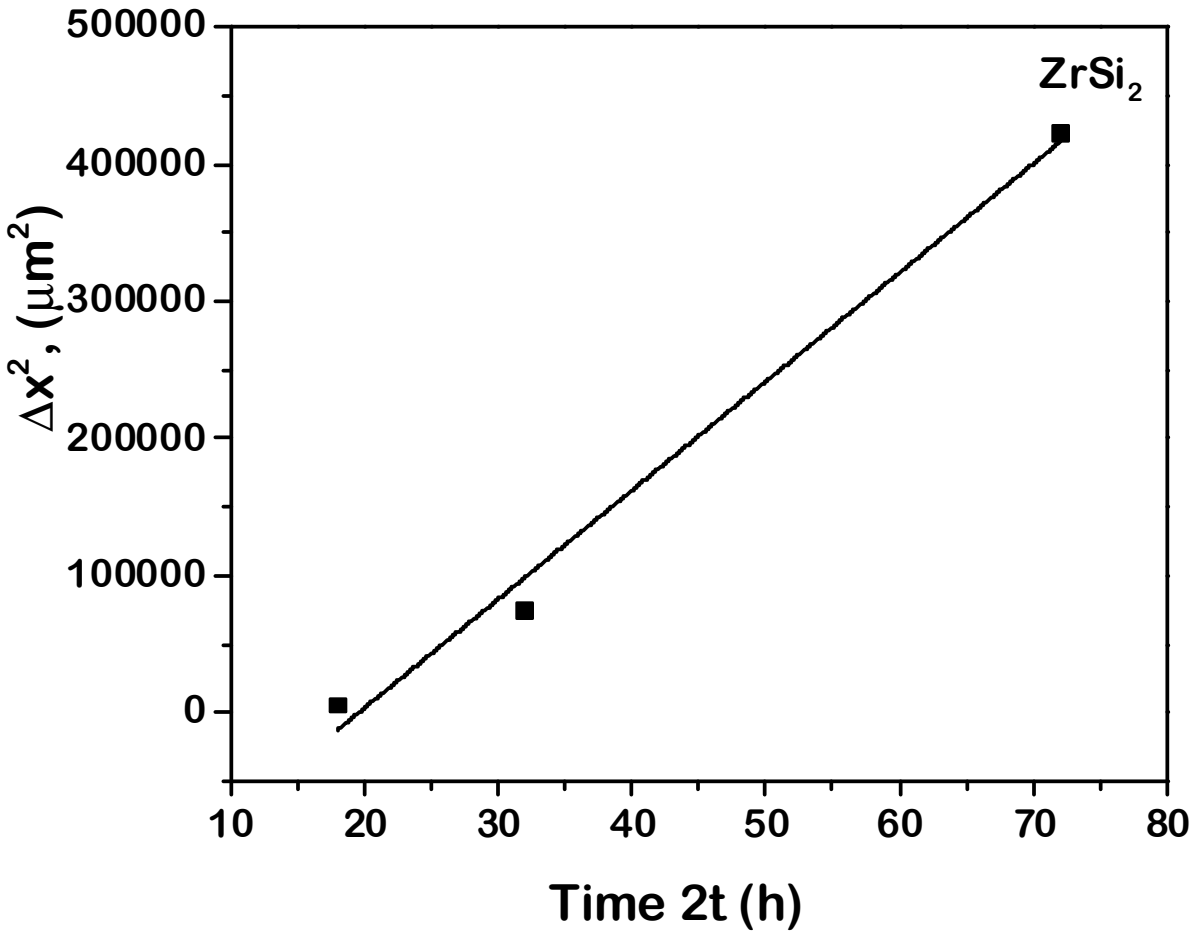

Figure 12 The time-dependent growth of ZrSi2 phase at 1200 °C [5].

The integrated diffusion coefficients are estimated following the Equation 2. The composition profile in Figure 13 indicates negligible diffusion profile in the pure end members and therefore can be neglected for the calculation. The molar volume of $ZrSi_2$ is 10.03 $cm^3$/mol. The estimated values are plotted with respect to the Arrhenius equation in Figure 14. The Kirkendall marker experiments by using inert markers could not be done in this system because of joining issue of the diffusion couple. However, the location of this plane could be detected by analysing the microstructure in the $ZrSi_2$ phase. It is expected to be present at one of the interfaces. Since Si is the faster diffusing species in the disilicides, we can safely consider the location of the plane at the Si/$ZrSi_2$ interface. As it is described in the $TiSi_2$ phase, the tracer diffusion coefficients in this phase also can be estimated using the thermodynamic parameters as listed in Table 4. The estimated tracer diffusion coefficients are plotted with respect to the Arrhenius equation in Figure 15.

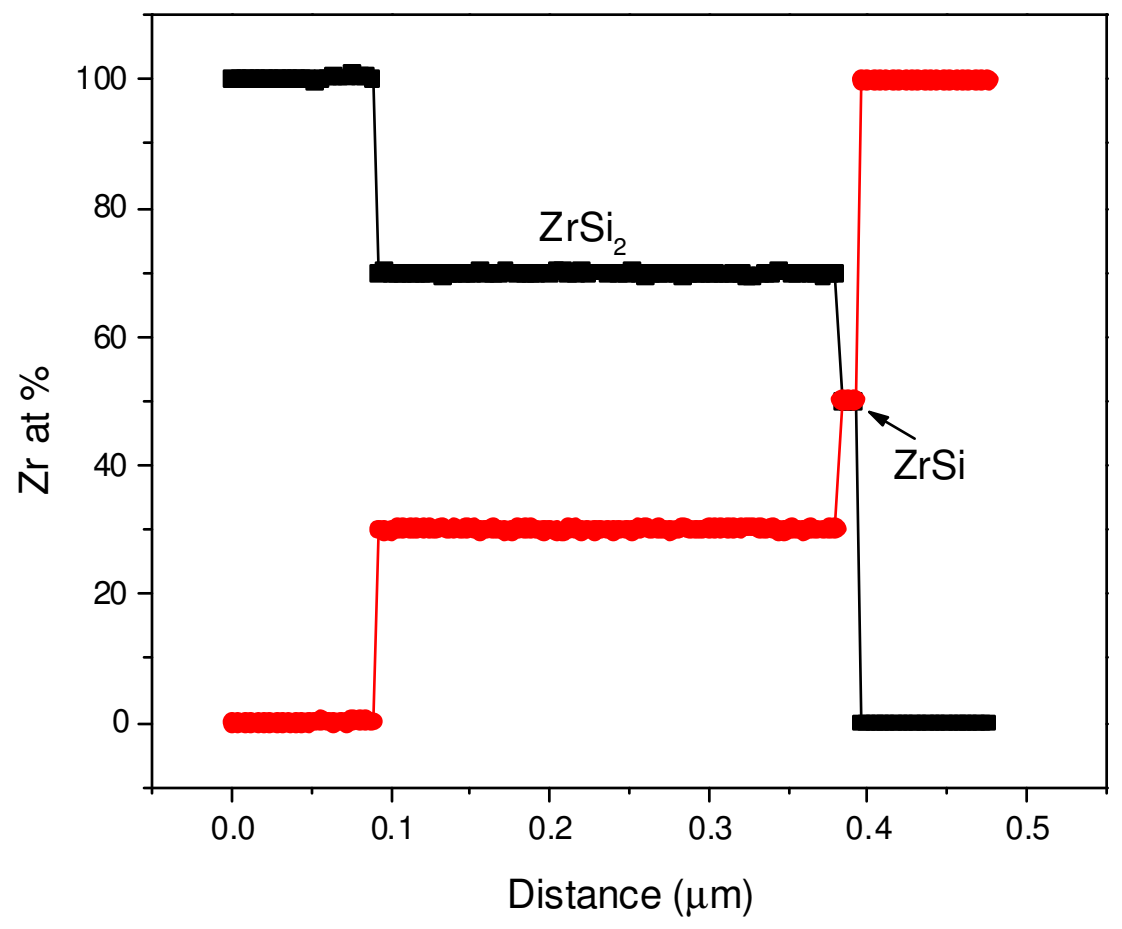

Figure 13 Composition profile of Zr/Si diffusion couple annealed at 1200 °C for 16 hrs [5].

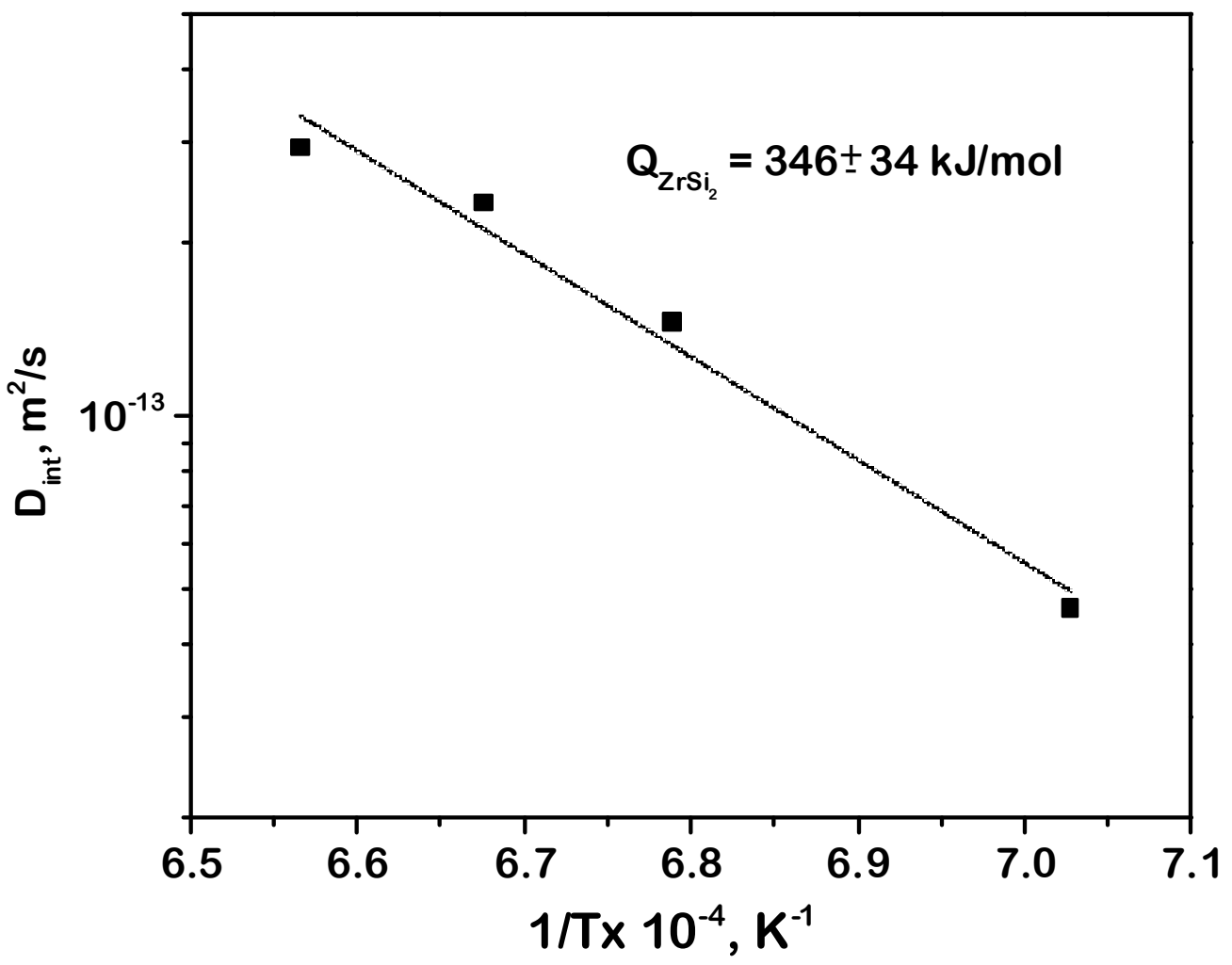

Figure 14 The integrated diffusion coefficients estimated for $ZrSi_2$ [5].

Table 4 Thermodynamic parameters related to $ZrSi_2$ [5].

| T (°C) | $\Delta G_{si}$ (diamond) J/mol | $\Delta G$ ($ZrSi_2$) J/mol | $\Delta G$ (ZrSi) J/mol | $\Delta_d G_{Si}$ J/mol |
|---|---|---|---|---|
| 1250 | -58329.1 | -306867.3 | -308440.38 | -245407 |
| 1225 | -56863.61 | -303516.33 | -305278.8 | -243136 |
| 1200 | -55410.1 | -300189.7 | -302141.56 | -240887 |
| 1150 | -52539.2 | -293610.6 | -295941.16 | -236424 |

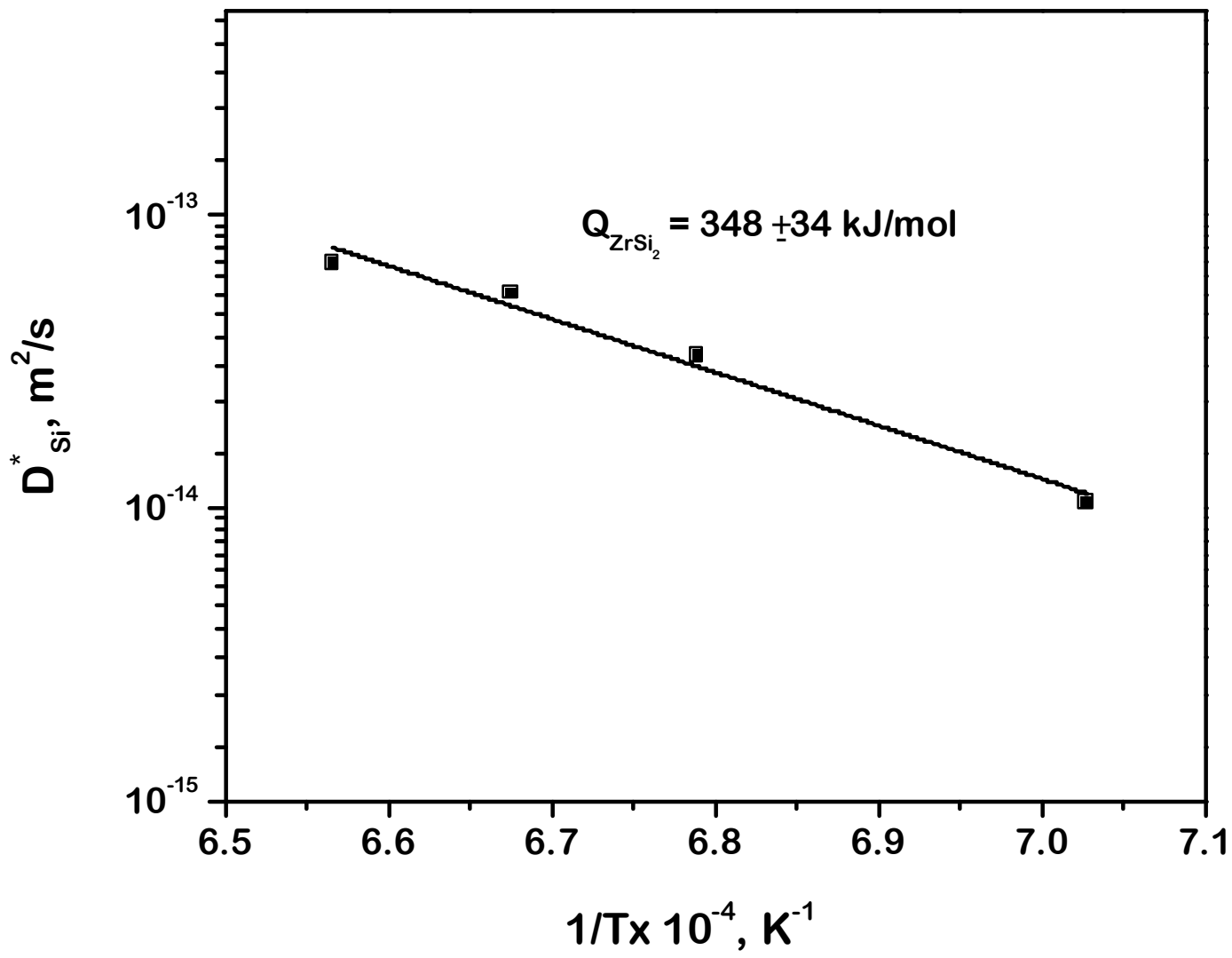

Figure 15 Tracer diffusion coefficients of Si in ZrSi2 [5].

## 2.3 Interdiffusion study in the Hf-Si system

Few studies on the growth of the phases in thin films are already been made in the Hf/Si system [20-23]. As shown in the phase diagram in Figure 16, the total five phases are present in this system and therefore all of them are expected to grow [24]. However, only two phases $HfSi_2$ and HfSi are reported to grow. Additionally, the sequential growth instead of simultaneous growth is also not so uncommon in thin film diffusion couples [25, 26]. The same phenomena are reported in this system also. The monosilicide phase HfSi is reported to grow first before the growth of the disilicide phase $HfSi_2$ at the interface of Si/$HfSi_2$ interface by nucleating on HfSi [23, 27]. The phase is found to grow parabolically with time indicating the diffusion controlled process [27]. Ziegler et al. [23] reported the activation energy for growth of the $HfSi_2$ phase as 2.5 eV (241 kJ/mol). They also reported Si as the faster-diffusing component. So et al. [27] studied the and calculated the activation energy as 3.5±0.3 eV. (337±29 kJ/mol). However, no diffusion studies are conducted in the bulk systems, which give better opportunities for quantitative analysis. Roy and Paul conducted [5] this study in detail and estimated the integrated diffusion coefficient of the phases and the tracer diffusion coefficient of Si in $HfSi_2$. In this section, these studies are summarized in detail.

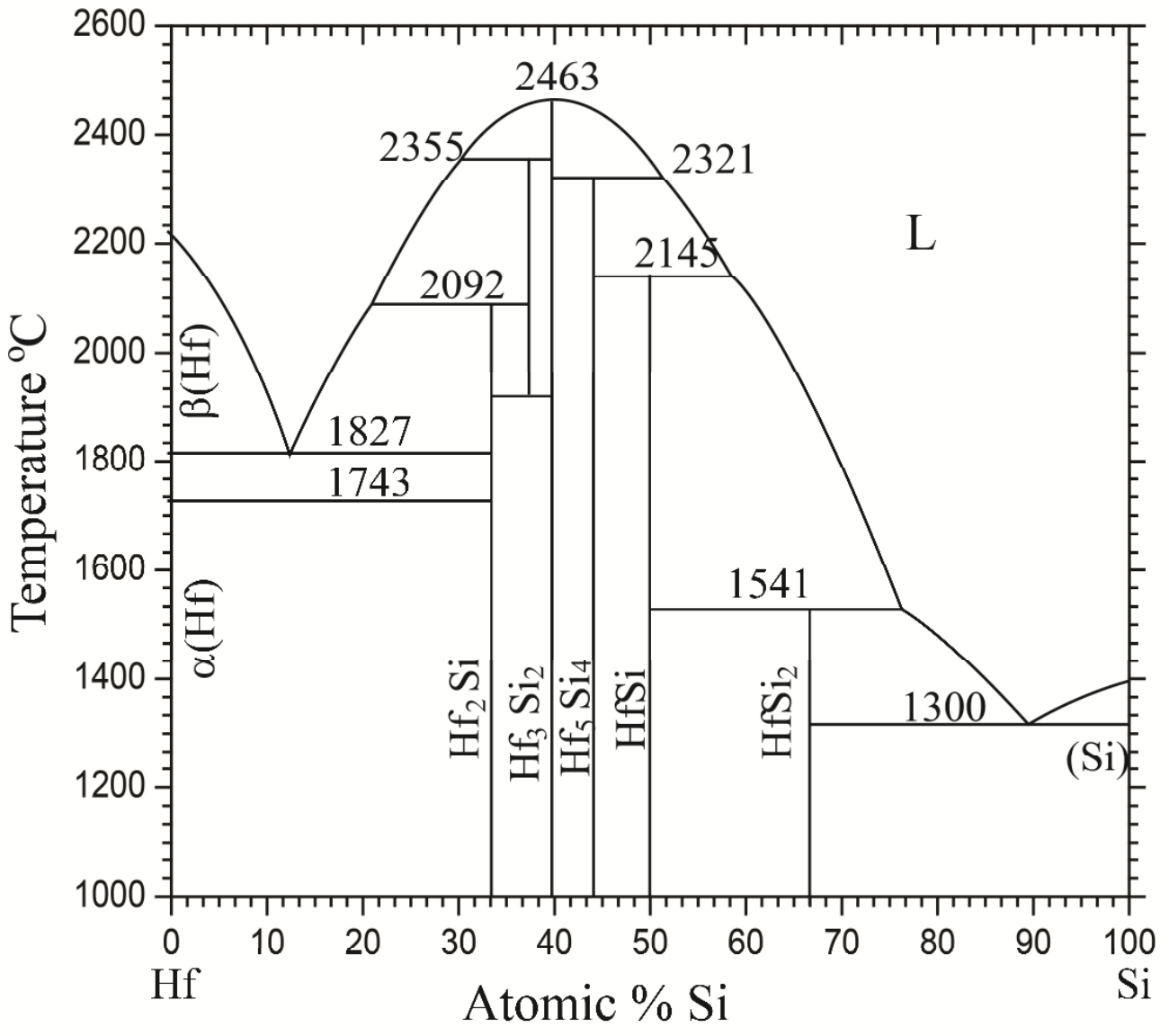

Figure 16 The Hf-Si phase diagram [5].

The BSE image of the interdiffusion zone of Hf/Si annealed at 1250 °C for 16 hrs is shown in Figure 17. The composition profile and X-ray diffractions identify the growth of two-phase $HfSi_2$ and HfSi, as shown in Figures 18 and 19. The disilicide phase grows with much higher thickness compared to the monosilicide phase. The location of the Kirkendall marker plane is identified at the Si/$HfSi_2$ interface. This indicates that the product phase grows by the diffusion of Si and the diffusion rate of Hf is negligible through $HfSi_2$. It should be noted here that we cannot comment on the relative mobilities of the components in the HfSi phase since we do not know the location of the Kirkendall marker plane in this phase.

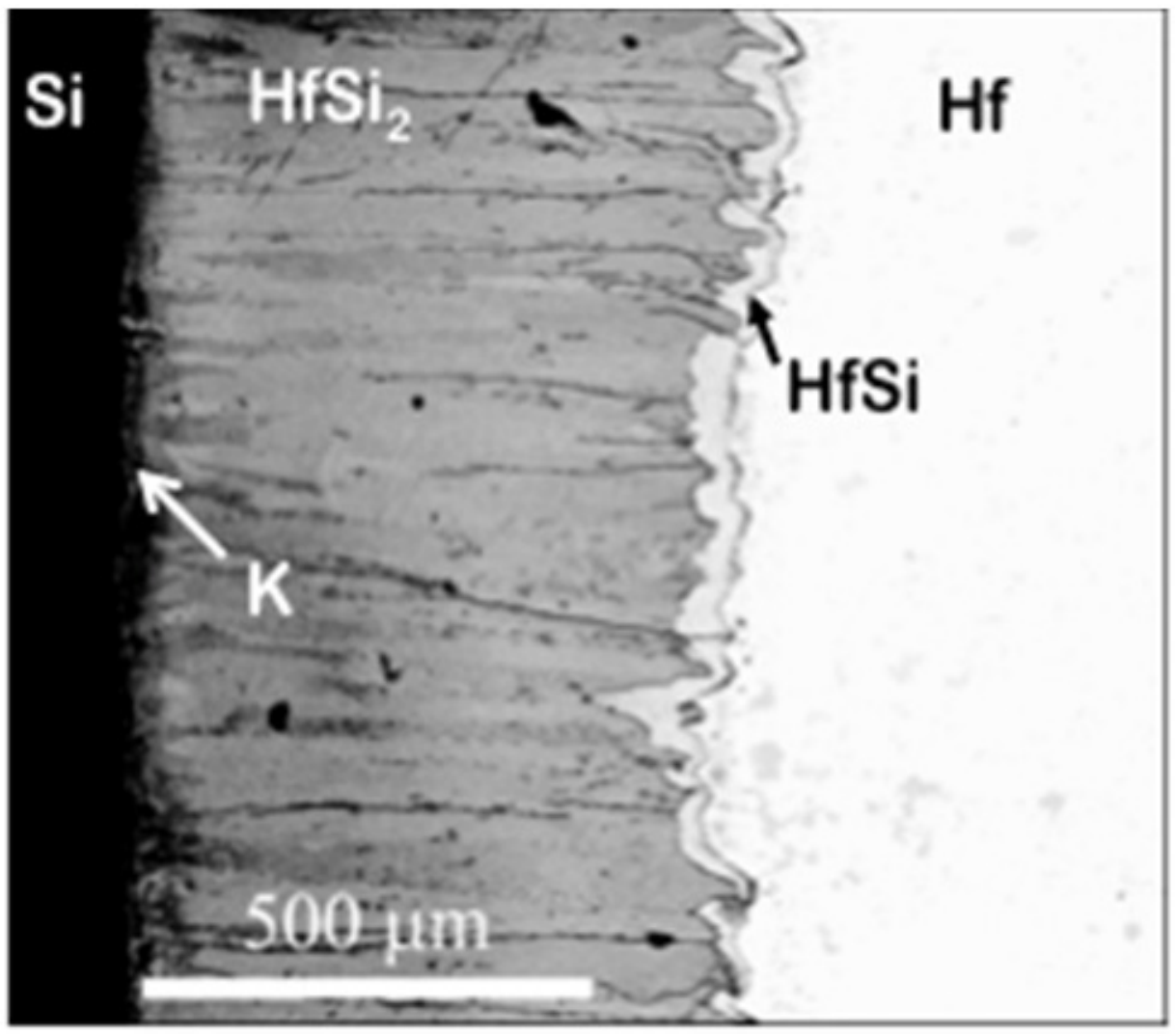

Figure 17 The interdiffusion zone of Hf/Si diffusion couple annealed at 1250 °C for 16 hrs [5].

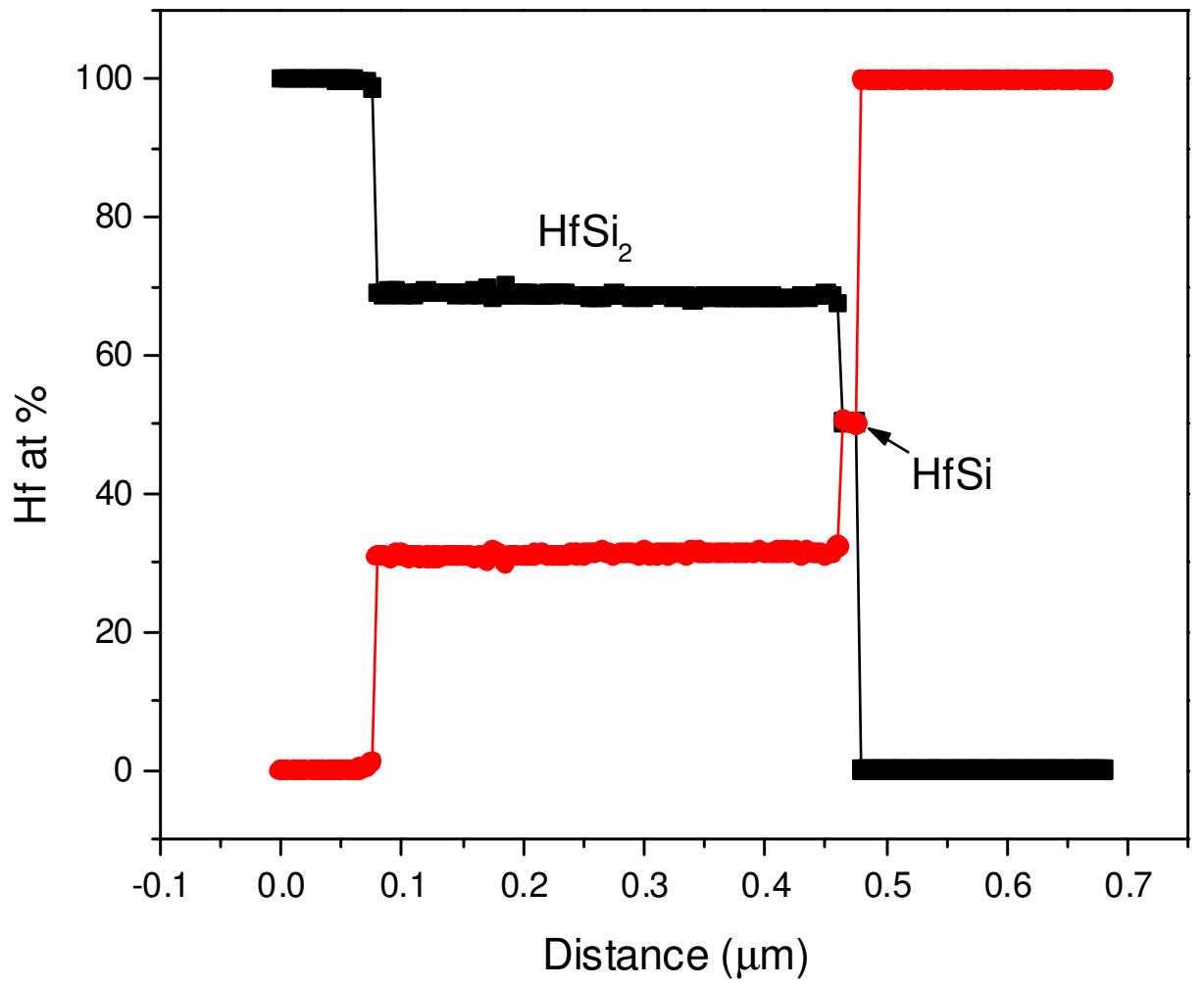

Figure 18 The composition profile of Hf/Si diffusion couple [5].

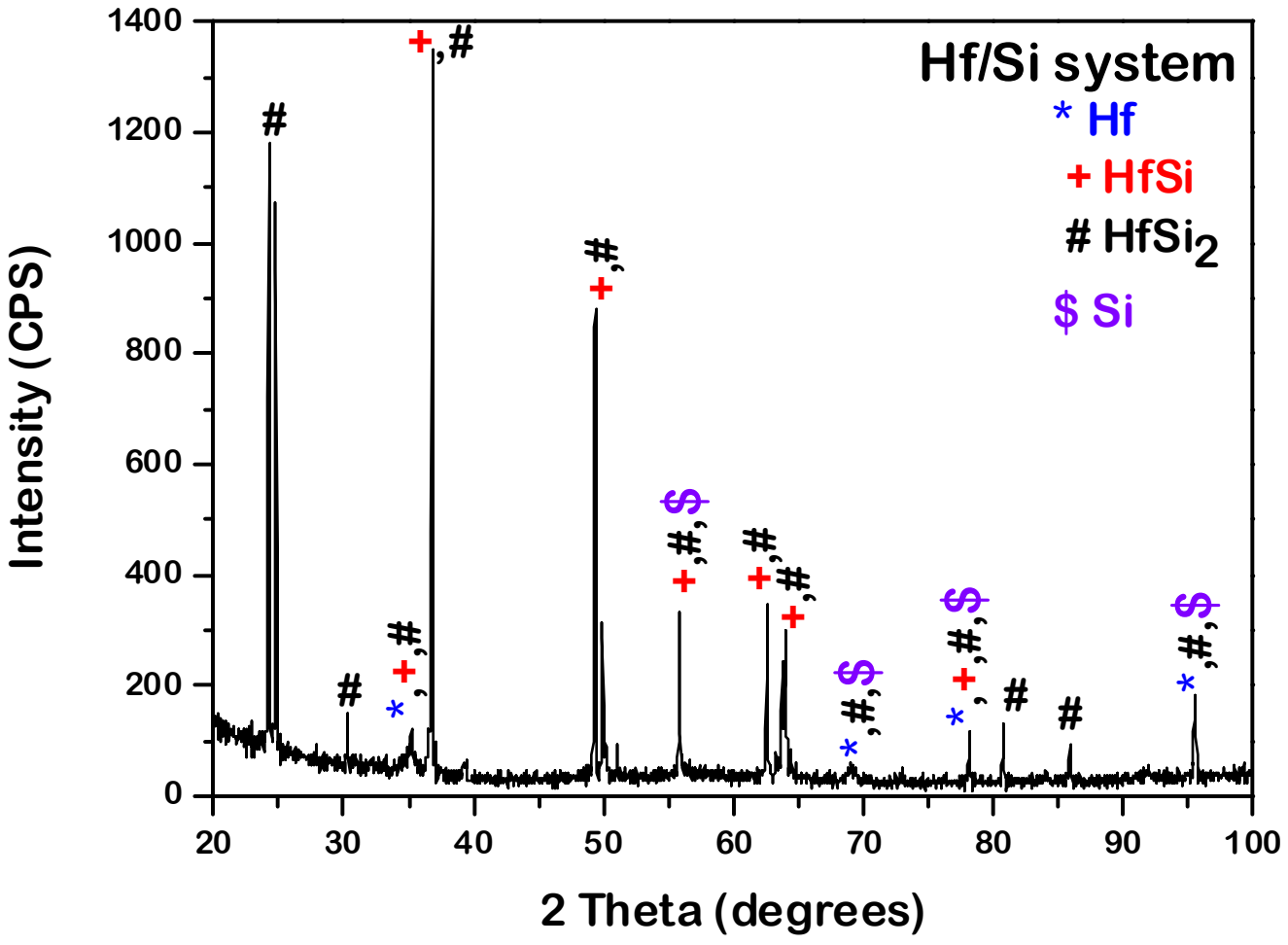

Figure 19 The X-ray diffraction pattern taken in the Hf/Si diffusion couple [5].

The integrated diffusion coefficients are estimated following Equation 2. Again, similar to the Zr-Si system, the composition profile in Figure 18 indicates negligible diffusion penetration in the Hf and Si end members. The time-dependent growth of the phases is shown in Figure 20, which indicates the diffusion-controlled the growth of both the phases. The molar volumes of $HfSi_2$ and HfSi used for the estimation of the diffusion coefficients are 9.77 and 9.97 $cm^3$/mol. The estimated integrated diffusion coefficients are plotted with respect to the Arrhenius equation in Figure 21. Following the tracer diffusion coefficients are estimated using the thermodynamic parameters as listed in Table 5. The estimated tracer diffusion coefficients are plotted with respect to the Arrhenius equation in Figure 22.

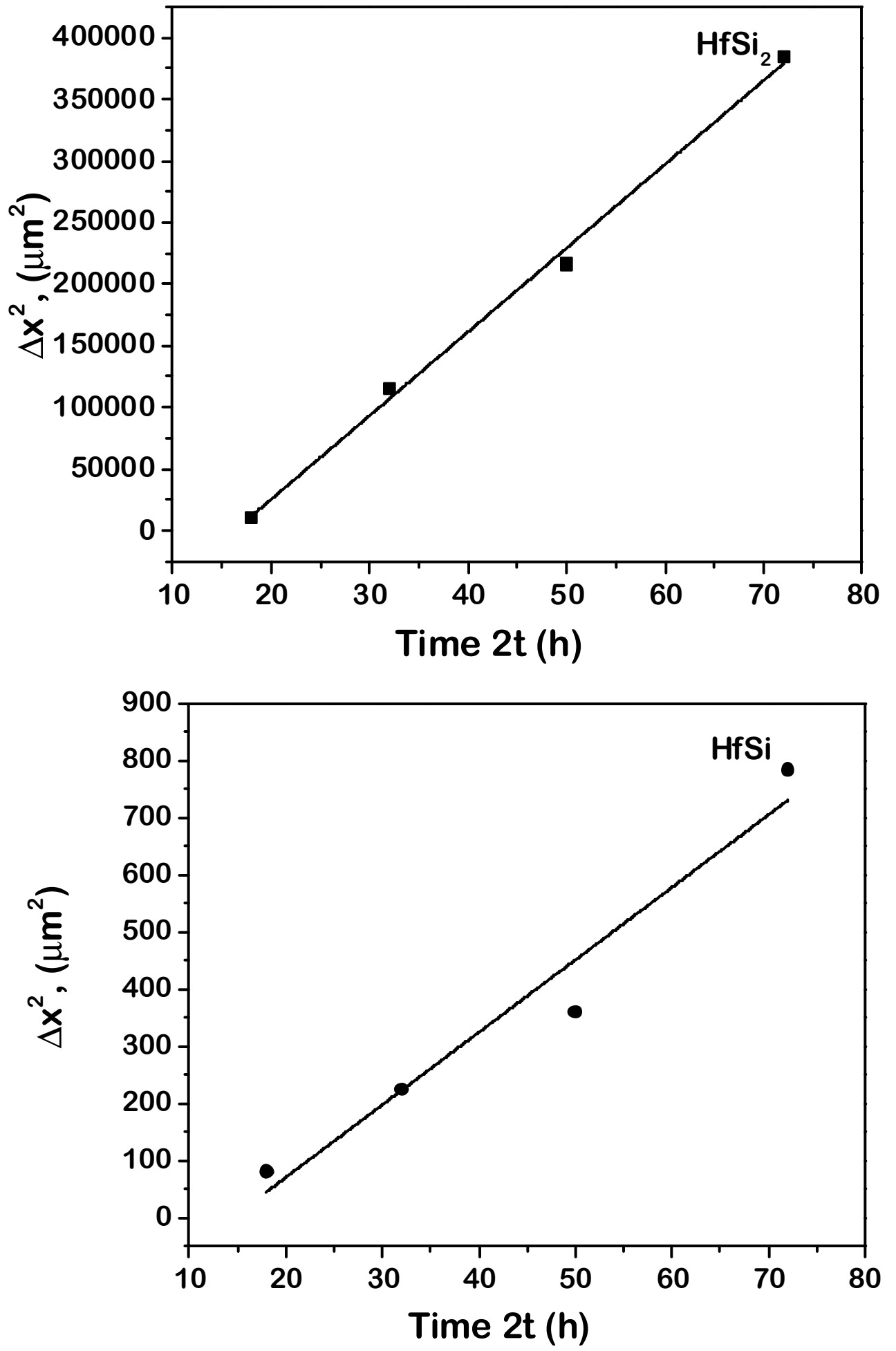

Figure 20 The time-dependent growth of the phases (a) $HfSi_2$ and (b) HfSi at 1200 °C [5].

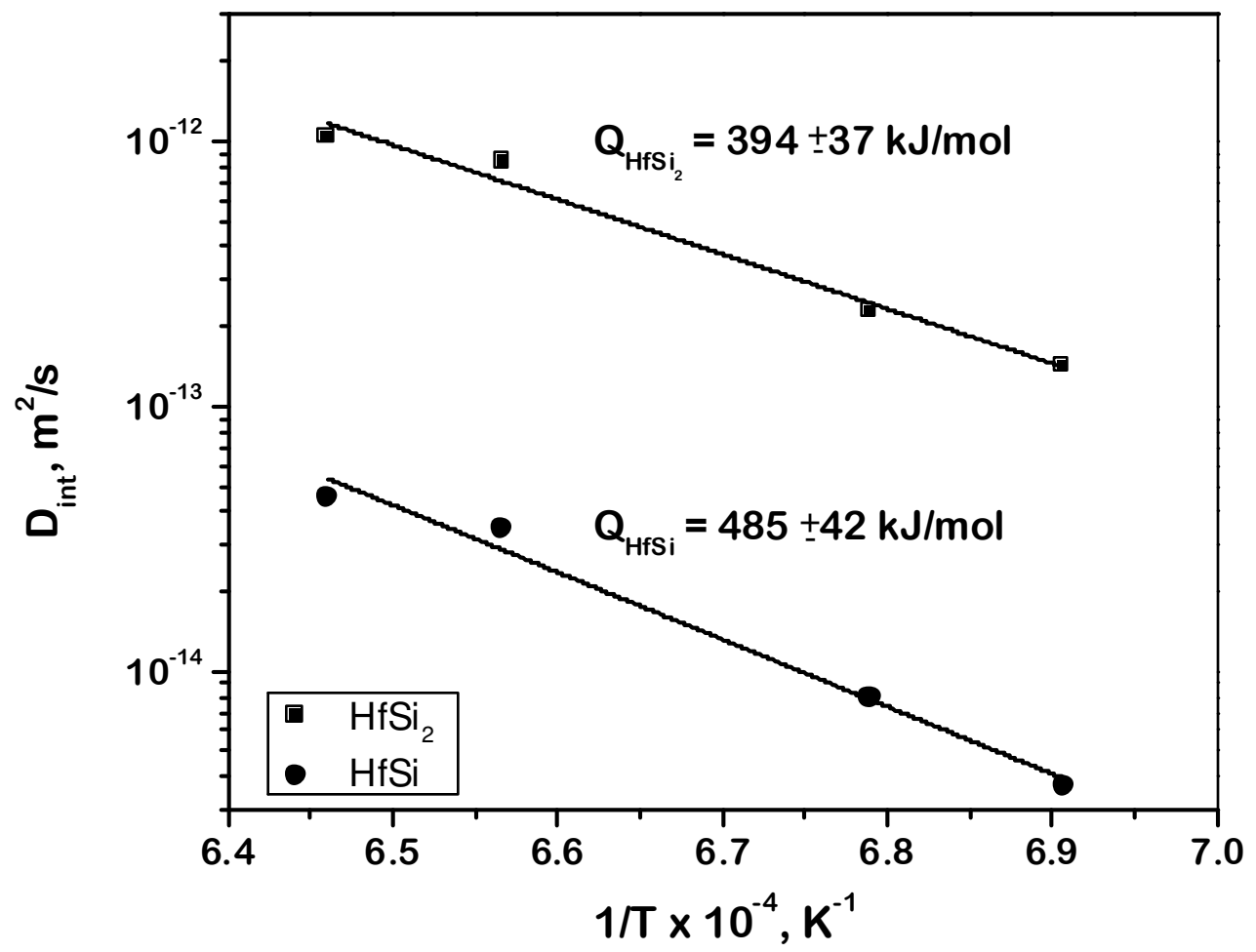

Figure 21 The estimated integrated diffusion coefficients of $HfSi_2$ and HfSi [5].

Table 5 The thermodynamic parameters related to $HfSi_2$ used for the estimation of the tracer diffusion coefficients [5].

| T (°C) | $\Delta G_{Si}$ J/mol | $\Delta G$ ($HfSi_2$) J/mol | $\Delta G$ (HfSi) J/mol | $\Delta_d G_{Si}$ J/mol |
|---|---|---|---|---|
| 1275 | -59806.3 | -127919.2 | -151724.1 | -20645.8 |
| 1250 | -58329.1 | -126435.94 | -149893.4 | -21332.5 |
| 1200 | -55410.1 | -123507.6 | -146271.6 | -22675.0 |
| 1175 | -53968.51 | -122062.8 | -144480.7 | -23392.7 |

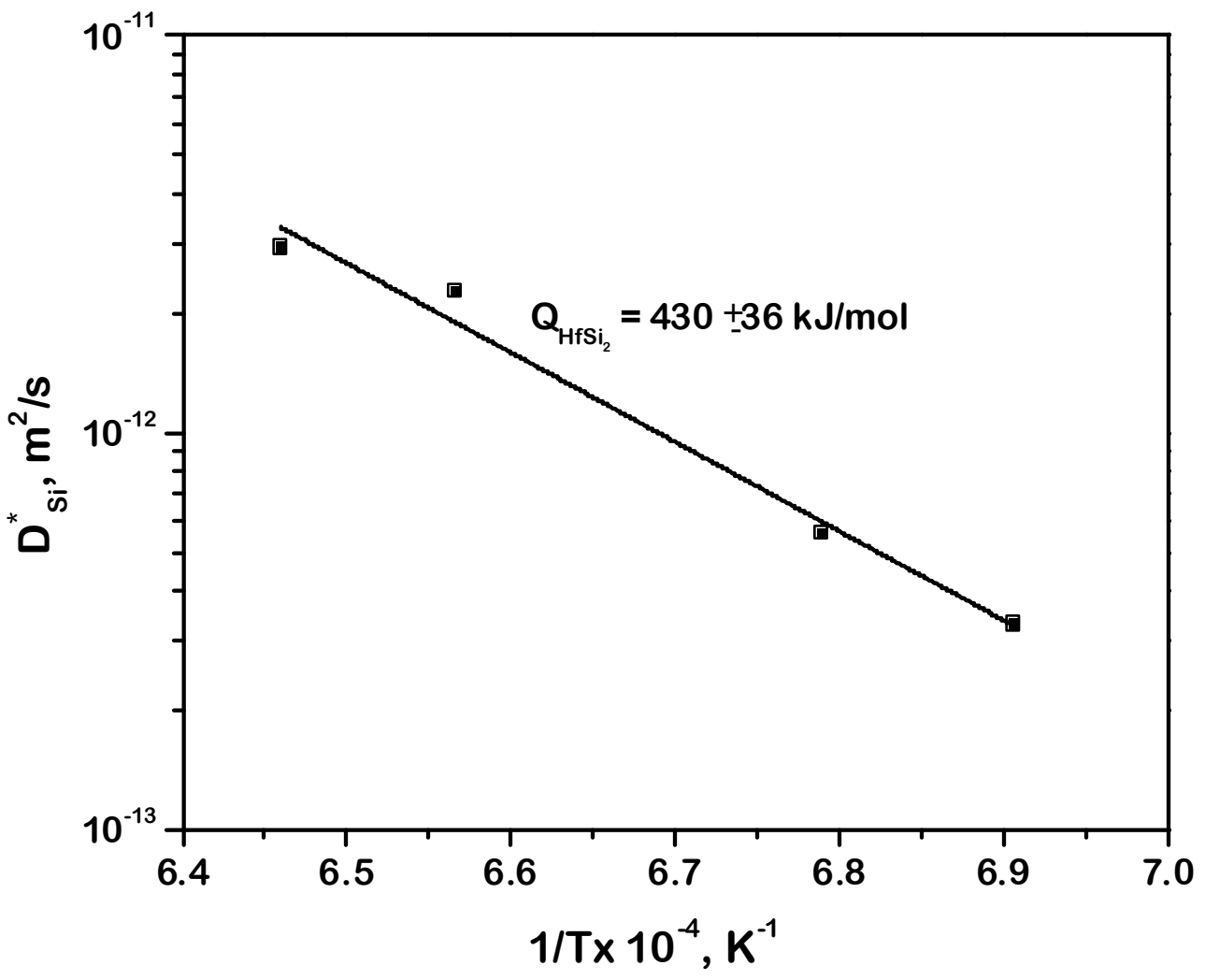

Figure 22 The estimated tracer diffusion coefficients in the $HfSi_2$ phase [5].

## 2.4 Summary of interdiffusion studies in Group IVB Metal – Silicon systems

The study on three metal silicon systems is reported here. In Ti-Si system many phase layers are found to grow depending on the temperature of interest. On the other hand, the diffusion-controlled growth behaviour in the Zr and Hf-Si systems are found to be similar to the growth of only two product phase. In fact, in both the systems, the disilicide and monosilicide phases are found to grow. However, there are mainly two common behaviours in all the systems:

(i) The growth rate of the disilicide phase is the highest of all the systems
(ii) This phase grows through the diffusion of Si. The diffusion rate of the metal component is very small.

To investigate it further, we compare the diffusion coefficients in these phases. It can be seen in Figure 23 that the integrated diffusion coefficient increases with the increase in an atomic number of the metal component. For example, it is the lowest for the $TiSi_2$ phase and the highest for the $HfSi_2$ phase. Additionally, there is an increase in the activation energy with the increase in atomic number indicating the increase in activation barrier with the atomic number. However, the increase in integrated diffusion coefficient because of the increase in the pre-exponential term reflects a higher number of jumps or exchange of positions between atoms and vacancies. The integrated diffusion coefficients are influenced by the presence of driving forces, which is different in different systems. Therefore, as shown in Figure 24, we compare the tracer diffusion coefficient of Si in the same phase. It can be seen that there is no change trend with respect to the diffusion coefficient and the activation energy.

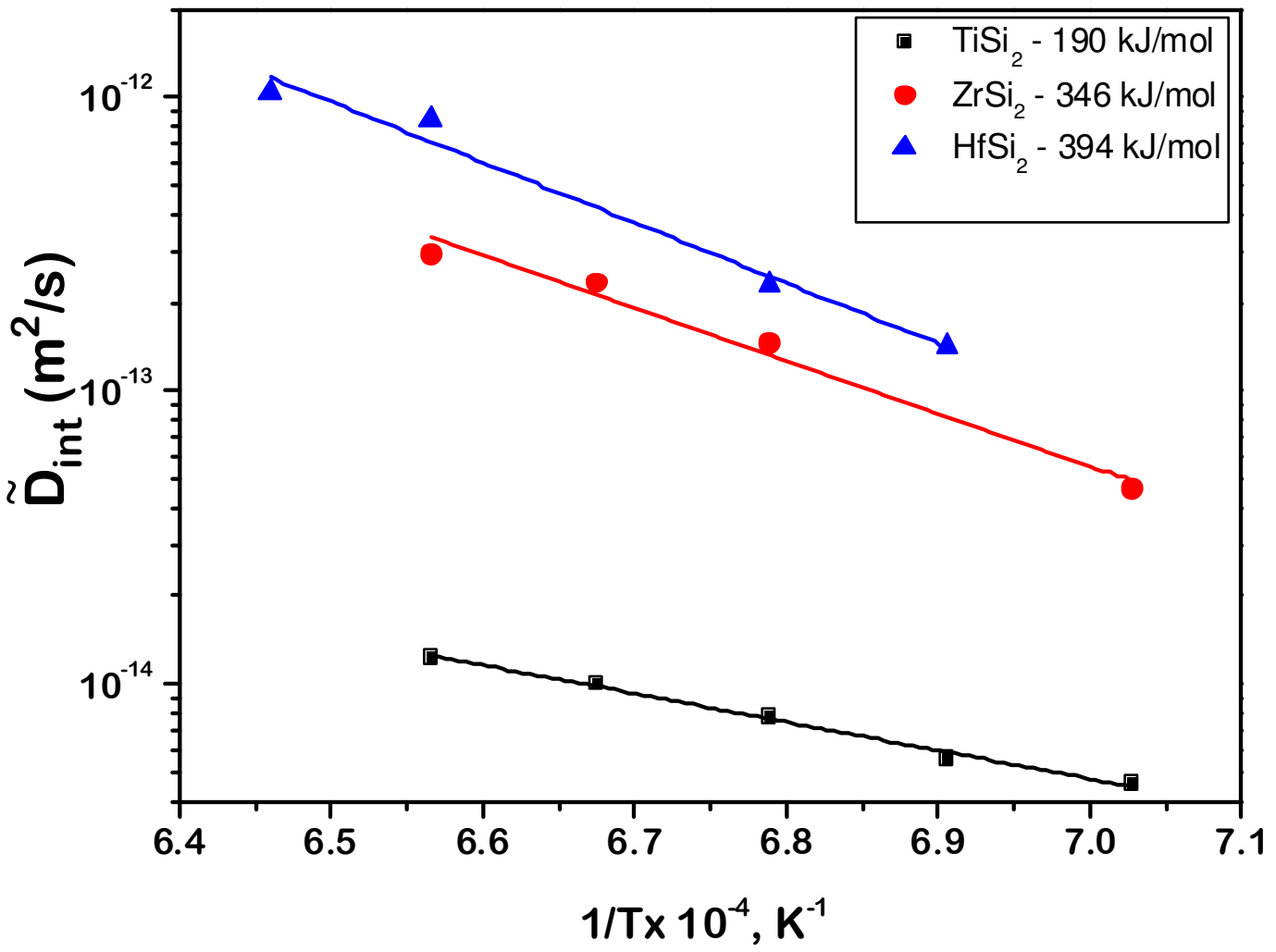

Figure 23 The comparison of the integrated diffusion coefficients in $TiSi_2$, $ZrSi_2$ and $HfSi_2$.

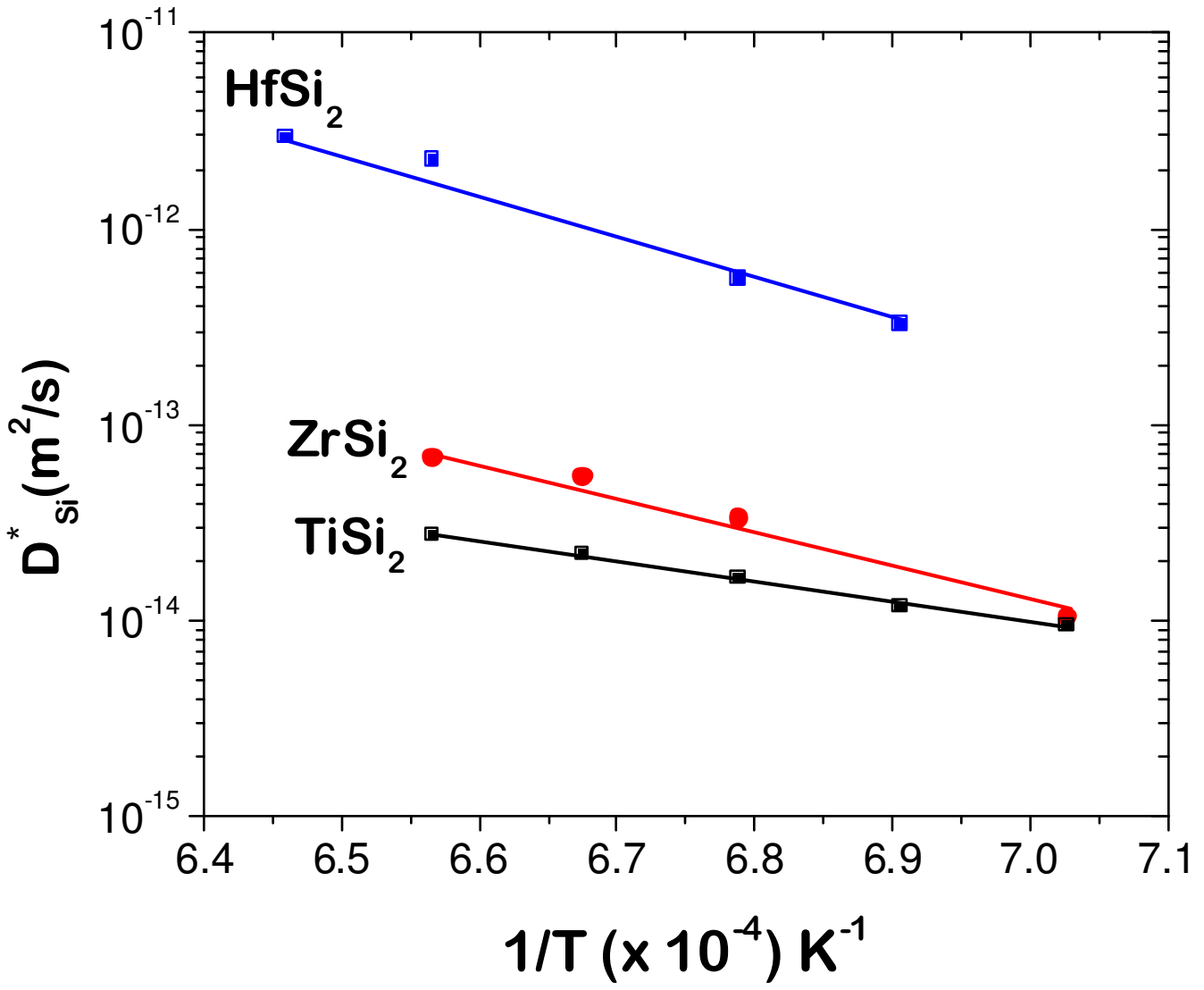

Figure 24 The comparison of the tracer diffusion coefficients in $TiSi_2$, $ZrSi_2$ and $HfSi_2$.

To understand the reason for the diffusion of Si only through the product phase, we need to know the types and concentration of defects in the structure. It should be noted here that in most of the intermetallic compounds, two types defects are present such as vacancy and antisites [1]. Only in few special intermetallics, triple defects are present [1, 28] leading to a different mechanism of diffusion. On the other hand, in most of the other intermetallics, the sublattice diffusion mechanism is proven to be operative. The details of this mechanism can be learnt from the books mentioned in the references [1, 29]. However, it is impossible to estimate the concentration of defects experimentally in an intermetallic compound since different sublattices will have different concentration of defects. Calculation of these defects theoretically is not necessarily easy because of unavailability of basic parameters required for these estimations. Therefore these data are available only in very few systems to establish the diffusion mechanism of components based on a theoretical calculation of defects and experimentally determined diffusion rates of components [29]. We can rather draw a benefit from these understandings. Since we have already estimated the diffusion coefficients, these data can indicate the relative amounts of defects present in a particular phase.

To understand this, we first need to understand the crystal structure and how different components are surrounded. Let us first discuss the diffusion process in the $MoSi_2$ phase, which is established well based on experimental data and calculation of defects. Two types of defects i.e. vacancies and antisites (the atoms on a "wrong" sublattice) in an intermetallic compound can be distinguished into four types. A particular type of defect is present with different concentrations on different sublattices. The diffusion rate of a particular component depends on various parameters such as connectivity of the location of a particular component via its own sublattice, the availability of a vacant lattice site that an atom can exchange the position, energy required for the formation of vacancies and antisites, energy required for crossing the migration barriers and the last but not the least the corresponding correlation effect [29]. Following the analysis in Ref. 30, we know that the concentration of vacancies is present mainly on the Si sublattice with a low concentration on Mo sublattice in $MoSi_2$ with tI6 crystal structure. Mo atom is surrounded by 10 Si atoms. On the other hand, the Si atom surrounded by 5 Mo and 5 Si. Therefore, Si can diffuse via its own sublattice. Mo cannot diffuse unless antisites are present The negligible diffusion rate of Mo indicates the absence of these defects. This must be the reason that Si has a higher diffusion rate compared to Mo by few orders of magnitude [31, 32].

As shown in Figure 25, the $TiSi_2$ phase has an orthorhombic face-centred structure (oF24) with 24 atoms in the unit cell. Ti atom is surrounded by 4 Ti and 10 Si. Si atoms are surrounded by 5 Ti and 9 Si. Therefore, both the components could diffuse over their own sublattice if vacancies are present on both the sublattices occupied by Ti and Si. Since the $TiSi_2$ phase grows by the diffusion of Si, it is evident that vacancies are present mainly on the Si sublattice with a negligible concentration of vacancies on Ti sublattice. This conclusion agrees with recent computations of the defect formation energies in $TiSi_2$ by tight binding molecular dynamics [33].

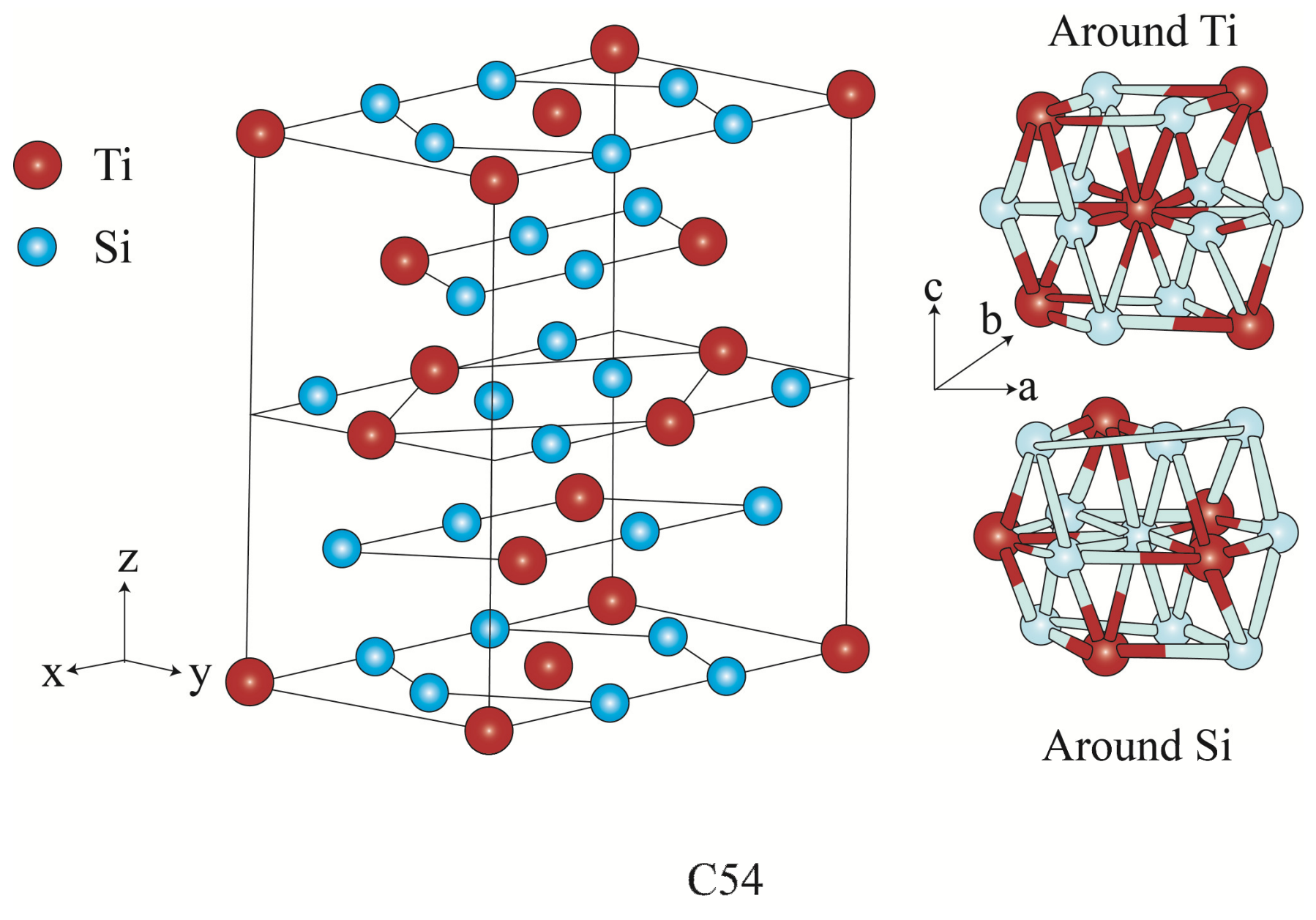

Figure 25 Crystal structure of C54 $TiSi_2$ Phase (oF24) showing the nearest neighbours of Ti and Si [4].

Both $ZrSi_2$ and $HfSi_2$ have the same crystal structure *i.e.* orthorhombic C49 (oC12) with 12 atoms per unit cell. This is shown in Figure 26. Metal atoms (M= Zr and Hf) have only one sublattice whereas Si has two, as designated by SiI and SiII. M is surrounded by 10 Si (4 SiI and 6 SiII) and 6 M. SiI is surrounded by 12 Si (8 SiI and 4 SiII). SiII is surrounded by 10 Si (6 SiI and 4

SiII) and 6 M. Therefore, both the components can diffuse via its own sublattice on the condition that vacancies are present on both the sublattices. Negligible diffusion of M compared to Si indicates that vacancies are mainly present on the Si sublattice with a negligible concentration of vacancies on the sublattice of M. Therefore, irrespective of different crystal structures, all the three compounds $TiSi_2$, $ZrSi_2$ and $HfSi_2$ have a similar character that both the components could diffuse. However, only Si can diffuse because of the presence of vacancies on the Si sublattice.

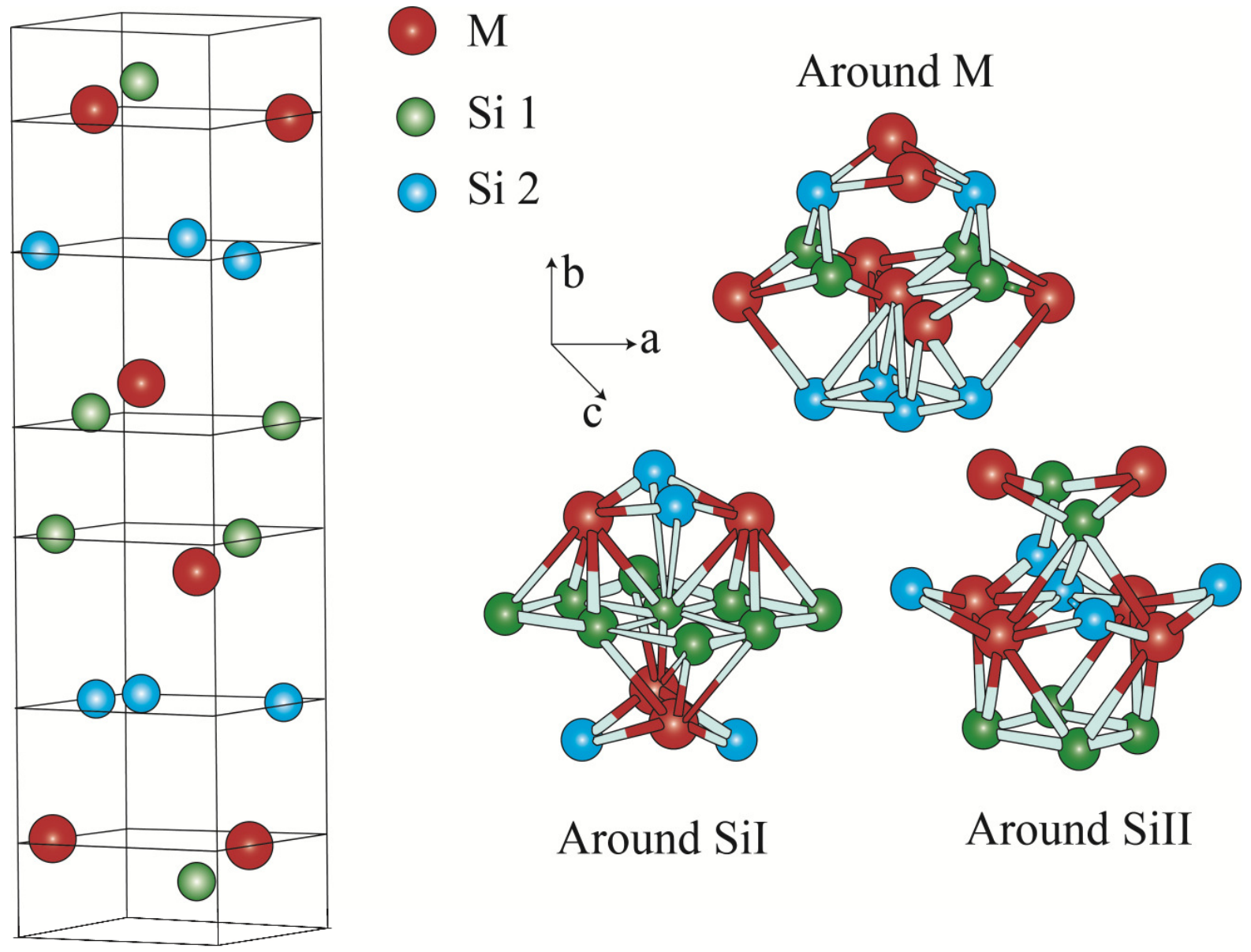

C49

Figure 26 The crystal structure C49 (oC12) of $ZrSi_2$ and $HfSi_2$. First and second nearest neighbours of Ti, SiI and SiII are shown [5].

## 3. Diffusion study in Group VB (V, Nb and Ta) Metal (M)-Silicon (Si) systems

### 3.1 Interdiffusion study in the V-Si system

Vanadium disilicide is used as a protective coating and as an Ohmic contact for Schottky diodes. $V_3Si$ phase is an intermetallic superconductor and is being explored for the application as a replacement of currently used $Nb_3Sn$ in various applications. Before the publication by Prasad and Paul [9], only one bulk interdiffusion study was reported in this system by Milanese *et al.* [34]. It can be seen in the phase diagram in Figure 27 that there are three intermetallic line compounds ($VSi_2$, $V_6Si_5$, $V_5Si_3$) and one phase ($V_3Si$) with narrow homogeneity range. Therefore, one cannot estimate the composition profile in an interdiffusion zone in this phase. This rules out the estimation of the interdiffusion coefficient $\tilde{D}$. One can rather estimate the integrated diffusion coefficient, which is already explained in the previous sections.

Milanese *et al.* [34] used an indirect method developed by Buscaglia *et al.* [35] to estimate $\tilde{D}$ instead of $\tilde{D}_{int}$ in this system. As we have seen already, the Wagner's method measures the integrated diffusion coefficient directly from the composition profile. However, following Buscagila's method, it is necessary first to estimate the parabolic growth constant of the second kind from the first kind, which are defined by Wagner [36]. The first kind of parabolic growth constant is estimated from the measured thickness of the product phase that is grown in a diffusion couple prepared with pure end members. The second kind of parabolic growth constant is estimated from the measured thickness of a product phase that is grown in an incremental diffusion couple such that a single phase layer grows between two neighbouring phases in a phase diagram.

Subsequently, they calculated $\tilde{D}$ using thermodynamic parameters. To understand the shortcomings of this approach, we need to have a look into the relationship developed by Buscaglia *et al.* which were used by Milanese *et al.*

$$\tilde{D}_i = k_i^{II}\left[\frac{v_i^2(v_{i-1}-v_{i+1})^2}{(v_{i-1}-v_i)^2(v_i-v_{i+1})^2}\frac{1+v_i}{v_i}\frac{|\Delta G_i^0|}{RT}\right]^{-1} \quad (6a)$$

$$\tilde{D}_i = \frac{1}{1+v_i}\overline{D}_{V,i} + \frac{v_i}{1+v_i}\overline{D}_{Si,i} \quad (6b)$$

where $k_i^{II}$ is the parabolic growth constant of the second kind, $\tilde{D}_i$ is the interdiffusion coefficient of the $i^{th}$ phase, $\overline{D}_{V,i}$ and $\overline{D}_{Si,i}$ are the self-diffusion coefficient of V and Si, respectively in the $i^{th}$ phase. $\Delta G_i^0$ is the Gibbs free energy change for the formation of 1 mole of phase $A_{v_i}B$. $v_i$ is the mole fraction of A of the $i^{th}$ phase $A_{v_i}B$. $v_{i-1}$ and $v_{i+1}$ are the mole fraction of element A of the adjacent phases in the phase diagram. This method unnecessarily complicates the estimation of a diffusion parameter without the need for the thermodynamic parameters, which is a large source of error. Therefore, Prasad and Paul [9] followed a standard procedure, which is explained in detail.

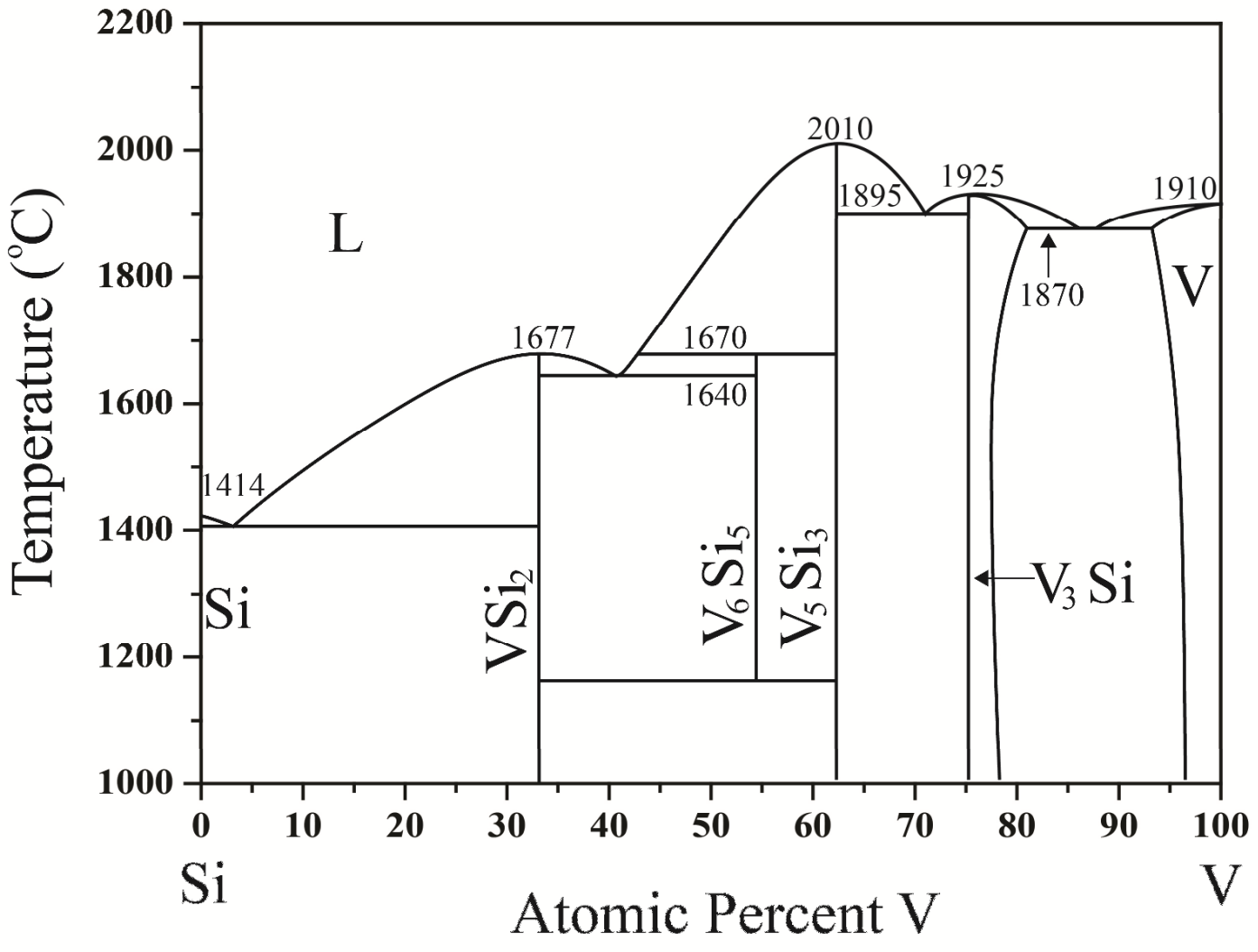

Figure 27 V-Si phase diagram [60].

Figure 28 shows the interdiffusion zone of V-Si diffusion annealed at 1200 °C for 16 hrs. Out of the four phases are present in the phase diagram, two-phase $V_5Si_3$ and $VSi_2$ phases grow with easily measurable layer thickness. $V_3Si$ has a very small thickness and $V_6Si_5$ has a negligible thickness. The presence of the thin phases is clearer in Figure 29, which shows a focused region of the interdiffusion zone. The average thickness of $V_3Si$, $V_5Si_3$ and $VSi_2$ phases are measured in the range of 1.6-4, 11.2 - 36.5 and 65.2 - 112 μm in the experimental annealing temperature range of 1150-1300 °C. The measured composition profile is shown in Figure 30. Additionally, the point analysis was conducted to estimate the composition range of the phases. Irrespective of the small thickness of two phases, Milanese *et al.* [34] estimated the diffusion coefficients in all the phases, which might need significant error in calculation. To explain this issue we estimated the data in $V_3Si$ phase and then compared with the data available which are estimated following the incremental diffusion couple experiments.

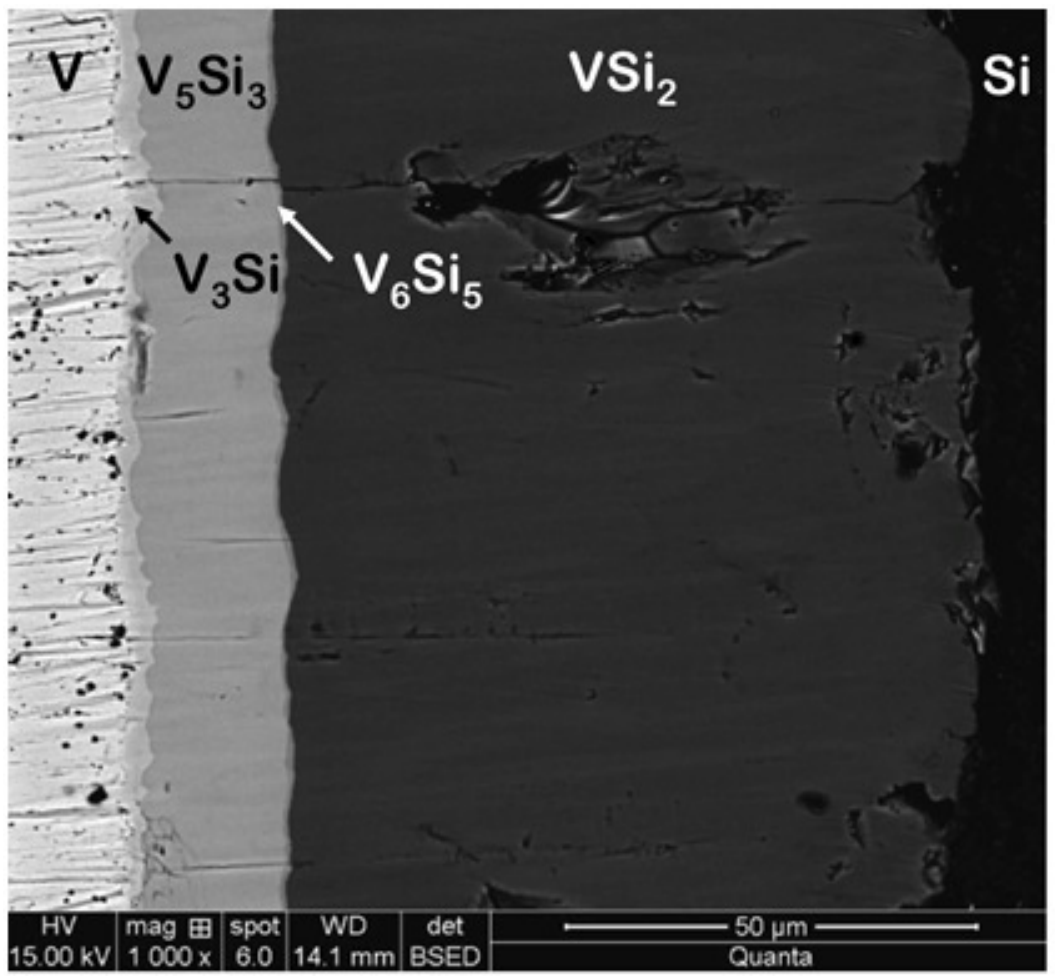

Figure 28 Backscattered electron image of V-Si diffusion couple annealed at 1200 $^{o}$C for 16h [9].

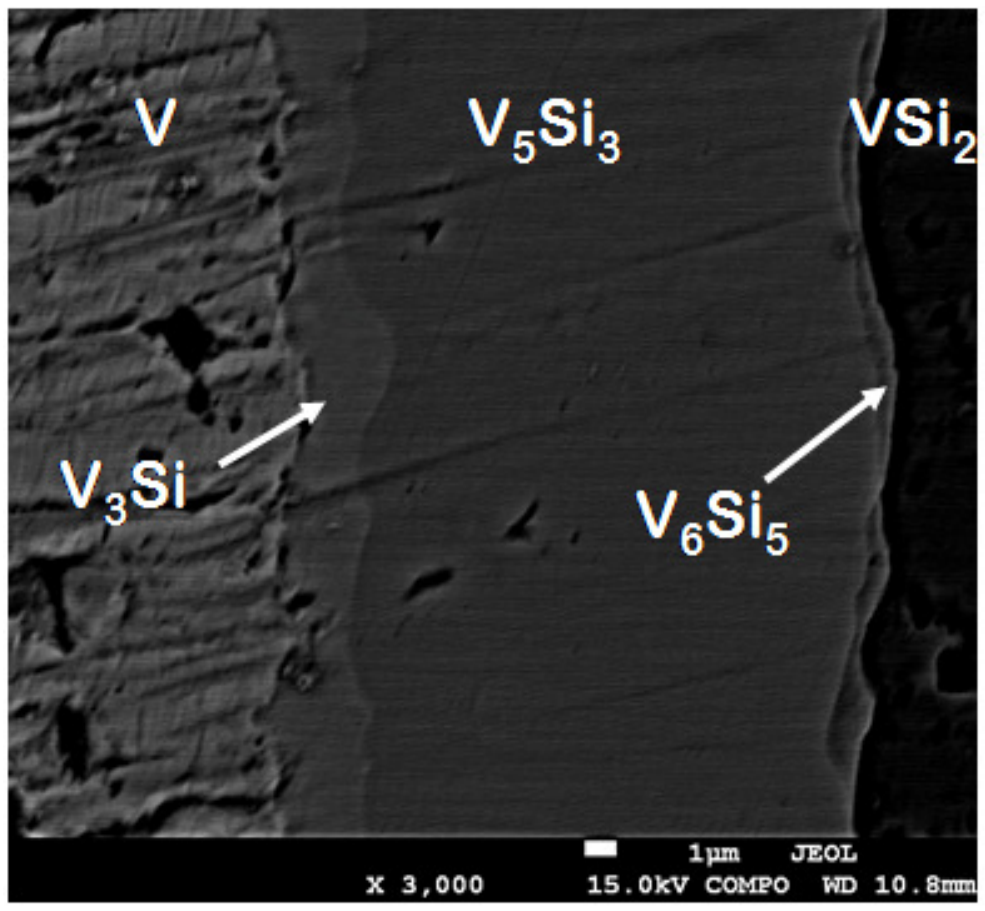

Figure 29 A focused region of the interdiffusion zone [9].

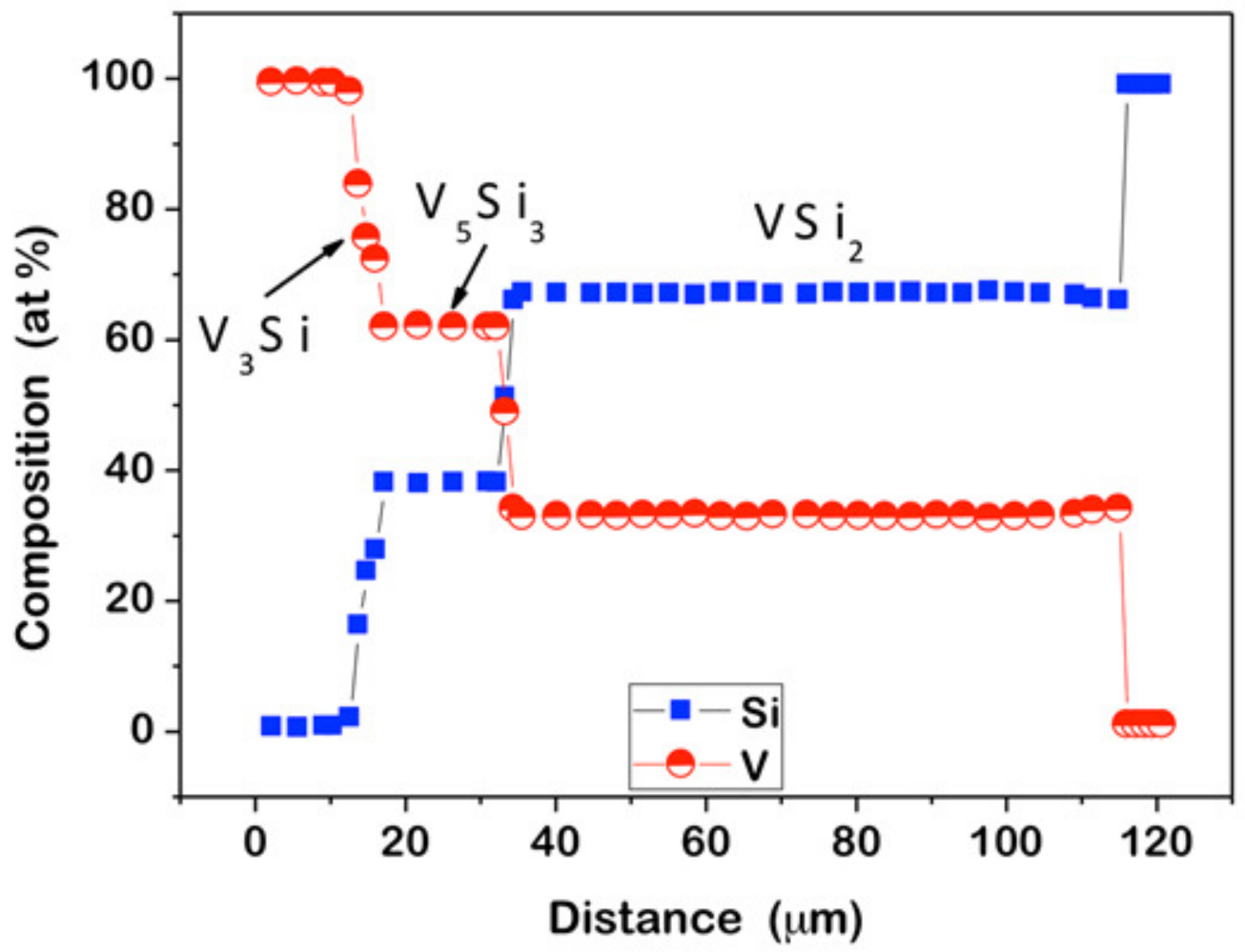

Figure 30 The measured composition profile of the V-Si diffusion couple annealed at 1200 $^{o}$C for 16h [9].

The parabolic growth constant for $V_3Si$, $V_5Si_3$ and $VSi_2$ phases are estimated and plotted with respect to the Arrhenius equation in Figure 31. The data reported by Milanese et al. [34] are included for comparison. We have not calculated the data for $V_6Si_5$ phase, which would otherwise introduce very large error because of very small thickness. In fact, there is a very good match in the data for $V_5Si_3$ and only a small difference is noticed in the case of $V_3Si$ and $VSi_2$ phases. The activation energies are estimated at 230 for $V_3Si$, 297 for $V_5Si_3$ and 130 kJ/mol for $VSi_2$. Milanese *et al.* have not reported these data. Since these are not material constants and depend on the end member compositions, we need to estimate the integrated diffusion coefficients.

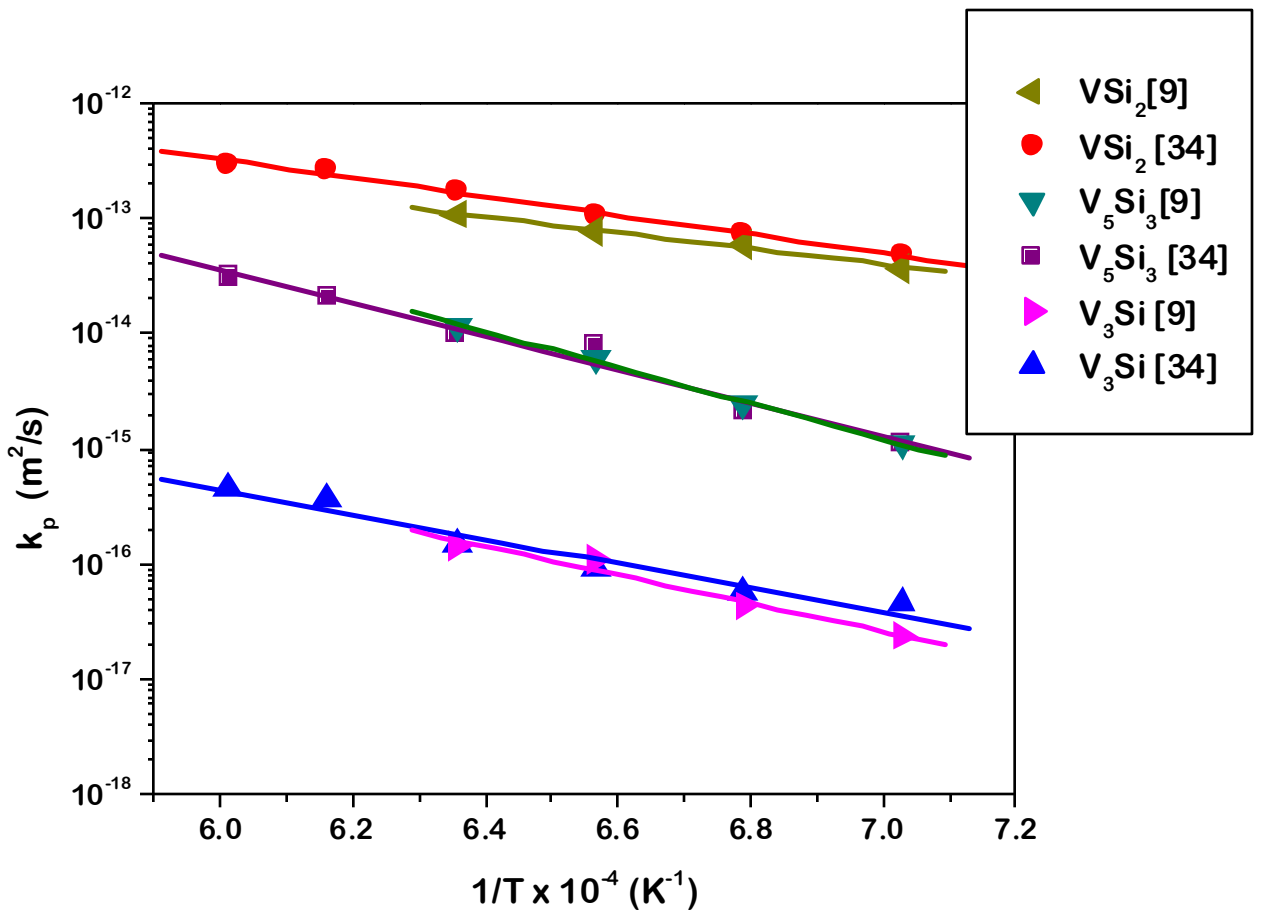

Figure 31 Arrhenius plot of parabolic growth constant for $V_3Si$, $V_5Si_3$ and $VSi_2$ phases [9, 34].

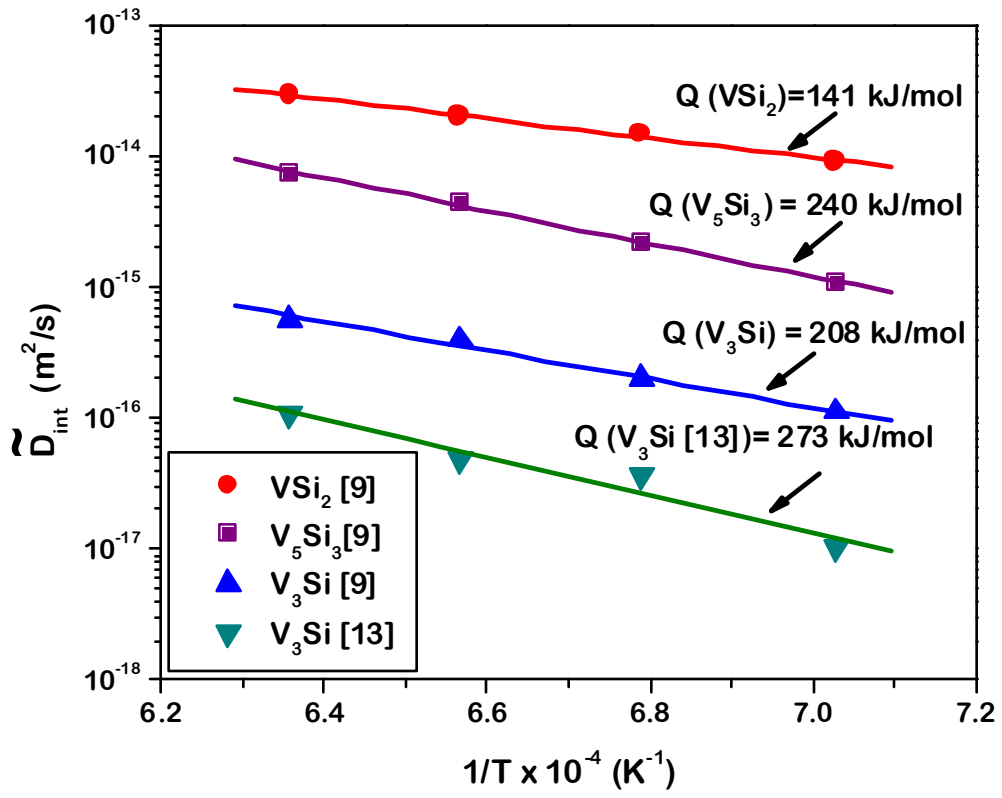

Figure 32 Arrhenius plot of the integrated diffusion coefficient, $\tilde{D}_{int}$ for $V_3Si$, $V_5Si_3$ and $VSi_2$ phases [9].

The estimated $\tilde{D}_{int}$ for the phases $V_3Si$, $V_5Si_3$ and $VSi_2$ are shown in Figure 32. For the estimation, the molar volume coefficients values are used as 7.95 $cm^3$/mol for $V_3Si$ (cP8), 7.9 $cm^3$/mol for $V_5Si_3$ (tI32) and 7.73 $cm^3$/mol for $VSi_2$ (hP9). The activation energies are estimated as 234, 240 and 141 kJ/mol for these phases. These values are the material constants and useful for understanding the diffusion controlled growth mechanism of the phases. The growth rate of a particular phase depends on the end member compositions. For example, in the case of an incremental diffusion couple in which only one product phase grows in the interdiffusion zone, will have a higher parabolic growth constant value compared to the value estimated in a diffusion couple prepared with end member compositions because of competition for growth among different phases in a multiphase interdiffusion zone. All these details can be understood better by the physicochemical approach [37, 38]. This method considers the growth-consumption of the phases at the interface unlike the standard method which considers only the composition profile.

$\tilde{D}_{int}$ calculated for the $V_3Si$ phase in a diffusion couple with pure end members and in an incremental couple [8] are shown in Figure 32. This phase grows with a very thin layer in the V-Si diffusion couple and therefore one should be careful during the estimation of the diffusion coefficients. There is always a chance to induce a very high error in calculation in such a situation. To clarify it further, we consider the error in the calculation of the thickness of the phases and as a result the error in the calculation of the diffusion parameters. The average values of the thickness of the phases $V_3Si$, $V_5Si_3$ and $VSi_2$ are estimated as 2.5, 17 and 84.6 μm with an error of 60, 9 and 3 percent. This leads to the error of 180, 19 and 7 percent in the calculation of the parabolic growth constant and a similar range in the calculation of the integrated diffusion coefficients. Therefore, we have introduced a very high error in the calculation of the integrated diffusion coefficient of the $V_3Si$ phase. Error in the calculation of diffusion coefficient can also produce a huge error in the calculation of activation energy for diffusion. For example, we found 20, 2.5 and 2 percent error in the calculation of activation energy values for $V_3Si$, $V_5Si_3$ and $VSi_2$ phases, respectively. In such a situation we should estimate the data of interest in a phase from an incremental diffusion couple as it is done in Ref. [13]. The data estimated are included in Figure 32. Because of a similar problem, the calculation of the diffusion parameter for $V_6Si_5$ by Milanese *et al.* [34] must have created a significant error in calculation.

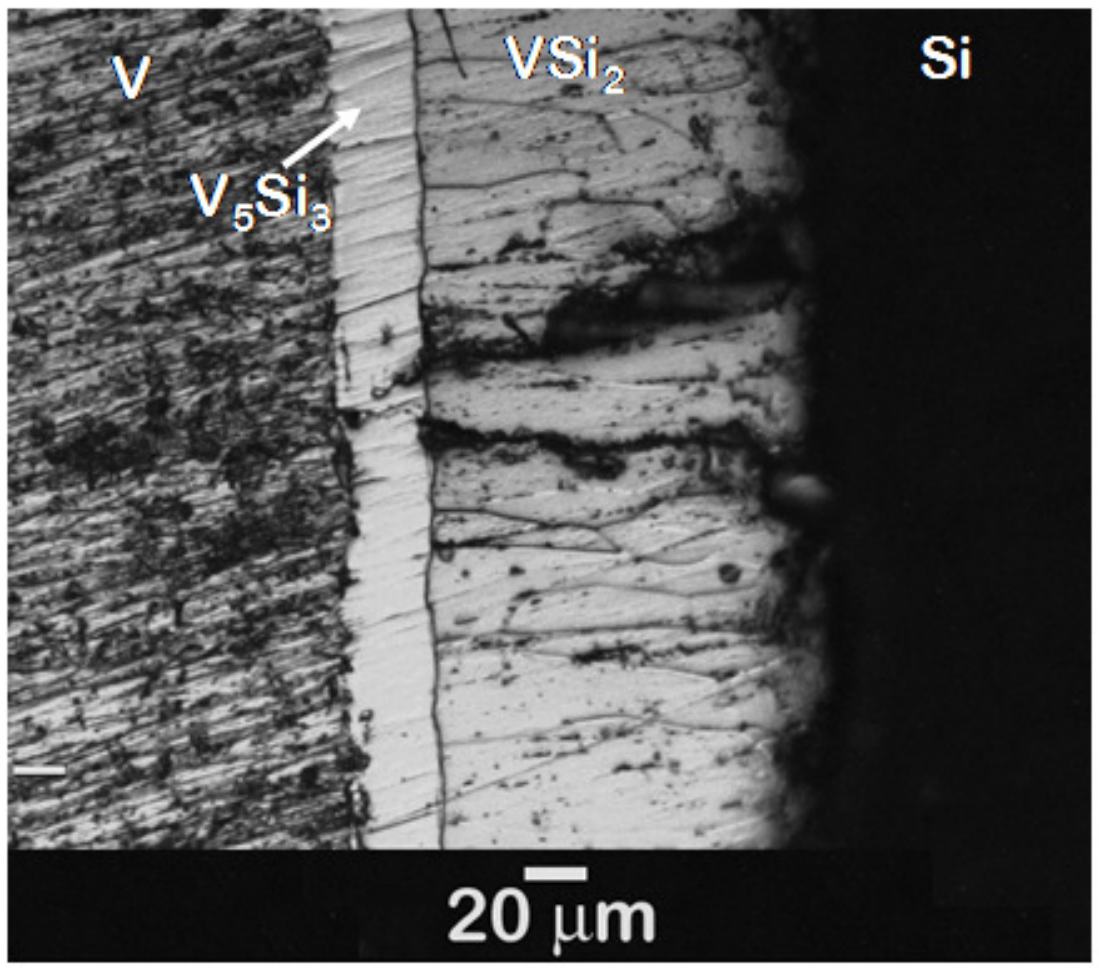

Figure 33 The optical image of the V-Si diffusion couple annealed at 1200 °C for 16 h [9].

Based on the image shown in Figure 33, it is clear that $VSi_2$ phase has columnar grains covering the whole interdiffusion zone. It indicates that this phase must have grown mainly by the diffusion of Si with the almost negligible diffusion of V. Since the $V_3Si$ phase is important for the application as an intermetallic superconductor, a detailed analysis was done for the estimation of the integrated and intrinsic diffusion coefficients by coupling V with an alloy V-29at%Si. The alloy is a phase mixture of $V_5Si_3$ and $V_3Si$. An example of this incremental diffusion couple is shown in Figure 34. The location of the marker plane was detected by using $TiO_2$ and $ZrO_2$ inert particles at the V/$V_3Si$ interface. It means that this phase is grown because of diffusion V and the diffusion rate of Si in this phase is negligible. At first, the integrated diffusion coefficients are estimated and then utilizing the thermodynamic parameters the tracer diffusion coefficients are estimated. These data are plotted together in Figure 35. Note that the same integrated diffusion coefficients are compared in Figure 32.

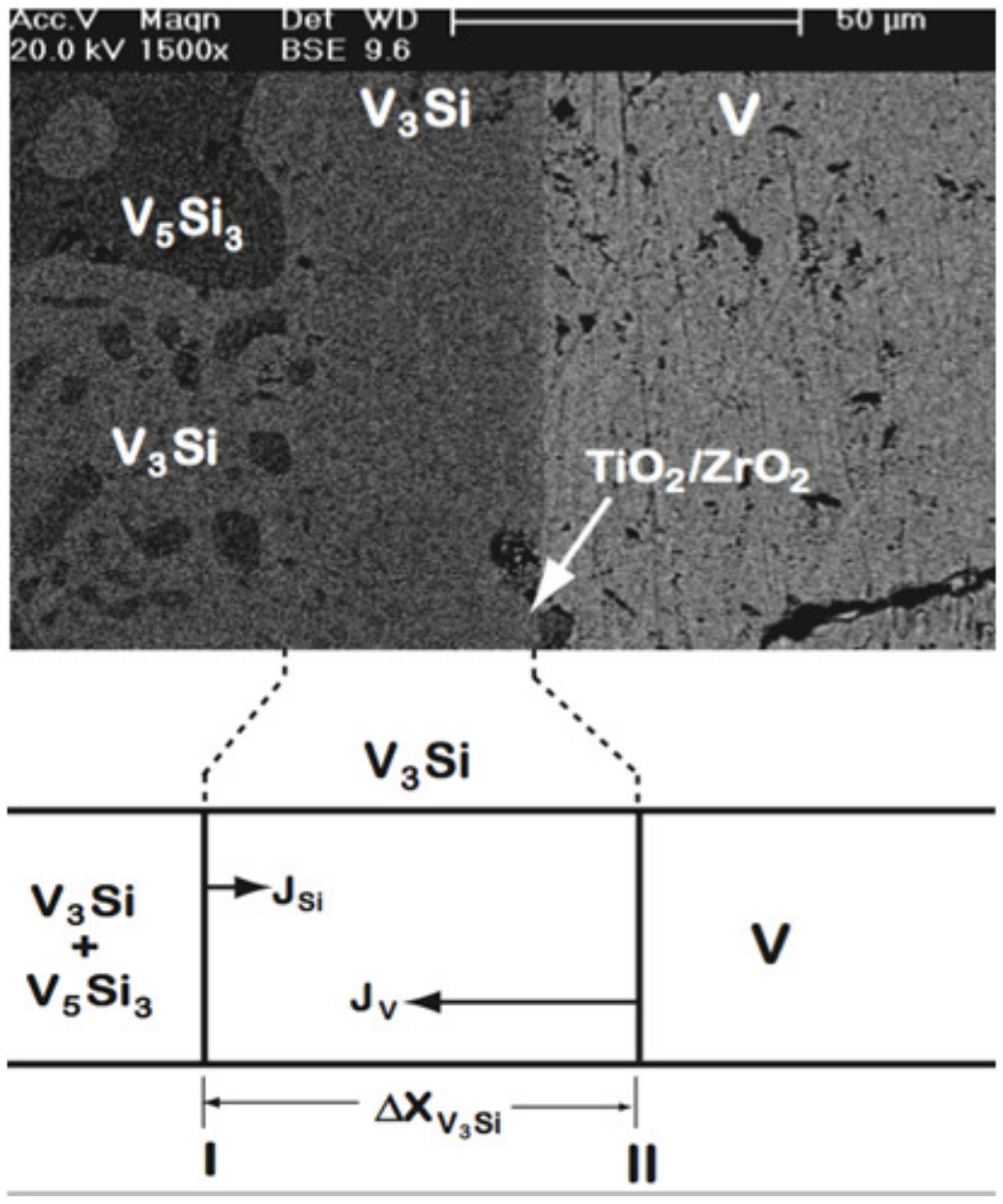

Figure 34 Incremental diffusion couple of V-29at.%Si and V annealed at 1350 $^{o}$C for 24 hrs is shown. The schematic diagram explains the diffusion direction of component V and Si [13].

Table 6 The thermodynamic parameters used for the estimation of the tracer diffusion coefficients in the $V_3Si$ phase.

| Temperature (K) | $\widetilde{D}_{int}^{V_3Si}$ (x10$^{-17}$ m$^2$/s) | $\Delta_r G_V^o$ (j/mole) | $D_V^*$ (x10$^{-17}$ m$^2$/s) |
|---|---|---|---|
| 1373 | 0.554 | -13274 | 2.542 |
| 1423 | 1.032 | -12932 | 5.035 |
| 1473 | 3.692 | -12591 | 19.151 |
| 1523 | 4.889 | -12249 | 26.953 |
| 1548 | 8.136 | -12079 | 46.236 |
| 1573 | 10.789 | -11909 | 63.193 |
| 1598 | 15.649 | -11740 | 94.448 |
| 1623 | 21.304 | -11570 | 132.511 |

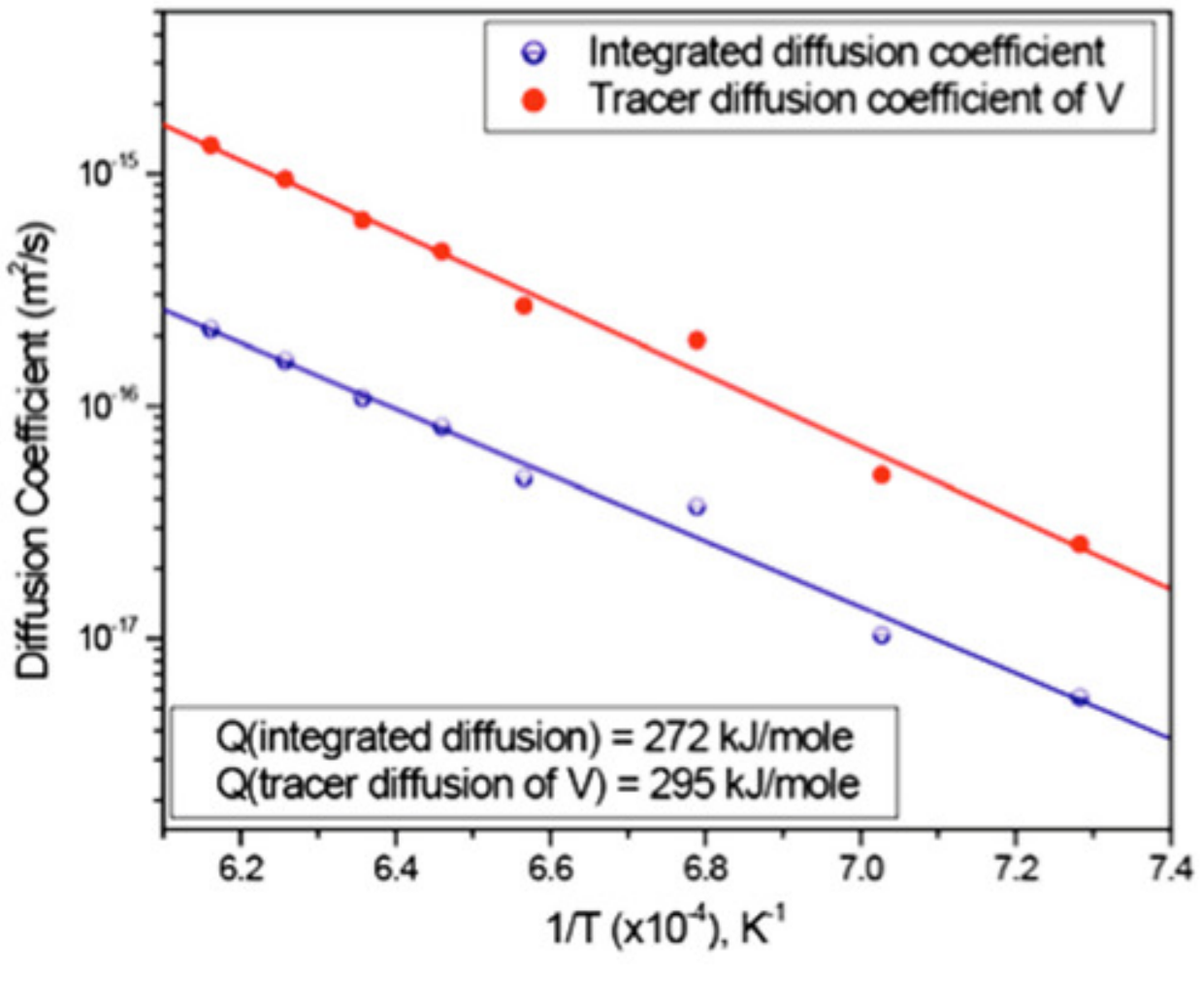

Figure 35 Integrated and tracer diffusion coefficients in the $V_3Si$ phase [13].

### 3.2 Interdiffusion study in the Nb-Si system

There is a constant search for an alternate material to replace Ni-base superalloys in high temperature applications, which can withstand even high operating temperature for the increase in fuel efficiency and for the reduction of emission of harmful gases. At a certain point, Nb-based silicides were considered seriously for such an option along with Mo based silicides. These have excellent creep resistance along with good oxidation resistance. Beside, an excellent strength to density ratio brings and additional advantage for the use at above 1350 °C in various applications. However, currently these are limited by their brittle nature. Several attempts are being made to compensate these shortcomings [39-43]. Study on diffusion is important for developing indepth understanding of many properties. As usual, many studies are conducted in this system as thin films as laminate structures [44-47]. On the other hand, diffusion studies in bluk materials are very important to avoid the effect of any unwanted parameters such as stress. One of the initial studies with bulk samples were conducted by Milanese *et al.* [48]. Subsequently, Prasad and Paul studied this system extensively [12]. These two studies are considered for the description of diffusion studies in this system in detail.

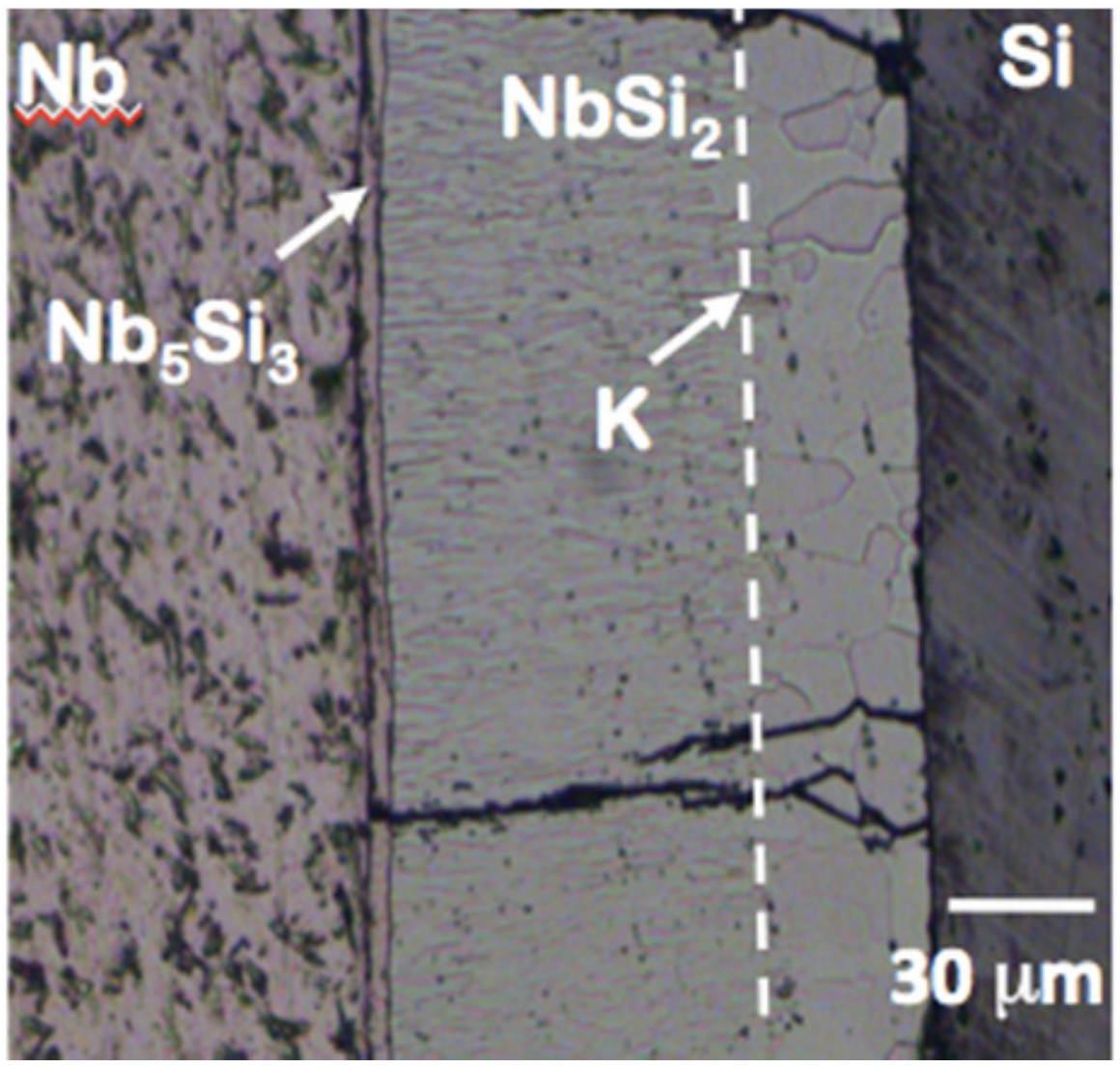

Figure 36 Optical micrograph of Nb/Si diffusion couple annealed at 1250 °C, 24 hrs [12].

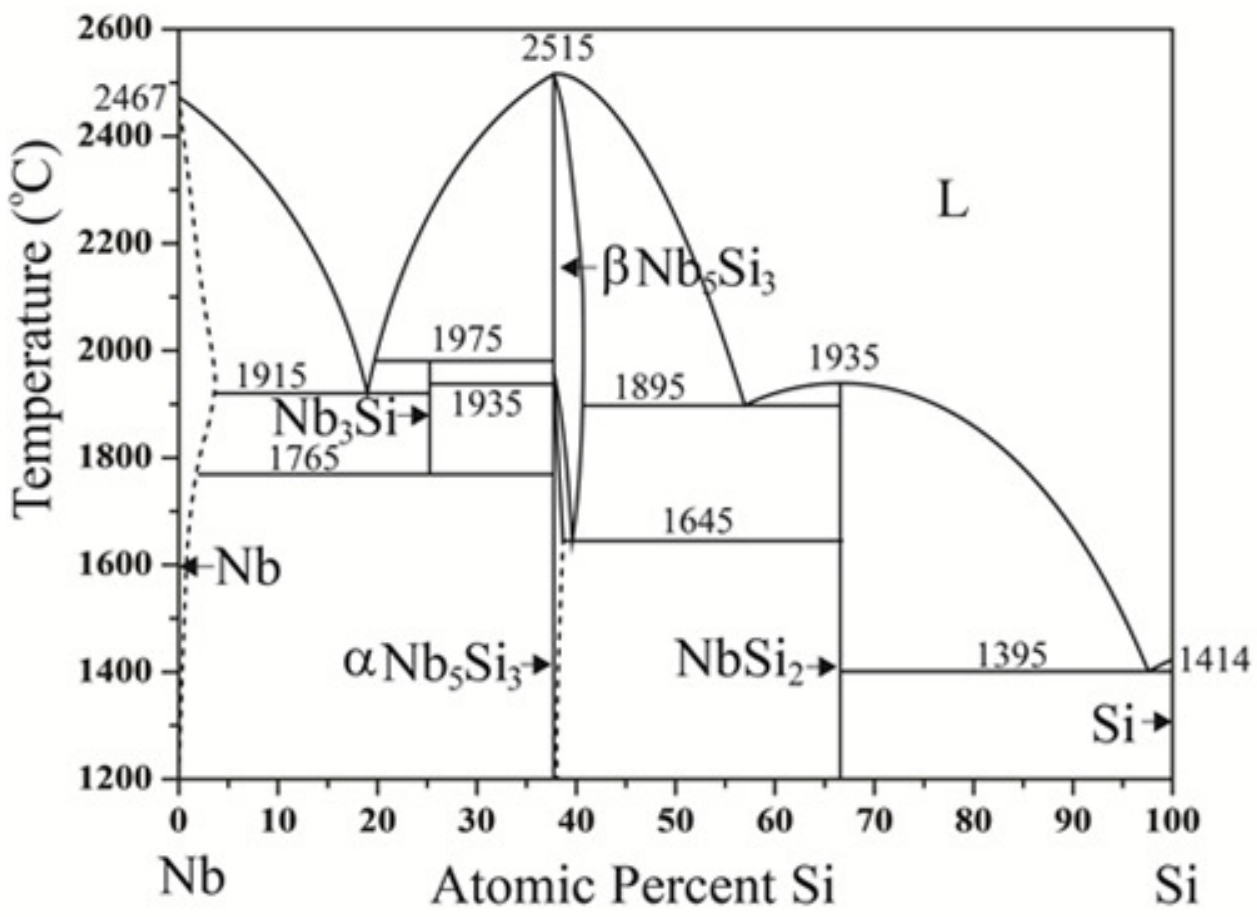

Figure 37 Nb-Si phase diagram [60].

Figure 36 shows an image of an interdiffusion zone of Nb/Si diffusion couple, which was annealed at 1250 °C for 24 h. Both the phases, which are supposed to be present according to the phase diagram as shown in Figure 37, is found in the interdiffusion zone. $NbSi_2$ phase is found to grow with much higher thickness compared to $Nb_5Si_3$. The microstructural evolution of the interdiffusion zone revealed with the help of an etchant shows an interesting feature in the $NbSi_2$ phase. A duplex morphology can be seen easily. Relatively columnar and fine grains seen in the sublayer that are grown from the $Nb_5Si_3$ phase side. Relatively equiaxed and bigger grains are found to grow from the Si end member. This kind of microstructural features helps to identify the location of the Kirkendall marker plane that is demarcated by the growth of sublayers differently from two opposite side of the interfaces [1].

Next, we estimate the integrated diffusion coefficients. The molar volume of the phases used for estimation of the data are considered as $V_m^{NbSi_2}$ = 8.7 and $V_m^{Nb_5Si_3}$ = 9.6 $cm^3$/mole for the $NbSi_2$ and $Nb_5Si_3$ phases, which are estimated utilizing the lattice parameter data available in the literature [19]. The estimated diffusion coefficients are shown in Figure 38. The data published by Milanese *et al.* [48] are included for comparison. The activation energies are estimated as 193 for the $NbSi_2$ phase and 250 kJ/mol the $Nb_5Si_3$ phase. The integrated diffusion coefficient and the activation energies reported by Milanese *et al.* [48] are close to the $Nb_5Si_3$ phase (263 kJ/mol). However, a lower value of the integrated diffusion coefficient and higher value of the activation energy in the $NbSi_2$ phase (304 kJ/mol) are reported by them.

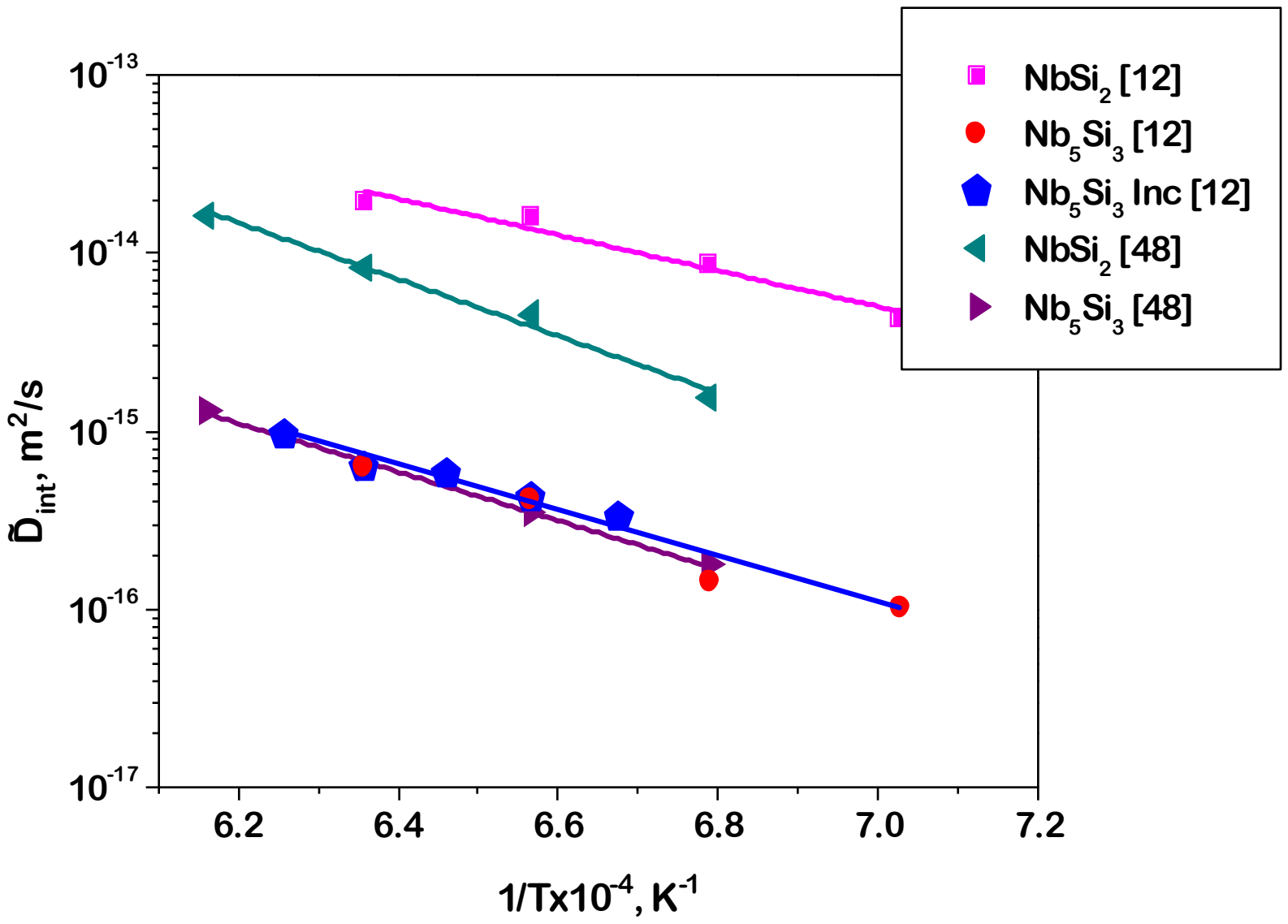

Figure 38 Arrhenius plot of the integrated diffusion coefficient of $NbSi_2$ and $Nb_5Si_3$ phases [12, 48].

It is already mentioned that the location of the Kirkendall marker plane is detected based on the morphological evolution, as indicated by K in Figure 36. The ratio of the tracer diffusivities at this position in the $NbSi_2$ phase is estimated as $D_{Si}^{*}/D_{Nb}^{*} = 4.8 \pm 1.4$. The same data could not be estimated since the marker plane is not present in the $Nb_5Si_3$ phase. Moreover, this phase is grown as a small thickness and therefore it is necessary to confirm the estimation following the incremental diffusion couple experiments in which the phase grows with higher thickness. The interdiffusion zone of Nb and (Nb - 57 at. pct. Si) alloy that is developed at 1325 °C after 24 hrs is shown in Figure 39. It can be seen that the (Nb - 57 at. pct. Si) alloy is actually a mixture of the $NbSi_2$ and $Nb_5Si_3$ phases. The interdiffusion zone of the product phase has distinctively two parts. In one part, the Kirkendall voids are present in high density and another part is free from these voids. $TiO_2$ particles were used to detect the Kirkendall marker plane as indicated in the figure, which is located at the close of the boundary of these two parts. This indicates that two sublayers, one with voids and another without voids grow differently from two different interfaces. Moreover, the formation of voids in the Si-rich side sublayer itself indicates that the diffusion rate of Si must be much higher than Nb. The thickness of the phase varies in the range of 21–36 μm in the temperature range of 1200-1325 °C. The estimated integrated diffusion coefficients are shown in Figure 38 and a good match with the data estimated in Nb-Si diffusion couple can be seen. The ratio of the tracer diffusion coefficients are estimated at the Kirkendall marker planes as $D^*_{Si}/D^*_{Nb} = 31 \pm 15$.

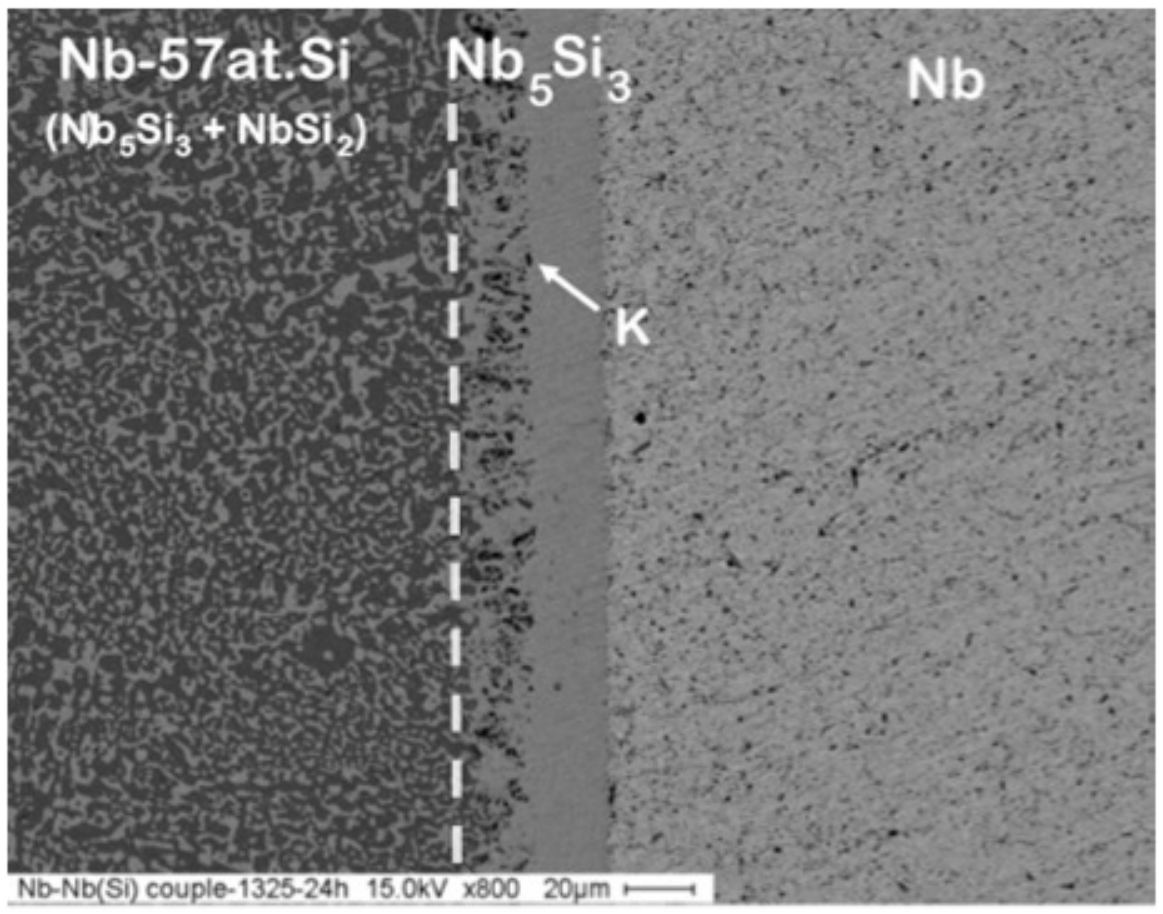

Figure 39 Image of the incremental diffusion couple annealed at 1325 °C for 24 h. Location of the Kirkendall marker plane is indicated by K [12].

Since the Nb-Si diffusion couple is found to grow with characteristic microstructural features, the physicochemical approach could be very useful to explain it, which was developed by Paul *et al.* [1, 37, 38]. This approach is phenomenological in nature and we do not need the details of the atomic mechanism of diffusion. It considers the possible reaction-dissociation equations at the interfaces and then combines with the flux of components. We consider the example of the diffusion couple annealed at 1300 °C for 16 h for this analysis as shown in Figure 36. The details of possible reaction and dissociation at different interfaces are shown in Figure 40. It should be noted here that virtually all the phases could have a Kirkendall marker plane. However, because of competition between growth and consumption, ultimately it may not present in all the phases. In most of the cases, it is present only in one phase. These are sometimes present in more than one phase. Such kind of experimental evidences are also found in many systems [1]. For the sake of physicochemical analysis, we need to consider the presence of this plane in both the product phases in the Nb-Si system. Following, it will be evident at which location(s) this is present ultimately.

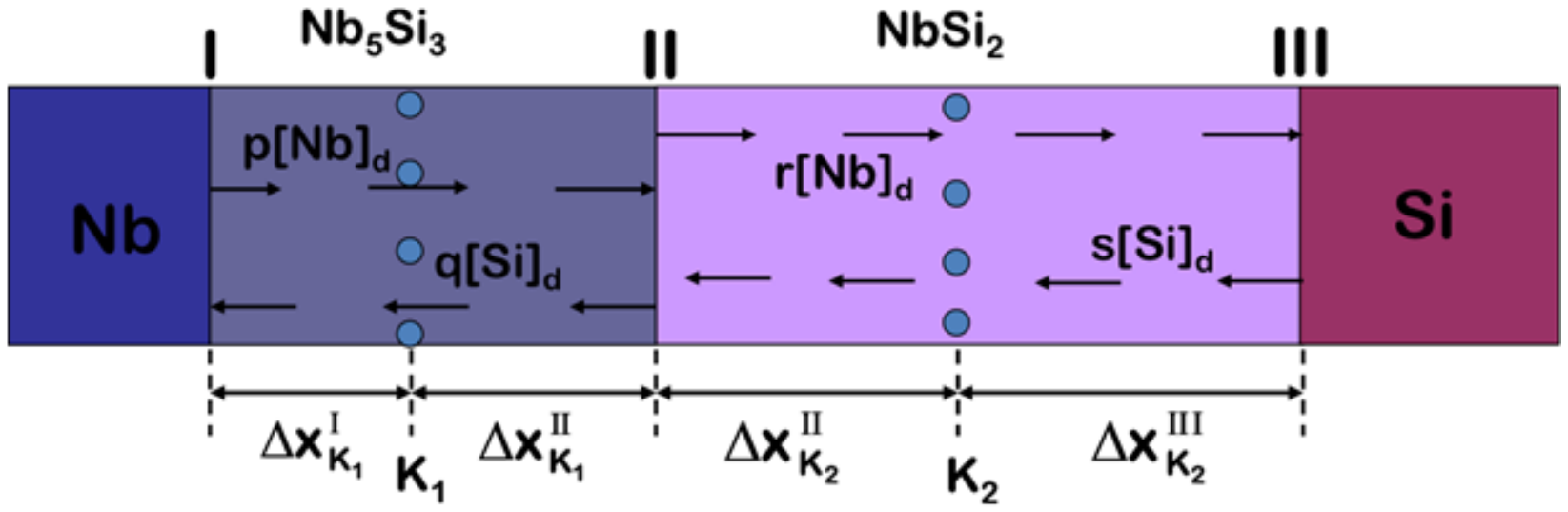

Figure 40 Schematic explanation of the reaction-dissociation in the Nb/Si diffusion couple for the growth of the product phases. The Kirkendall marker positions in the product phases are indicated

by $K_1$ and $K_2$. $\Delta X^{I}_{K_1}$, $\Delta X^{II}_{K_2}$, $\Delta X^{II}_{K_2}$ and $\Delta X^{III}_{K_2}$ are the sublayers, which are grown on different sides of the Kirkendall marker planes [12].

As described in Figure 40, Nb dissociates from the interface I which diffuse through the product phase $Nb_5Si_3$ phase to react at the interface II. At the same time, Si dissociates from $NbSi_2$ to produce $Nb_5Si_3$. This diffuse through the production phase to react with Nb for the production of the same phase. $NbSi_2$ phase also grows similarly. Nb first dissociates from $Nb_5Si_3$ at the interface II for the production of $NbSi_2$. This then diffused through the same product phase $NbSi_2$ and react with Si at the interface III for the production of $NbSi_2$. At the same time, Si dissociates from the Si end member diffuse through the product phase and react with $Nb_5Si_3$ for the production of $NbSi_2$. Therefore, it is clear that both the phases could grow from two opposite sides differently and the location of the Kirkendall marker plane is placed identifying the thickness of the sublayers. The growth process at the interface II is very complicated. A particular product phase grows by consuming the other phase and at the same time, the same phase gets consumed by the other phase. The overall thickness of the product phases and the thickness of the sublayers depend on the diffusion coefficients of the components. The reaction-dissociation equations at different interfaces can be expressed as

*Interface I*

$$\frac{5}{3}q\,Nb + q[Si]_d \Rightarrow \frac{8}{3}q\,Nb_{5/8}Si_{3/8}$$

*Interface II* ($Nb_5Si_3$ side)

$$\frac{9}{7}p\,Nb_{1/3}Si_{2/3} + p[Nb]_d \Rightarrow \frac{16}{7}p\,Nb_{5/8}Si_{3/8}$$

$$\frac{15}{7}q\,Nb_{1/3}Si_{2/3} \Rightarrow q[Si]_d \Rightarrow \frac{8}{7}q\,Nb_{5/8}Si_{3/8}$$

*Interface II* ($Nb_5Si_3$ side)

$$\frac{8}{7}s\,Nb_{5/8}Si_{3/8} + s[Si]_d \Rightarrow \frac{15}{7}s\,Nb_{1/3}Si_{2/3}$$

$$\frac{16}{7}r\,Nb_{5/8}Si_{3/8} \Rightarrow r[Nb]_d + \frac{9}{7}r\,Nb_{1/3}Si_{2/3}$$

*Interface III*

$$2r\,Nb + r[Nb]_d \Rightarrow 3r\,Nb_{1/3}Si_{2/3}$$

The terms $p$ and $q$ express the total amount of components Nb and Si in moles per unit area diffused through the product phase $Nb_5Si_3$. Similarly, the terms $r$ and $s$ express the total amount in moles per unit area of Nb and Si diffuse through the product phase $NbSi_2$. These are the total amount of fluxes after complete annealing time $t$. It can be seen that $\left(\frac{16}{7}p + \frac{8}{7}q\right)$ moles of $Nb_5Si_3$ grows at the interface II by consuming $NbSi_2$. At the same time, $\left(\frac{8}{7}s + \frac{16}{7}r\right)$ moles of the same phase gets consumed by $NbSi_2$. At the same interface, $\left(\frac{15}{7}s+\frac{9}{7}r\right)$ moles of $NbSi_2$ grows on the other side of the interface II by consuming $\left(\frac{15}{7}q+\frac{9}{7}p\right)$ moles of $Nb_5Si_3$. At interface I and III, $Nb_5Si_3$ and $NbSi_2$ grow from pure end members without getting consumed. Therefore, we can write

$$\frac{8}{3}q\times V_m^{Nb_5Si_3} = \Delta x_{k_1}^{I} \tag{7a}$$

$$\left[\left(\frac{16}{7}p+\frac{8}{7}q\right)-\left(\frac{8}{7}s+\frac{16}{7}r\right)\right]\times V_m^{Nb_5Si_3} = \Delta x_{K_1}^{II} \tag{7b}$$

$$\left[\left(\frac{15}{7}s+\frac{9}{7}r\right)-\left(\frac{15}{7}q+\frac{9}{7}p\right)\right]\times V_m^{NbSi_2} = \Delta x_{K_2}^{II} \tag{7c}$$

$$3r\times V_m^{NbSi_2} = \Delta x_{K_2}^{III} \tag{7d}$$

The amount of the product expressed in moles in the reaction-dissociation equations are equalized with the thickness of the sublayers by considering the molar volume of the phase. $V_m^{Nb_5Si_3}$ is the molar volume of the $Nb_5Si_3$ phase and $V_m^{NbSi_2}$ is the molar volume of the $NbSi_2$ phase. Following, the integrated diffusion coefficients of the phases are expressed as [1]

$$\tilde{D}_{\text{int}}^{Nb_5Si_3} = \frac{V_m^{Nb_5Si_3}}{2t}\left[N_{Nb}^{Nb_5Si_3}\,q + N_{Si}^{Nb_5Si_3}\,p\right]\left(\Delta x_{K_1}^{I} + \Delta x_{K_1}^{II}\right) \tag{7e}$$

$$\tilde{D}_{\text{int}}^{NbSi_2} = \frac{V_m^{NbSi_2}}{2t}\left[N_{Nb}^{NbSi_2}\,s + N_{Si}^{NbSi_2}\,r\right]\left(\Delta x_{K_2}^{II} + \Delta x_{K_2}^{III}\right) \tag{7f}$$

The tracer diffusion coefficients as the ratio can be expressed as [1]

$$\left.\frac{D_{Si}^*}{D_{Nb}^*}\right|_{Nb_5Si_3} = \frac{q}{p} \tag{7g}$$

$$\left.\frac{D_{Si}^*}{D_{Nb}^*}\right|_{NbSi_2} = \frac{s}{r} \tag{7h}$$

We further determine the values by solving the above equations as

$q = 2.45\ mol/m^2$ and $p = 0.08\ mol/m^2$

$s = 5.071\ mol/m^2$ and $r = 1.385\ mol/m^2$

$\Delta x_{K_1}^{I} = 62.7\mu m$ and $\Delta x_{K_1}^{II} = -58.4\mu m$

$\Delta x_{K_2}^{II} = 63.45\mu m$ and $\Delta x_{K_2}^{III} = 36.16\mu m$

It can be seen that both $\Delta x^{II}_{K_2}$ and $\Delta x^{III}_{K_2}$ have positive values indicating the presence of two sublayers in the $NbSi_2$ phase on either side of the Kirkendall marker plane. In the $Nb_5Si_3$ phase, $\Delta x^{II}_{K_1}$ is estimated as negative and $\Delta x^{II}_{K_2}$ as the positive value. The velocity diagram construction provides additional insights. These can be estimated as [1]

$$V_m^{Nb_5Si_3}\left(q-p\right)=2t\,v_K^{Nb_5Si_3}=\left(x_K^{Nb_5Si_3}-x_o\right) \tag{8a}$$

$$V_m^{NbSi_2}\left(s-r\right)=2t\,v_K^{NbSi_2}=\left(x_K^{NbSi_2}-x_o\right) \tag{8b}$$

$v_K^{Nb_5Si_3}=\dfrac{\left(x_K^{Nb_5Si_3}-x_o\right)}{2t}$ and $v_K^{NbSi_2}=\dfrac{\left(x_K^{NbSi_2}-x_o\right)}{2t}$ are the expressions of the velocity of the Kikendall planes in $Nb_5Si_3$ and $NbSi_2$ phases. $x_K^{Nb_5Si_3}$ and $x_K^{NbSi_2}$ are a distance of these planes from the initial contact planes $x_o=0$. These are estimated as

$v_K^{Nb_5Si_3}=1.975\times10^{-10}\,m/s$ and $v_K^{NbSi_2}=2.78\times10^{-10}\,m/s$

$x_K^{Nb_5Si_3}=22.75\,\mu m$ and $x_K^{NbSi_2}=32.1\,\mu m$

Utilizing these values the velocity diagram is drawn and shown in Figure 41. It can be seen that $2tv_K = x_K$ line hits the velocity curve only in the $NbSi_2$ phase indicating the presence of the Kirkendall marker plane in this phase. The location of the Kirkendall plane which could be present if the sublayer at the interface would not have consumed us indicates as $K_{virtual}$.

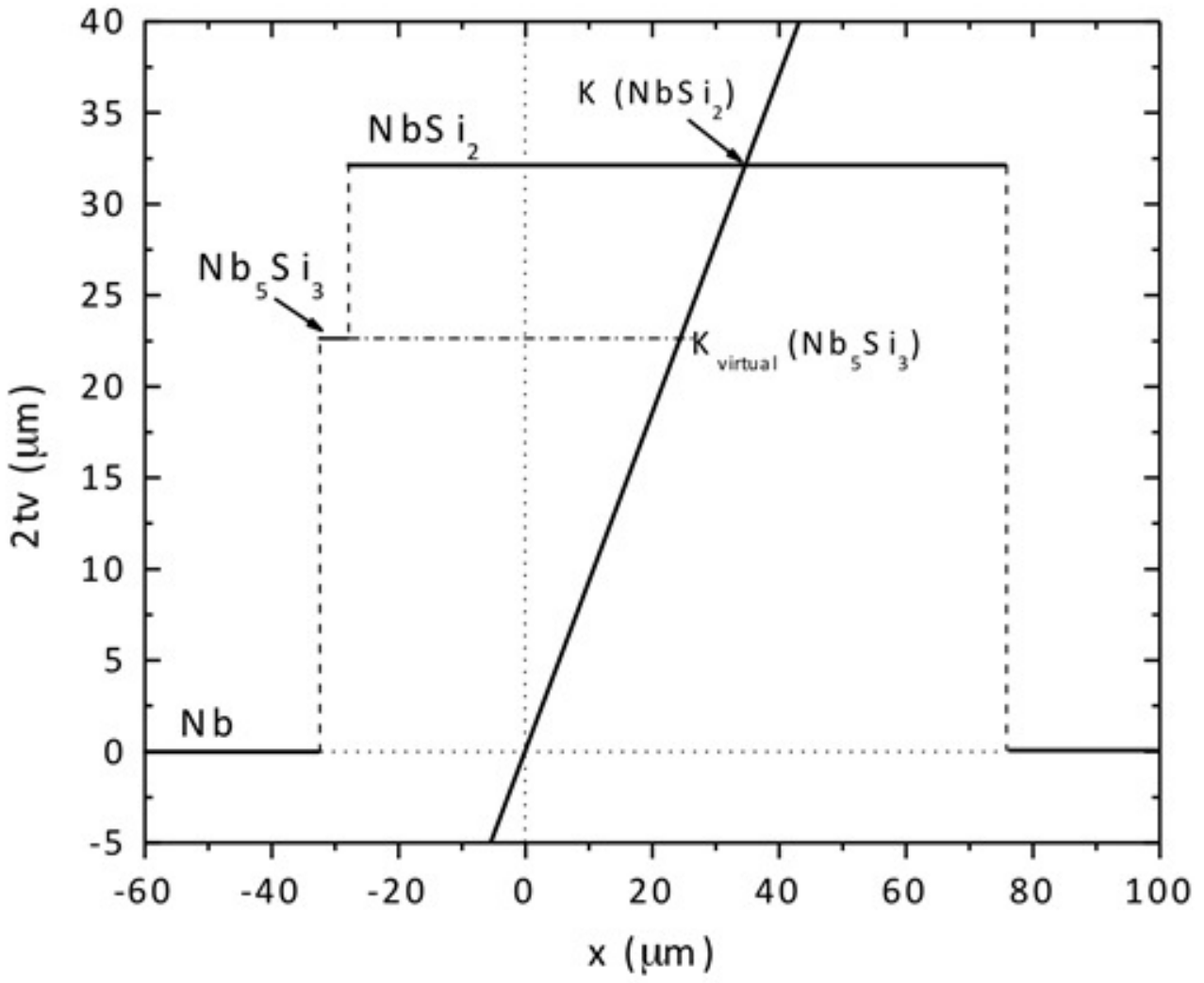

Figure 41 Velocity diagram in the Nb/Si diffusion couple that is annealed at 1300 °C for 16 hrs. K indicates the location of the Kirkendall marker plane in the $NbSi_2$ phase and $K_{virtual}$ is the virtual Kirkendall plane in the $Nb_5Si_3$ phase [12].

### 3.3 Interdiffusion study in the Ta-Si system

Tantalum silicide is important in the microelectronics industry for the use as interconnects and contacts because of thermal stability and low contact resistance [49]. This is produced by reaction-diffusion at the metal–silicon interface in the solid state. In Cu interconnects Ta was proposed to use as the diffusion barrier layer between Cu and Si/$SiO_2$ [50-53]. Cu is immiscible in Ta and therefore it does not leak in a high amount to contaminate with Si, although a small amount still could diffuse by the grain boundaries [54]. Additionally, the tantalum silicides also suitable for the high-temperature applications because of low density, high melting point and high specific strength [55, 56]. $Ta_5Si_3$ phase has higher oxidation resistance compared to other 5:3 silicides.

Interdiffusion studies reported in this system are very limited. Both bulk and thin film studies are conducted [56–58]. Christian et al. [56] conducted the incremental diffusion couple studies for the growth of $Ta_5Si_3$ in Ta/$TaSi_2$ in which the $TaSi_2$ end member was deposited by chemical vapour deposition. Milanese et al. [57] conducted Ta/Si diffusion couple experiments in which $TaSi_2$ and $Ta_5Si_3$ phases were grown. They calculated the parabolic growth constants and the average interdiffusion coefficients. However, the method followed by them for the estimation of the diffusion coefficients is erroneous. With the need for correct estimation of the data and also for the calculation of the tracer diffusion coefficients, Roy and Paul [8] studied this system extensively.

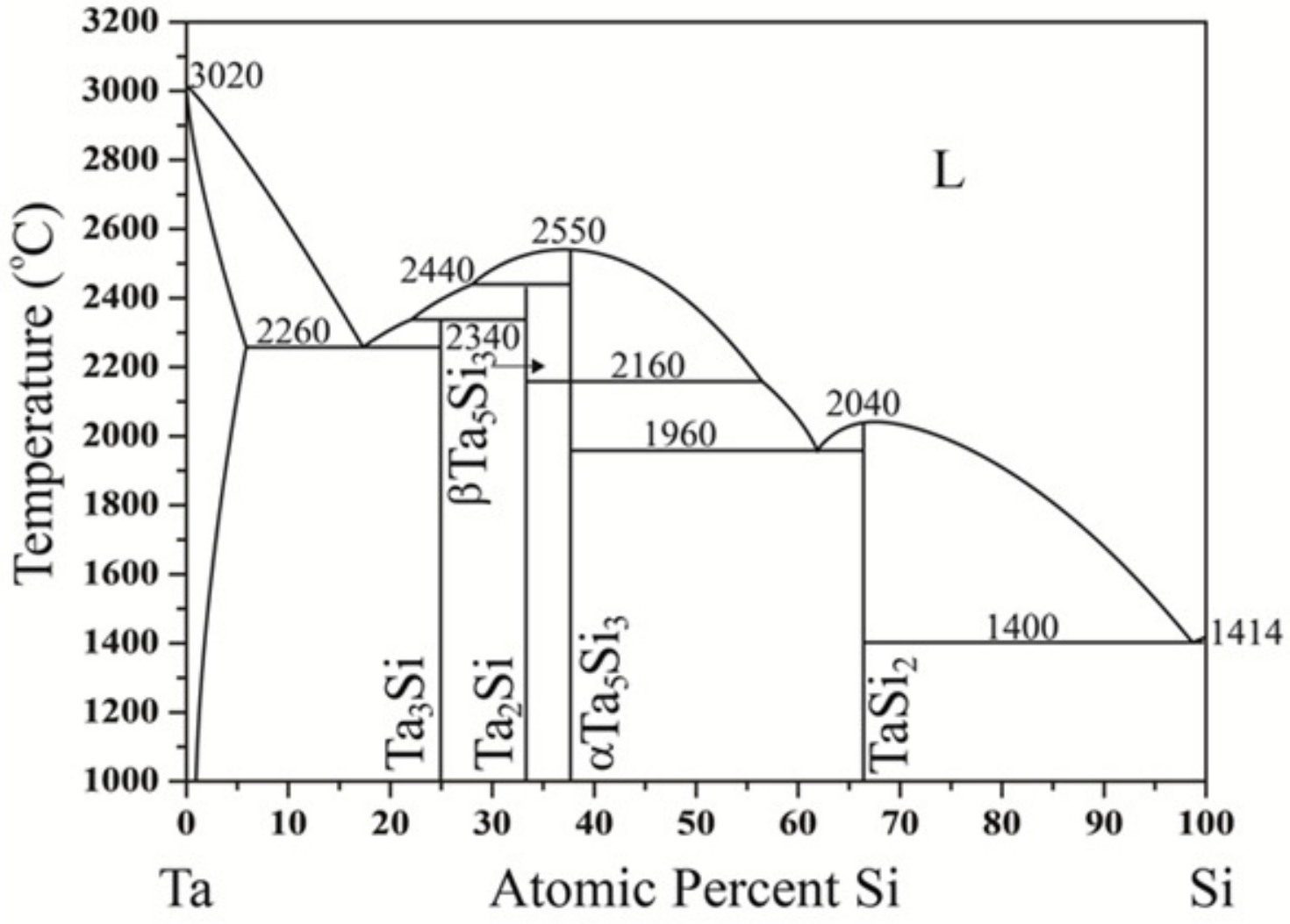

Figure 42 Ta-Si phase diagram [60].

It can be seen in the Ta-Si phase diagram in Figure 42 that there are four phases are present and therefore all these phases are expected to grow in a Ta/Si diffusion couple. Figure 43 shows the image of the Ta/Si diffusion couple annealed at 1250 °C after 9 hrs. It is evident that almost the whole interdiffusion zone is covered by the disilicide phase. A very thin layer of $Ta_5Si_3$ is also found. Other two phases $Ta_2Si$ and $Ta_3Si$ could not be found in the interdiffusion zone. It is well possible that these phases have very sluggish growth and not possible to detect in SEM. $TaSi_2$ phase thickness varies in the range of 52–145 μm in the experimental temperature range of 1200–1275 °C. On the other hand, the thickness of $Ta_5Si_3$ varies in the range of 0.9–4 μm. A duplex morphology is revealed after etching in the $TaSi_2$ phase. Fine grains have grown from the $Ta_5Si_3$ phase and a relatively coarse grain morphology has grown from the Si side. It can be noticed that the $Ta_5Si_3$ has grown with fine microstructure and therefore TaSi2 also grows with fine microstructure. When the same phase grows from Si single crystal, it has coarse microstructure. The Kirkendall marker plane demarcates these two types of structure, which are grown from different interface differently. The location of this plane is also evident by the presence of a line of pores.

The estimated integrated diffusion coefficients in the $TaSi_2$ phase is shown in Figure 44. The molar volume of the product phases are calculated as $V_m^{TaSi_2} = 8.71$ and $V_m^{Ta_5Si_3} = 9.48$ $cm^3$/mol from the lattice parameter data available in the literature [58, 59]. The activation energy is estimated following the Arrhenius equation as 550 ± 70 kJ/mol. The ratio of tracer diffusivities is estimated as $D_{Si}^{*} / D_{Ta}^{*}$ = 1.1-1.3. We have not estimated any data in the Ta/Si diffusion couple since this grows with the very thin layer. Therefore, incremental diffusion couple experiments are conducted to study diffusion in this phase.

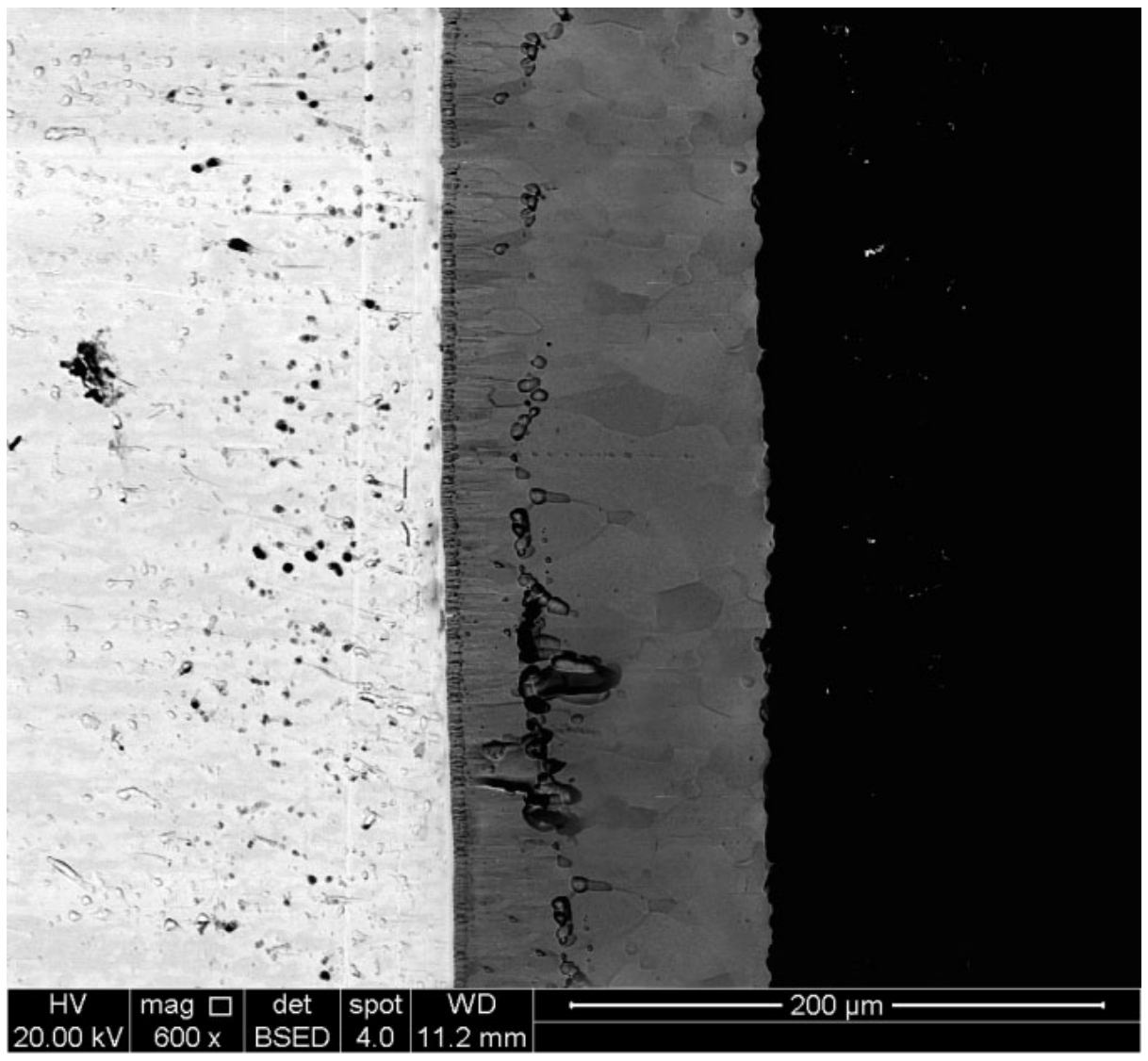

Figure 43 Interdiffusion zone of Ta/Si diffusion couple annealed at 1250 °C for 9 hours. K indicates the location of the Kirkendall marker plane [8].

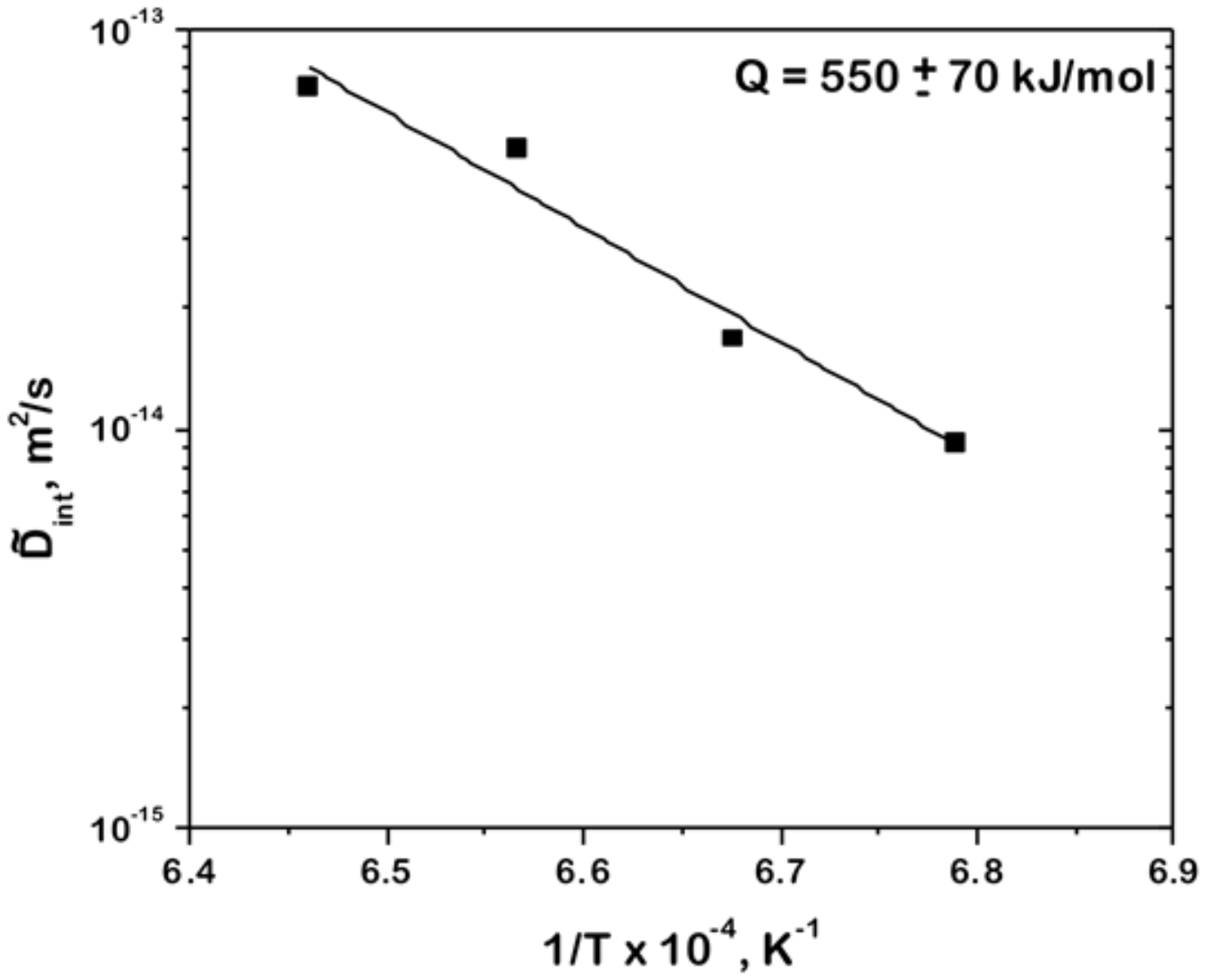

Figure 44 Integrated diffusion coefficients in the $TaSi_2$ phase [8].

The incremental diffusion couple experiments were conducted by coupling $TaSi_2$ and Ta. This is prepared by first producing the Ta/Si diffusion couple and then by removing Si end member. The initial phase layer thickness of $Ta_5Si_3$ was measured as 0.9 μm that were already present before the Ta/$TaSi_2$ diffusion couple annealing. The interdiffusion zone that was developed at 1350 °C after 16 h of annealing is shown in Figure 45. An additional $Ta_2Si$ phase is also found to grow in the interdiffusion zone, which was missing in the Ta/Si diffusion couple. Therefore, it is evident that the growth rate of this phase is much smaller compared to the other phases. Again the location of the Kirkendall marker plane can be detected by the presence of duplex morphology. The integrated diffusion coefficient in an incremental couple can be estimate utilizing only the first part of the Equation 2 i.e.

$$\tilde{D}_{\mathrm{int}}^{\beta} = \frac{\left(N_i^{\beta} - N_i^{-}\right)\left(N_i^{+} - N_i^{\beta}\right)}{N_i^{+} - N_i^{-}} \frac{\Delta x_{\beta}^{\ 2}}{2t} \tag{9}$$

Since a thin layer of the product phase was already present, the integrated diffusion coefficients are estimated using $\left[\Delta x_{\beta}^{2} - \left(\Delta x_{\beta}^{o}\right)^{2}\right]/2t$ instead of $\Delta x_{\beta}^{2}/2t$, where $\Delta x_{\beta}^{o}$ is the initial layer thickness. The estimated integrated diffusion coefficients are plotted with respect to the Arrhenius equation in Figure 46. The activation energy is calculated as 410 ± 39 kJ/mol. Subsequently, the ratio of the tracer diffusion coefficients is estimated, which are listed in Table 7.

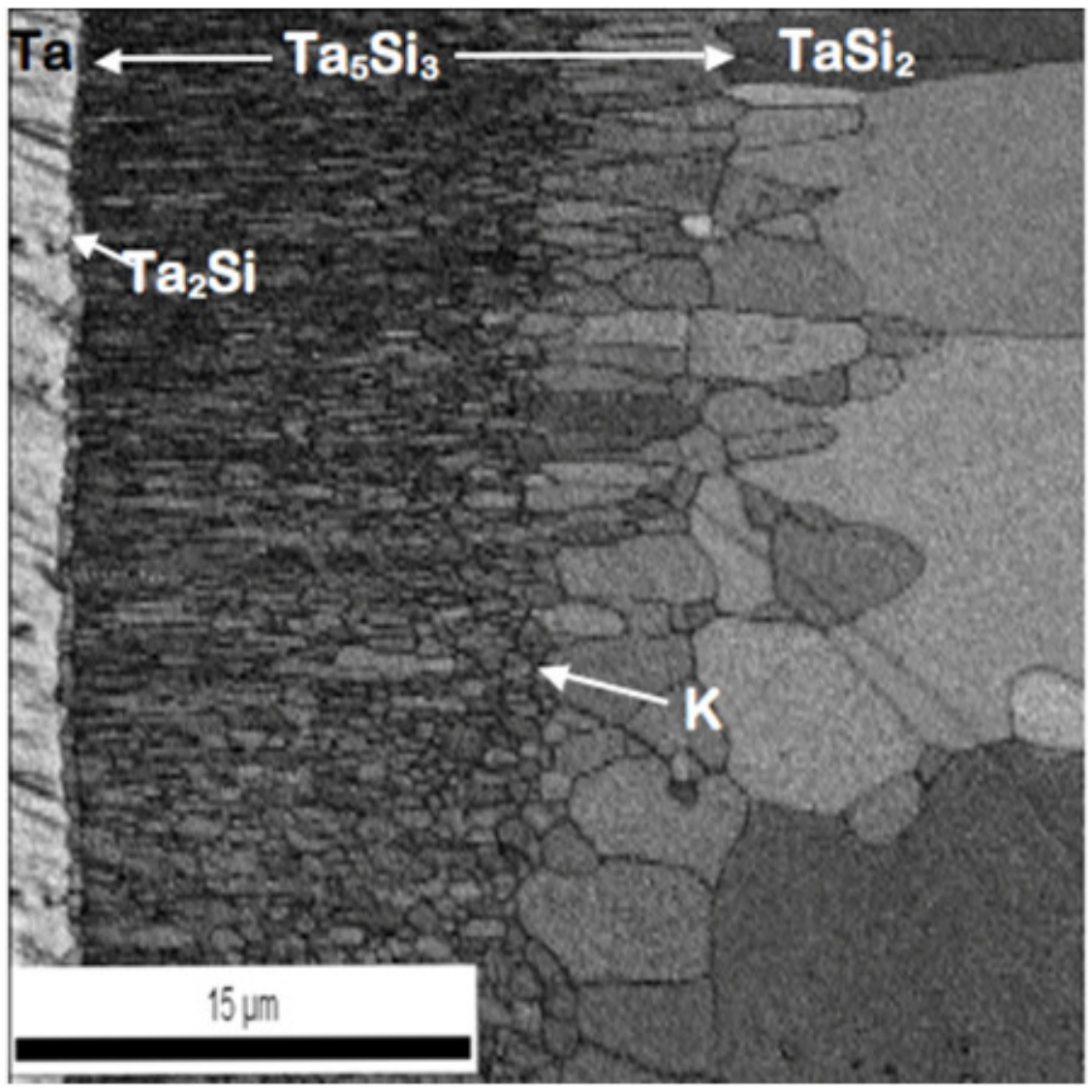

Figure 45 Incremental diffusion couple of Ta/TaSi2 annealed at 1350 °C for 16 h [8].

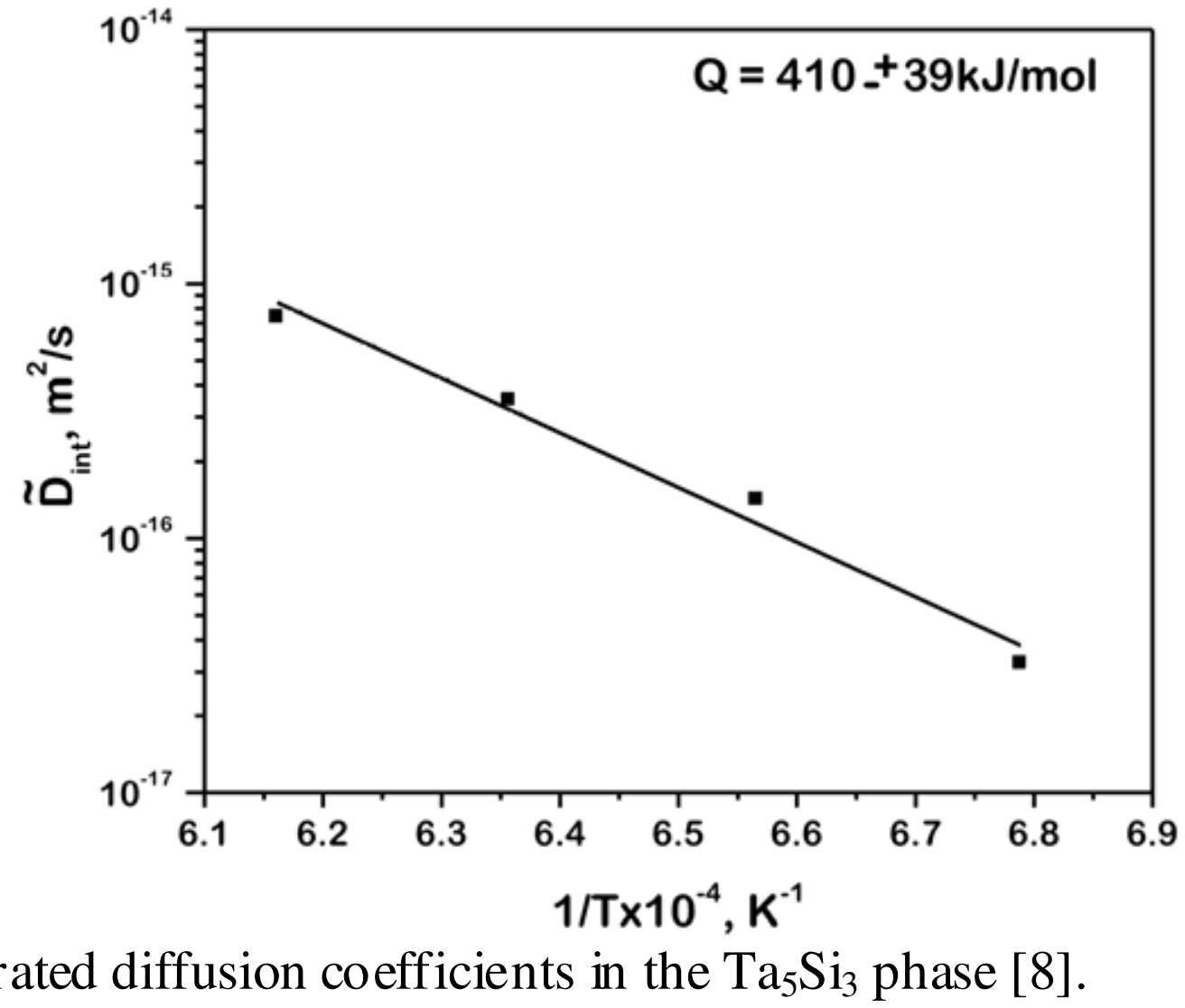

Figure 46 Integrated diffusion coefficients in the $Ta_5Si_3$ phase [8].

Table 7 The ratio of the tracer diffusion coefficients in the $Ta_5Si_3$ phase [8].

| Temperature ( $^oC$) | $D^*_{Si} / D^*_{Ta}$ in the $Ta_5Si_3$ phase |
|---|---|
| 1200 | 2.9 |
| 1250 | 5.8 |
| 1300 | 8.9 |
| 1350 | 10.8 |

As already discussed, $TaSi_2$ has a duplex morphology indicating the location of the Kirkendall marker plane in the $TaSi_2$ phase. We can do the additional analysis for understanding the growth mechanism of phases. The growth mechanism with respect to the diffusion of components is described in Figure 47 considering the Ta/Si diffusion couple at 1250 °C. Based on our analysis, we have seen that both the components diffuse in both the phases $TaSi_2$ and $Ta_5Si_3$. Following the schematic representation, Ta dissociates from the interface I and diffuse through the $Ta_5Si_3$ phase to react with the $TaSi_2$ phase at the interface II for the production of the $Ta_5Si_3$ phase. At the same interface, Si first dissociates from $TiSi_2$ and produce the $Ta_5Si_3$ phase. The same dissociated Si diffuses through the product phase $Ta_5Si_3$ to react with Ta at the interface I for the production of the $Ta_5Si_3$ phase. On the other side of the interface II, Ta dissociates from the $Ta_5Si_3$ phase such that $TaSi_2$ is produced. The same dissociated Ta diffuses through the $TaSi_2$ phase and reacts with Si to produce $TaSi_2$ phase at the interface III. Si dissociates from Si end member at the interface III diffuse through the $TaSi_2$ product phase and reacts with $Ta_5Si_3$ at the interface II to produce $TaSi_2$. Therefore, the growth process is complicated at the interface II at which $Ta_5Si_3$ grows by consuming $TaSi_2$ but also gets consumed by the growth of the $Tasi_2$ phase. In the beginning, we consider that both the product phases grow with two sublayers, as shown in Figure 47. This further means that we consider the presence of the Kirkendall marker plane in both the phases. We consider the sublayer thicknesses as $u_1$ and $u_2$ in $Ta_5Si_3$ and $v_1$ and $v_2$ in $TaSi_2$.

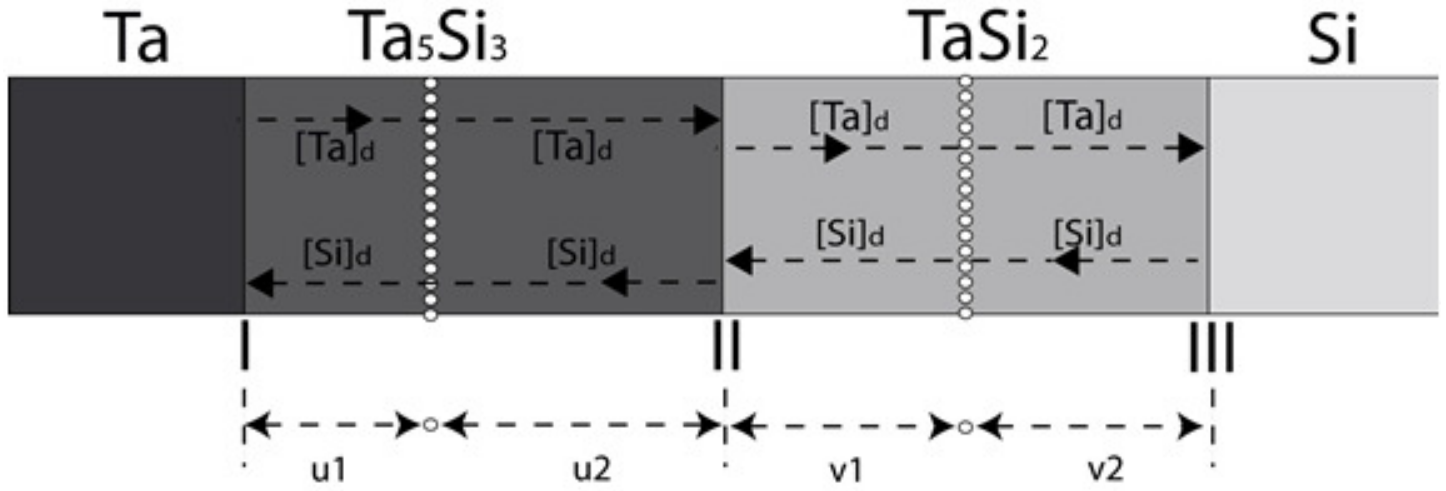

Figure 47 Schematic representation of the growth of the phases in the Ta/Si diffusion couple [8].

The integrated diffusion coefficients of the phases with respect to the thickness of the sublayers can be expressed following the physicochemical approach as

$$\tilde{D}_{int}^{Ta_5Si_3} = 1.41\times 10^{-16} m^2/\text{s} = \frac{1}{t}\left[0.117u^2 + 0.068uv\right]\times 10^{-12} \quad (10a)$$

$$\tilde{D}_{int}^{TaSi_2} = 5.02\times 10^{-14} m^2/s = \frac{1}{t}\left[0.111v^2 + 0.057uv\right]\times 10^{-12} \quad (10b)$$

Further, the ratio of the tracer diffusivities at 1250 °C can be related as

$$\left.\frac{D_{Si}^*}{D_{Ta}^*}\right|_{Ta_5Si_3} = 5.8 = \left[\frac{\frac{3}{8}u1}{\frac{1}{3}v\frac{9.48}{8.71}+\frac{5}{8}u2}\right] \quad (10c)$$

$$\left.\frac{D_{Si}^*}{D_{Ta}^*}\right|_{TaSi_2} = 1.1 = \left[\frac{\frac{3}{8}u\frac{8.71}{9.48}+\frac{2}{3}v1}{\frac{1}{3}v2}\right] \quad (10d)$$

where $u_1 + u_2 = u$ and $v_1 + v_2 = v$ (10e)

Considering the previously estimated integrated diffusion coefficients of the phases and the ratio of the diffusivities at 1250 °C, we estimate the thickness of the sublayers as $u_1$ = 64.41, $u_2$ = − 63.51, $u_3$ = 42 and $u_4$ = 79 μm. The negative value of $u_2$ indicates that the $Ta_5Si_3$ phase is present as single sublayer since one part is consumed by the neighbouring $TaSi_2$ phase. The velocity is estimated as

$$v_\beta = \frac{V_{Ta}}{V_m}\frac{\frac{D_{Si}}{D_{Ta}}-1}{\left(\frac{V_{Ta}D_{Si}}{V_{Si}D_{Ta}}\right)N_{Ta}+N_{Si}}\frac{\tilde{D}_{\text{int}}}{\Delta x_\beta} \quad (11)$$

We do not know the variation of the lattice parameter in a narrow composition range of a line compound and therefore we consider the partial molar volumes equal to the molar volume of the phase such that $V_i = V_m$. Therefore, we have the ratio of the intrinsic diffusion coefficients equal to the ratio of the tracer diffusion coefficients [1]. From Equation 11, we estimate the velocities as $1.87\text{x}10^{-10}$ *m/s* for the $Ta_5Si_3$ and $4.02\text{x}10^{-11}$ *m/s* for the $TaSi_2$ phases. These are plotted with respect to 2tv in Figure 47. Another line passing through the initial contact plane $x_o$ can be drawn utilizing $v_K = (x_K - x_o)/2t$. Since the line $2tv_K = (x_K - x_o)$ intersects the velocity line of $TaSi_2$ only, it is evident that only one Kirkendall marker plane is present in the Ta/Si diffusion couple. The virtual Kirkendall marker plane position is also indicated for the $Ta_5Si_3$ phase, which could be present if the sublayer that could be grown from $TaSi_2$ would not have consumed. Therefore, a duplex morphology is found in the $TaSi_2$ phase and a single morphology is found in the $Ta_5Si_3$ phase.

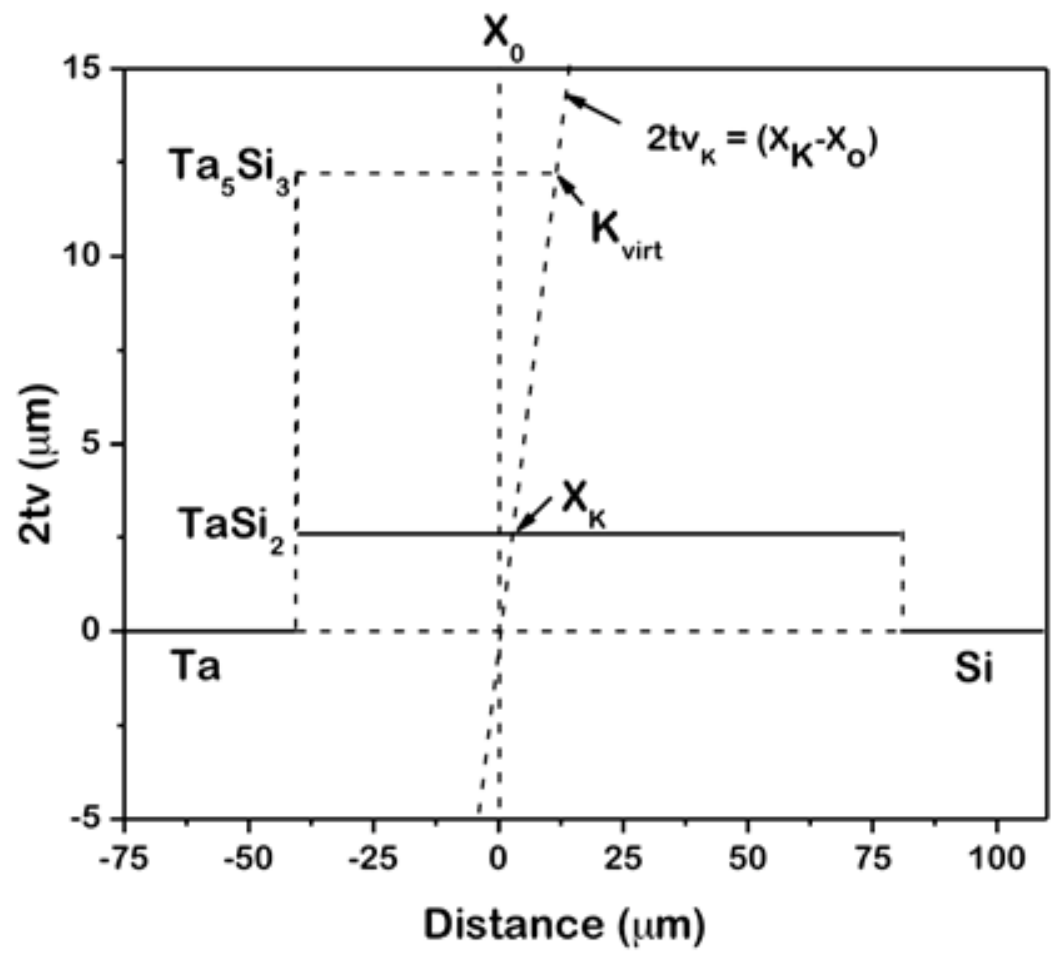

Figure 47 The velocity diagram in the Ta/Si diffusion couple [8].

### 3.4 Summary of interdiffusion studies in Group VB Metal – Silicon systems

In this section, we have discussed the interdiffusion studies in V-Si, Nb-Si and Ta-Si systems. In the V-Si system, four intermetallic compounds grow simultaneously. In the other two systems mainly the disilicide and 5:3 silicide grow in the interdiffusion zone. Therefore, we can compare the integrated diffusion coefficients of these two phases in different systems. First, we plot these parameters for the disilicide phases, as shown in Figure 48.

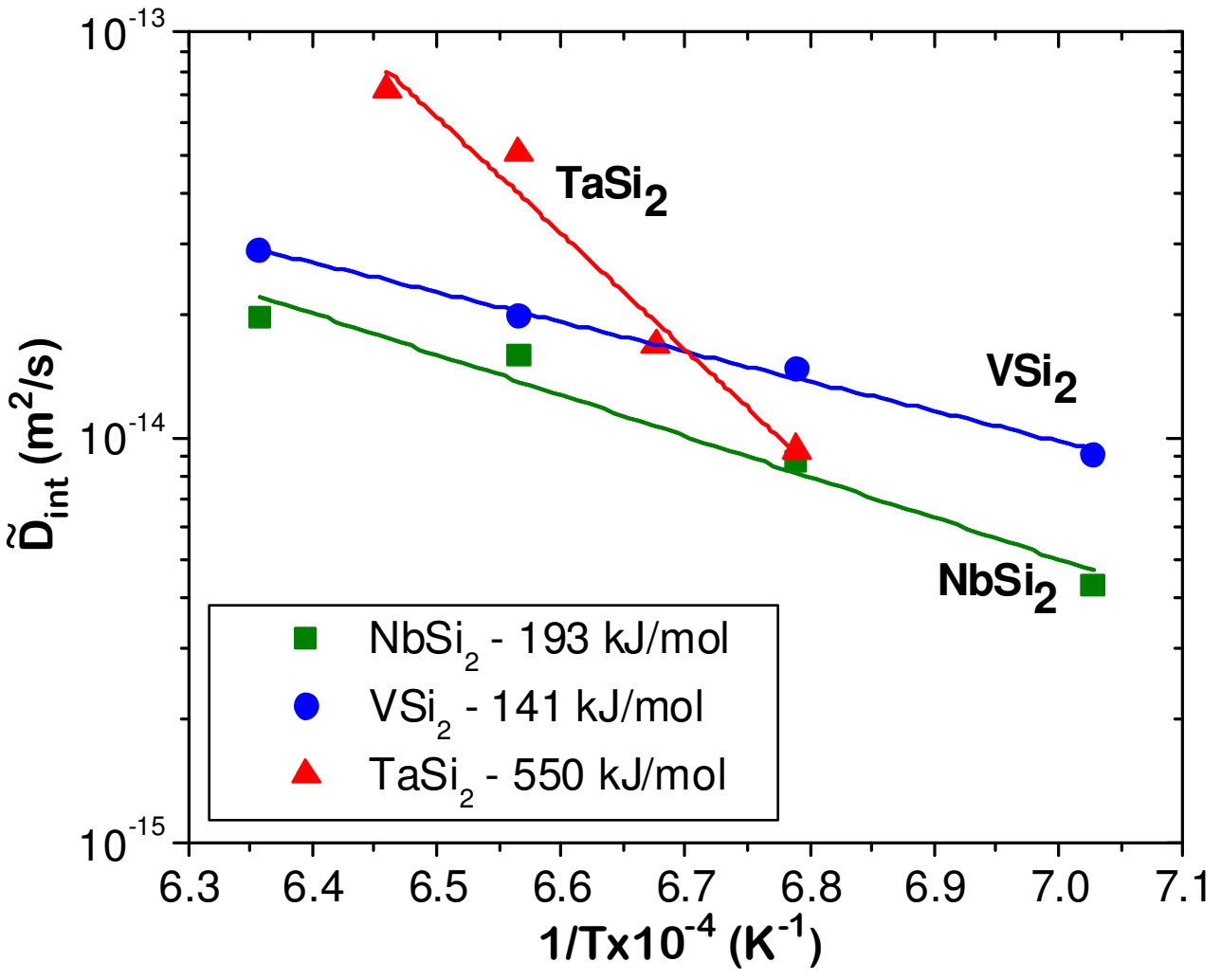

**Figure 48** Arrhenius plot of the integrated diffusion coefficients of the disilicide phases [6, 60].

It can be seen that there is an increase in the activation energy with the increasing atomic number of the refractory component. This is very much similar to the effect we have seen previously in the group IV refractory metal – silicon systems. However, it seems there is no pattern in the change in values of the integrated diffusion coefficients with the change in atomic number. At this point, it should be noted here that this comparison should be based on the homologous temperature of the compounds. Figure 49 shows the plot of the integrated diffusion coefficients with respect to the $T_m/T$, where $T_m$ is the melting point of the phase of interest and T is the experimental temperature. It can be clearly seen now that the integrated diffusion coefficients increase with the increase in an atomic number of the refractory components.

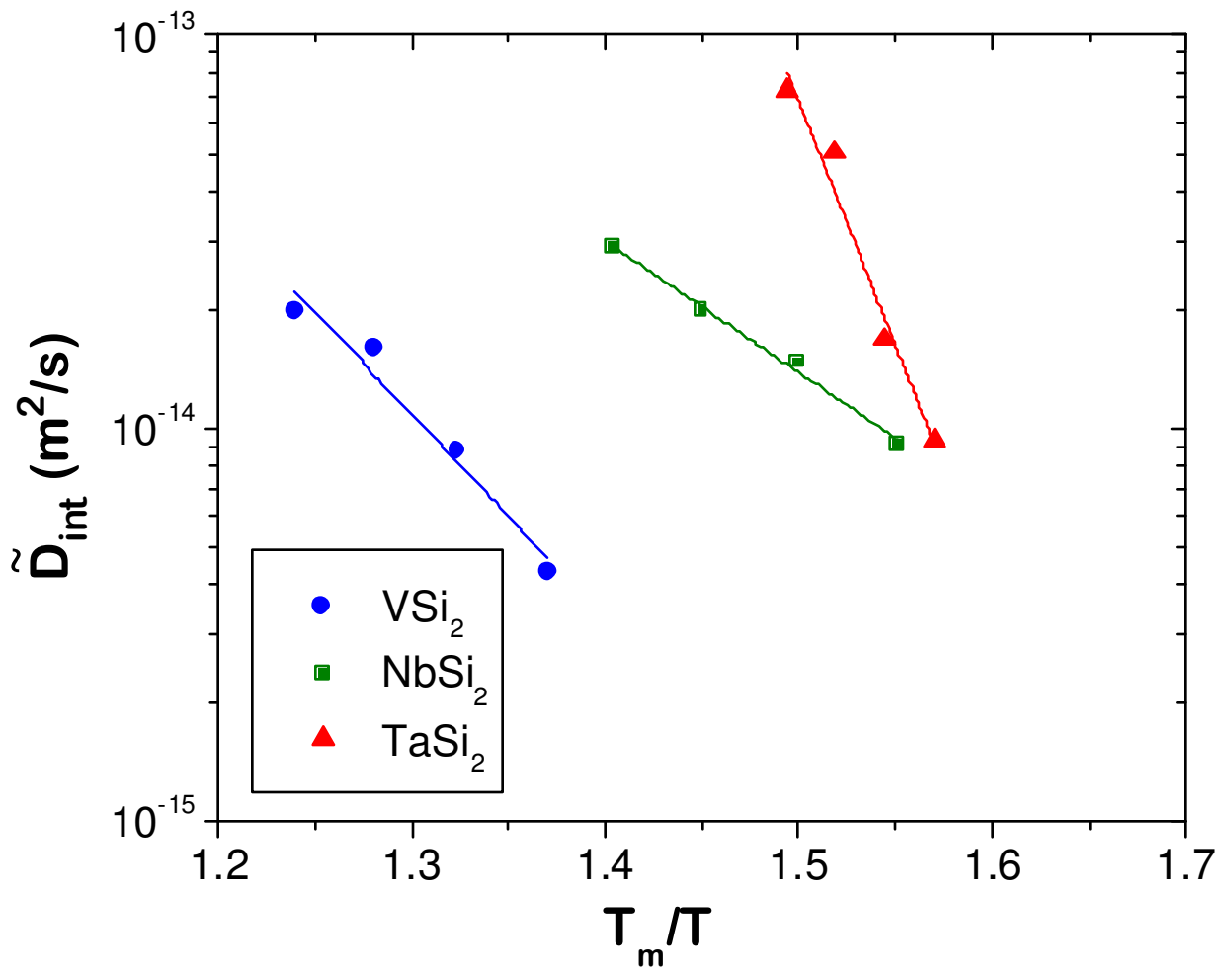

**Figure 49** Homologous temperature normalized Arrhenius plot of the integrated diffusion coefficients of the disilicide phases [6, 60].

Now we compare the integrated diffusion coefficients for the 5:3 silicides, as shown in Figure 50. When these are plotted with respect to the Arrhenius equation, it feels that the integrated diffusion coefficients decrease with the increase in atomic number of the refractory component. However, when we plot them with respect to the normalized homologous temperature, as shown in Figure 51, it is clear that integrated diffusion coefficients increase with the increase in atomic number of the refractory component. This comparison is fair considering difference in melting points of phases.

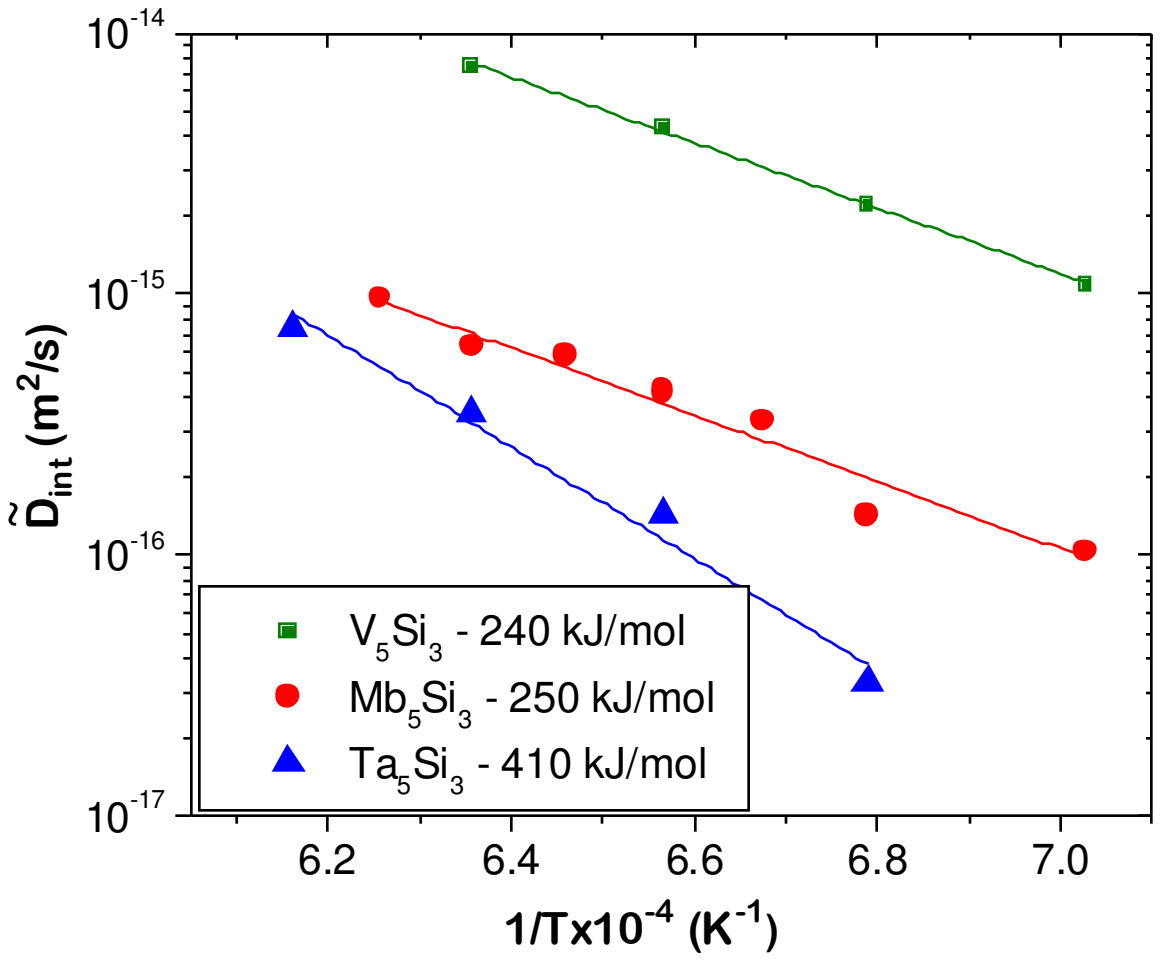

Figure 50 Integrated diffusion coefficients of the 5:3 silicides of group V refractory metal – silicon systems [6, 60].

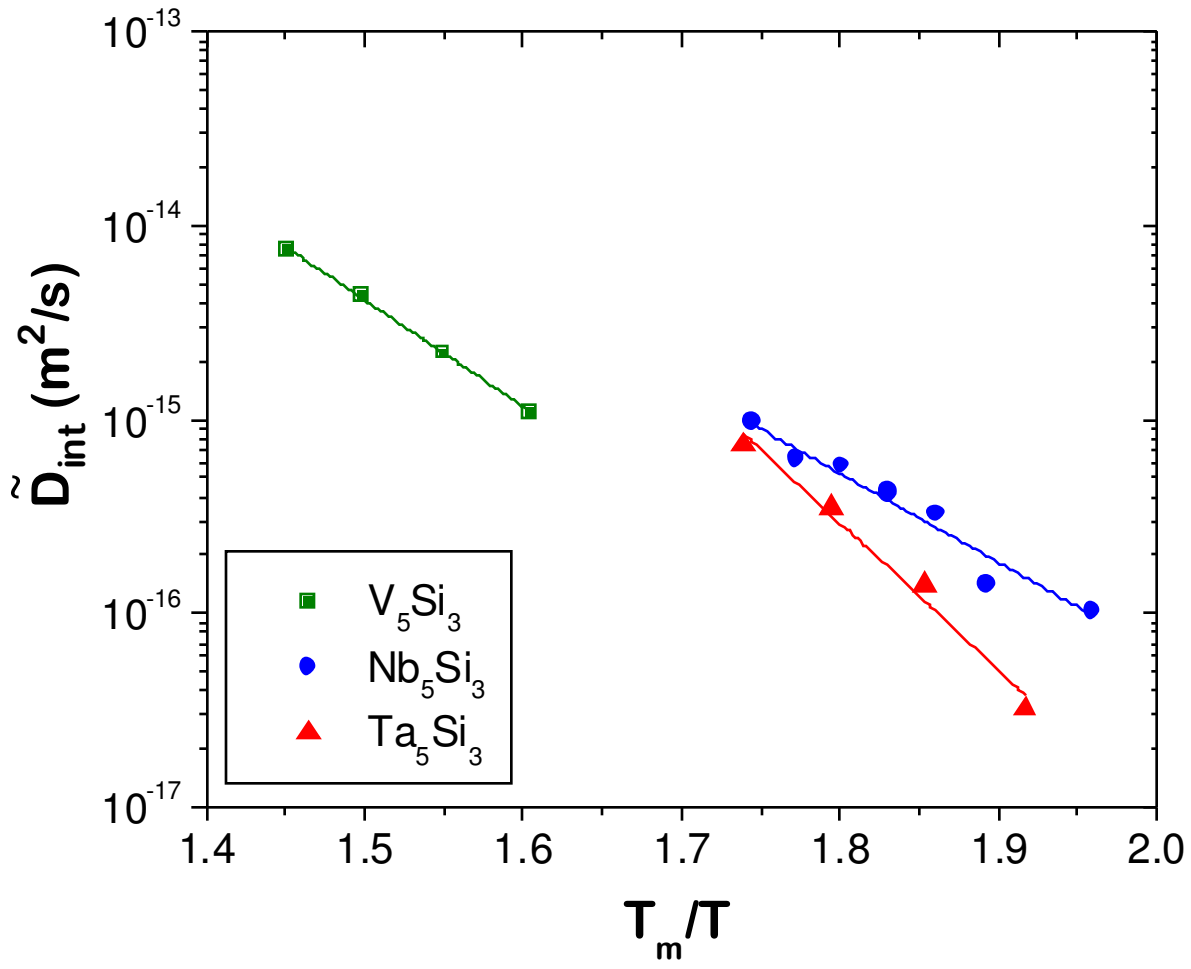

Figure 51 Homologous temperature normalized integrated diffusion coefficients of the 5:3 silicides of group V refractory metal – silicon systems [6, 60].

Now we compare the diffusion rates of components in different phases. We have already measured the ratio of the tracer diffusion coefficients in the disilicide phases as $D^*_{Si}/D^*_V = \infty$ in $VSi_2$, $D^*_{Si}/D^*_{Nb} = 4.8 \pm 1.4$ in $NbSi_2$ and $D^*_{Si}/D^*_{Ta} = 1.1 - 1.3$ in $TaSi_2$. So it is apparent that the diffusion rate of V is negligible compared to Si in the $VSi_2$ phase, the diffusion rate of Nb is a few times lower than Si in the $NbSi_2$ phase and the diffusion rate of Ta and Si are comparable in the $TaSi_2$ phase. To instigate it further let us examine these facts based on crystal structure and possible defects on different sublattices.

As a matter of fact, as shown in Figure 52, all these disilicides have the same crystal structure hP9. In this metal, atoms are surrounded by 5 Si at the next neighbour position. On the other hand, the Si atom is surrounded by 5 metal atom and 5 Si. Therefore, following the sublattice diffusion mechanism [1, 2], Si can diffuse easily via its own sublattice if vacancies are present on the same sublattice. Presence of Si antisites will enhance the diffusion rate further. On the other hand, metal components cannot diffuse even if the vacancies are present on the same sublattice or Si sublattice. Since the exchange of position with next neighbour position will bring the metal atom to the wrong sublattice i.e. sublattice of Si, which is not allowed unless metal atom antisites are present. These atoms can occupy the Si sublattice and therefore can help for direct diffusion of the metal component. These antisites must be present in all the compounds but with negligible to reasonable concentration depending on different compounds. Since the diffusion rate of V in $VSi_2$ phase is negligible, we can state that the concentration of V antisites must be negligible in this phase. In the $NbSi_2$ phase since Nb could diffuse with reasonable rate compared to Si, it is evident that Nb antisites are present to facilitate diffusion of this component. Since in the $TaSi_2$ phase Ta diffuse with comparable rate compared to Si, it is evident that the concentration of Ta antisites is reasonably high in this phase.

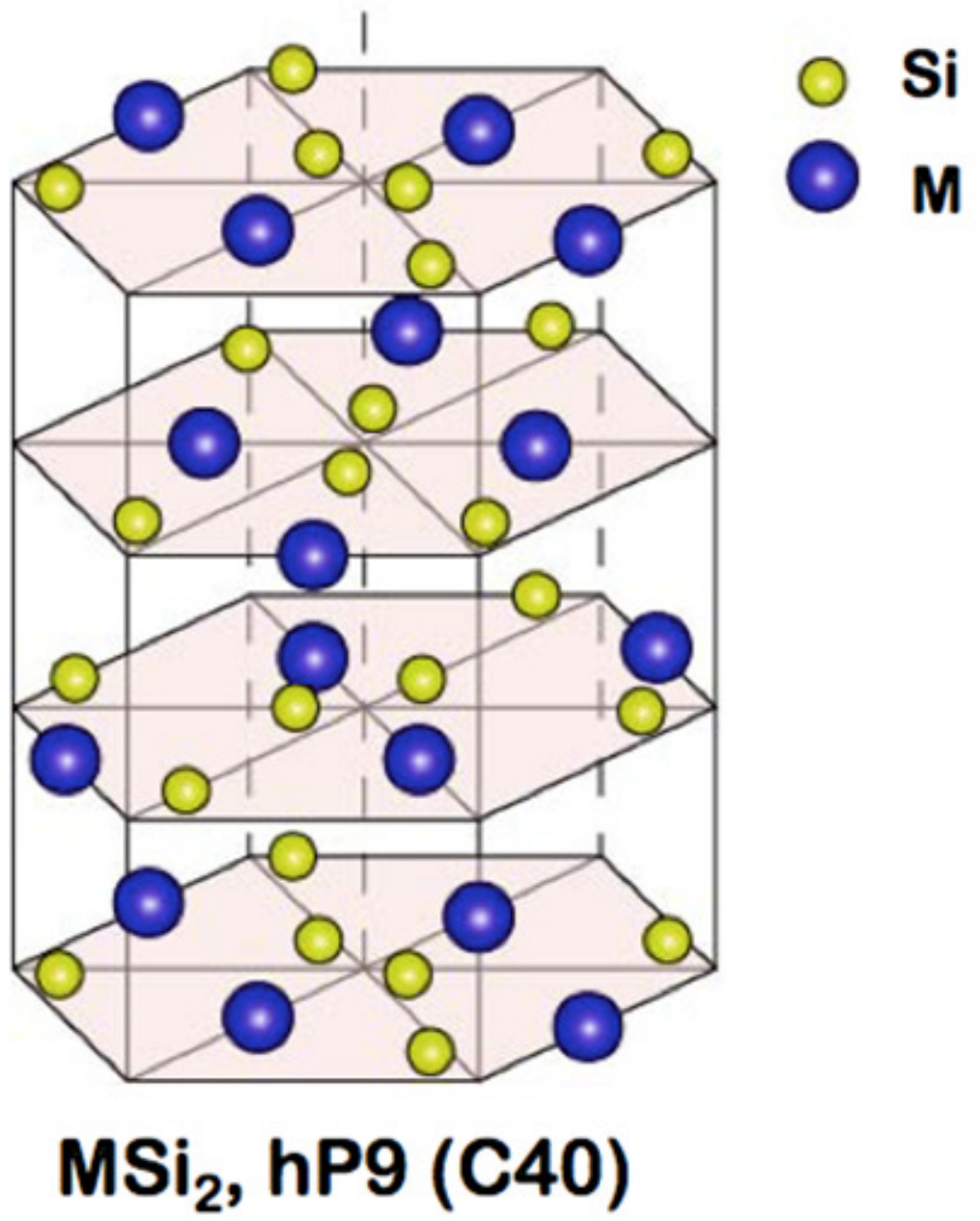

Figure 52 The crystal structure of $MSi_2$ (M = V, Nb and Ta) [6, 60].

Now we compare the diffusion rates of components in the 5:3 silicide. These are estimated only in $Nb_5Si_3$ and $Ta_5Si_3$ compounds. The data are estimated as $\frac{D^*_{Si}}{D^*_{Nb}} = 31 \pm 15$ in the $Nb_5Si_3$ phase and $\frac{D^*_{Si}}{D^*_{Ta}} = 3-11$ in the $Ta_5Si_3$ phase depending on the temperature of diffusion annealing. Therefore, Si has a much higher diffusion rate compared to Nb in the $Nb_5Si_3$ phase. Whereas, Ta also diffuse with significant rate compared to Si in the $Ta_5Si_3$ phase. Both the phases have the same crystal structure of with 32 atoms (tI32) in the unit cell, as shown in Figure 53. It has two sublattices ($\alpha, \beta$) for the metal component and two sublattices for Si ($\gamma, \delta$). The $\alpha$ sublattice of metal component (M) is surrounded by 8 M and 6 Si atoms. The $\beta$ of metal component (M) is surrounded by 11 M and 5 Si atoms. The $\gamma$ of Si is surrounded by 10 Ta atoms and $\delta$ of Si is surrounded by 8 Ta and only 1 Si atom. Therefore, M atoms are surrounded by many M and Si atoms. Therefore, these metal atoms can easily exchange positions on their own sublattice on the condition that vacancies are present on these sublattices. On the other hand, Si is surrounded by only one Si. Therefore, in the presence of only thermal vacancies, the diffusion rate of Si would be lower than the diffusion rate of the metal components. However, since Si has a higher diffusion rate in both the compounds, it is evident that Si antisites are present with high concentration.

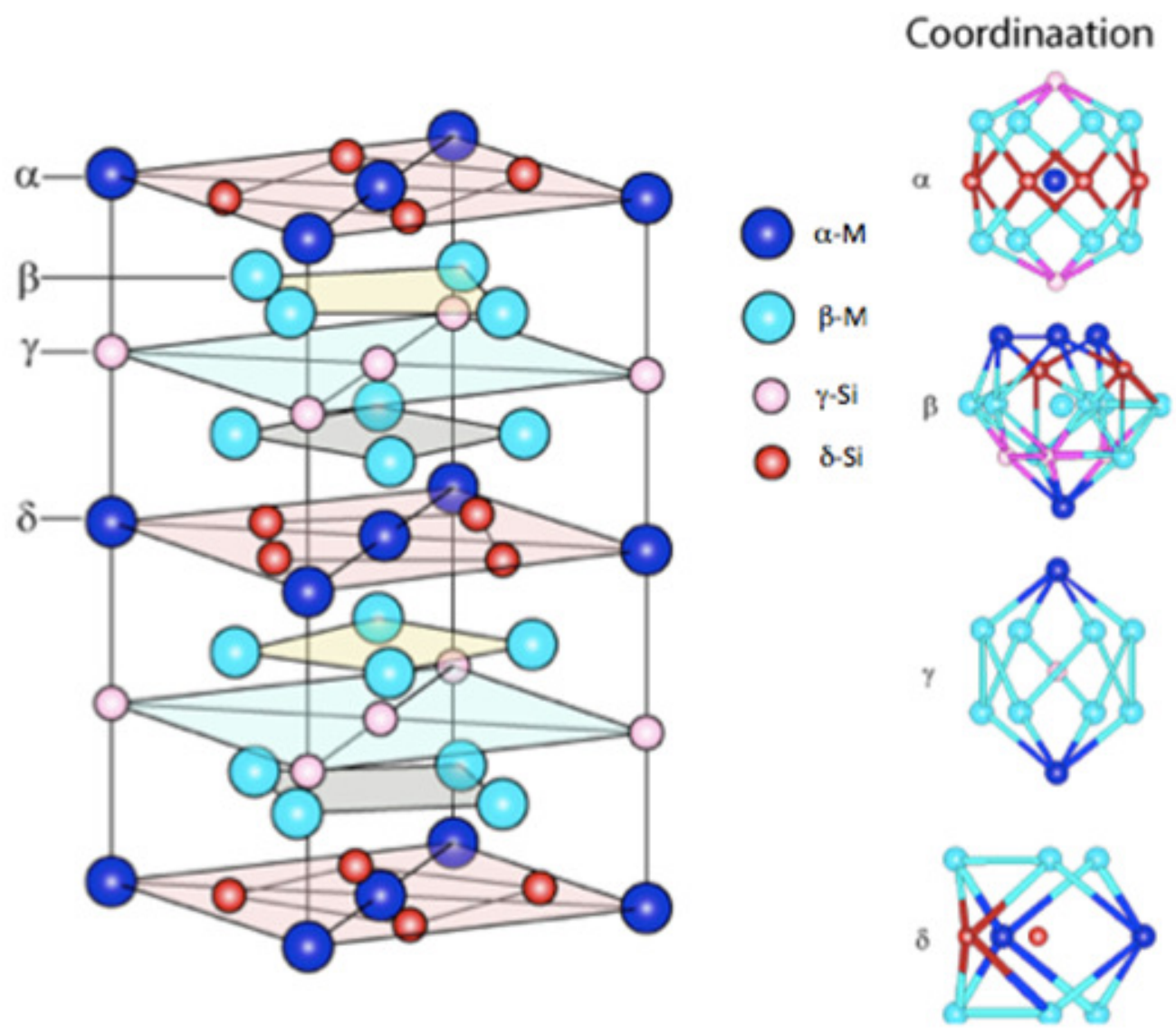

Figure 53 The $M_5Si_3$ (M = Nb and Ta) crystal structure [6, 60].

Additionally, we can make a similar discussion for $V_3Si$ with the A15 crystal structure, as shown in Figure 54. In a perfect crystal, every A (V) is surrounded by 10 A (V) and 4 B (Si), whereas, each B (Si) is surrounded by 12 A (V). Therefore, V can diffuse via its own sublattice; whereas, Si cannot diffuse unless antisites are present. The negligible diffusion rate of Si is evident from the location of the Kirkendall marker plane at the $V_3Si$/V interface, as shown in Figure 34. Therefore, it is evident that Si antisites concentration is negligible in this compound.

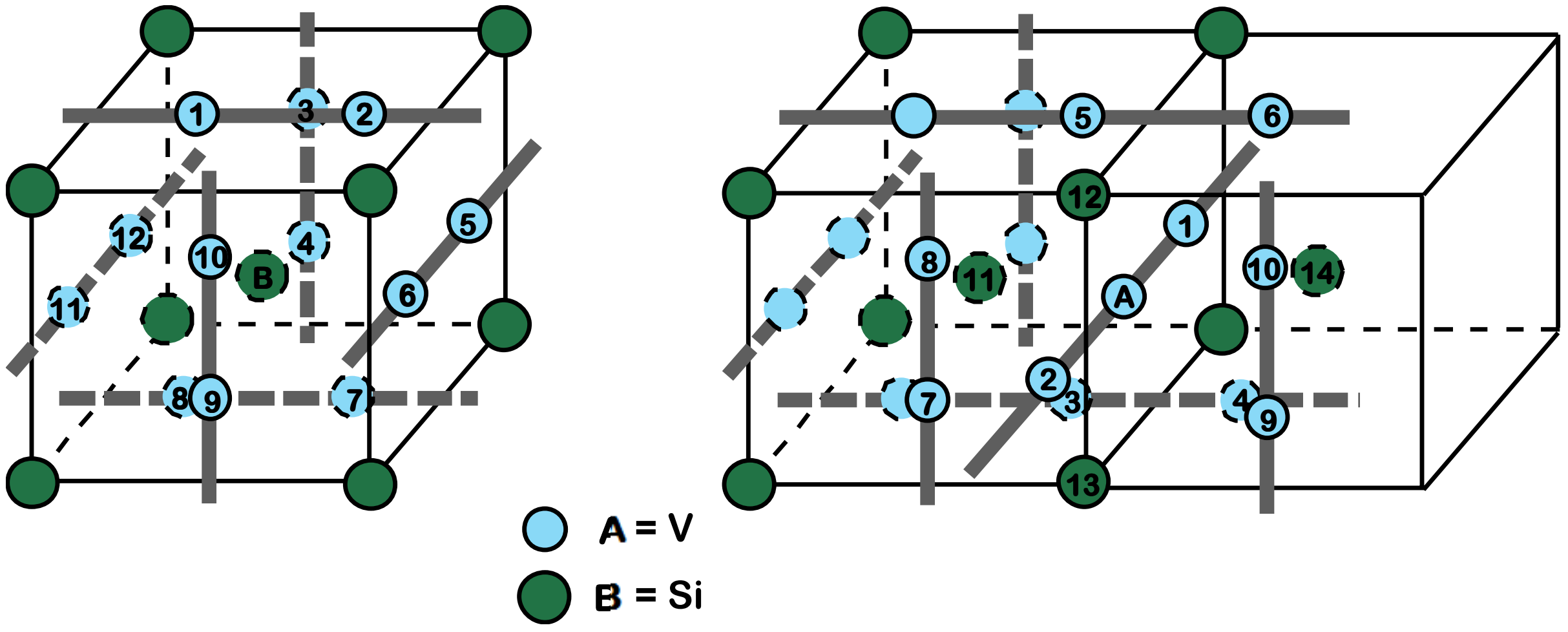

Figure 54 Atomic arrangements of $V_3Si$ with A15 crystal structure [13].

**4. Diffusion study in Group VIB (Mo and W) Metal (M)-Silicon (Si) systems**

Interdiffusion studies are conducted by many mainly in the Mo-Si and W-Si systems. There are no studies available in the Cr-Si especially in the bulk condition in which one can estimate the bulk diffusion parameters. Therefore, one couple, which was reported in the thesis of Roy [60], will be used for the comparison. Following, the diffusion studies in the Mo-Si and W-Si systems are discussed next in detail.

### 4.1 Interdiffusion study in the Cr-Si system

The Cr-Si phase diagram is shown in Figure 55. Following the phase diagram, four intermetallic compounds are expected to be found in the interdiffusion zone if all of them have the comparable growth rate. The interdiffusion zone of the Cr/Si diffusion couple annealed at 1250 °C for 16 hrs is shown in Figure 56. It can be seen that $CrSi_2$ and $Cr_5Si_3$ phases grow with a comparably high thickness in the interdiffusion zone. CrSi phase grows with very small thickness and $Cr_3Si$ phase could not be detected, which might have grown with much smaller thickness and difficult to detect in SEM. The morphological features in the $CrSi_2$ phase indicate that the Kirkendall marker plane must be present at the $CrSi_2$/Si interface. This further indicates that the diffusion rate of Si compared to Cr is infinitely (by few orders of magnitude) higher.

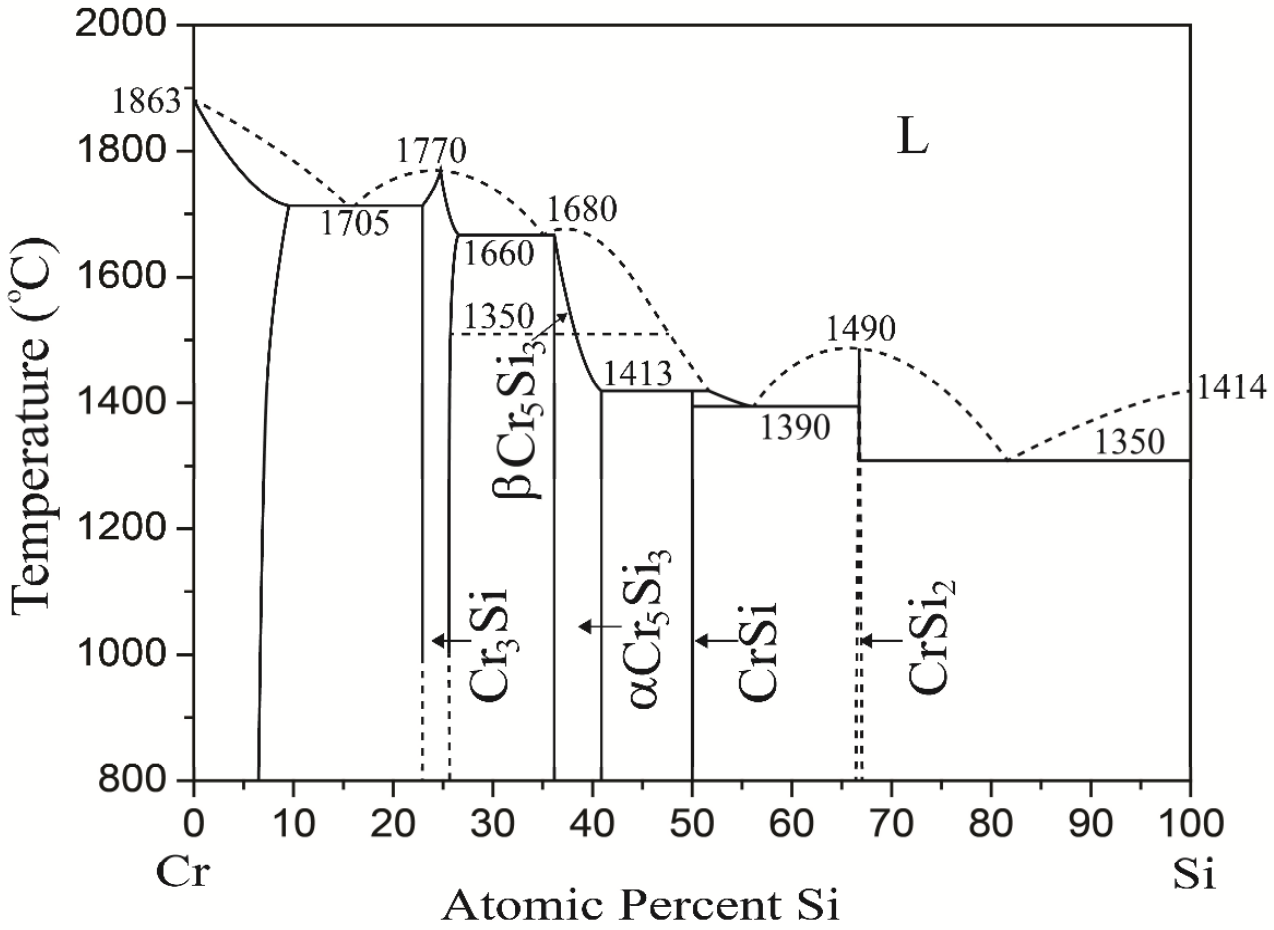

Figure 55 The Cr-Si phase diagram [60].

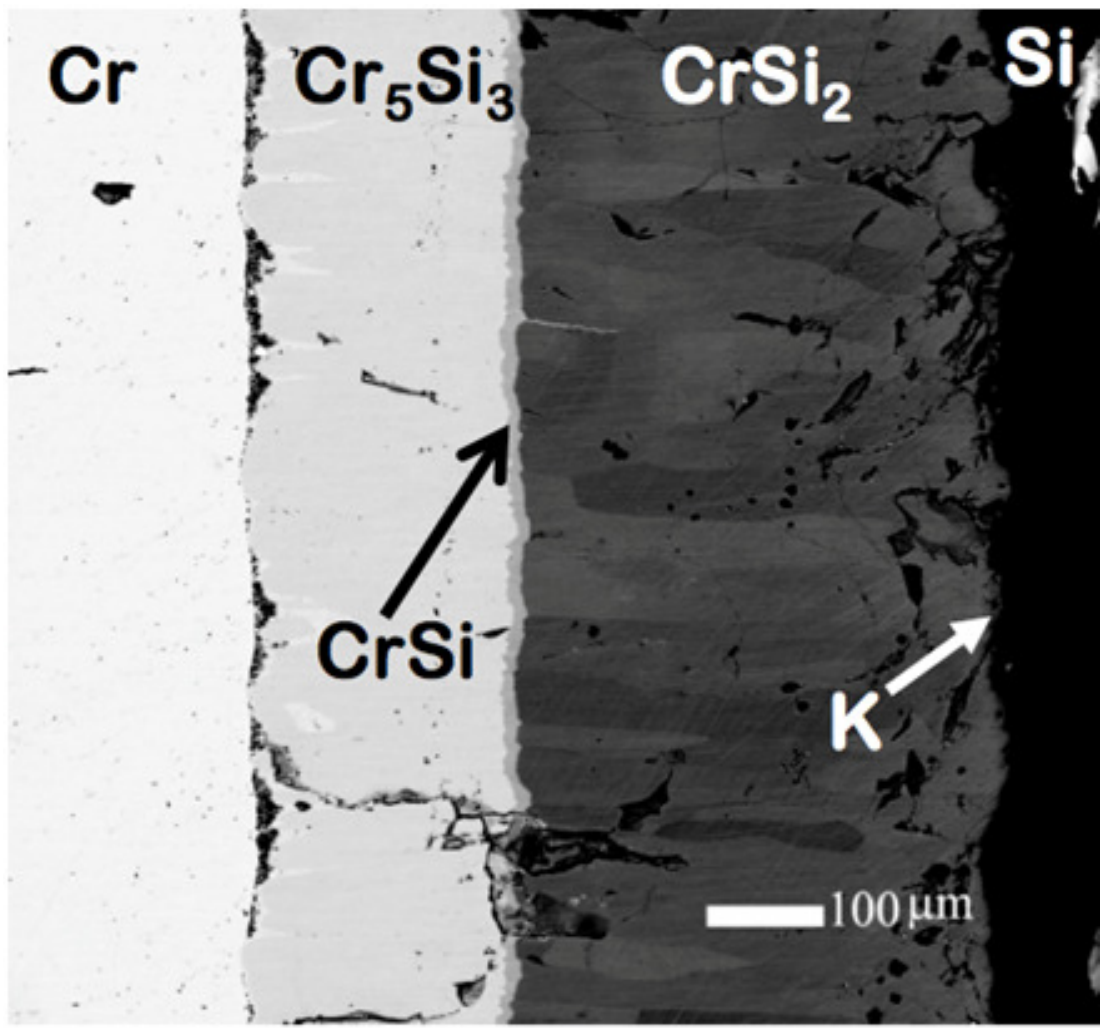

Figure 56 The Cr-Si interdiffusion zone annealed at 1250 °C for 16 hrs [60].

### 4.2 Interdiffusion study in the Mo-Si system

Mo-Si system is one of the most used systems in various applications. This is widely used as Schottky contacts in microelectronics industries [61-63]. This is also used in interconnects in VLSI (very large scale integration). Even for the fabrication of VSLI, this is used as a photomask material [64, 65]. Soft X-ray mirror is produced utilizing these materials [66-68]. Mo is successfully used as an interlayer for diffusion bonding of $Si_3N_4$ in which Mo-silicides grow at the interface. One of the oldest use of $MoSi_2$ is as the heating elements because of excellent oxidation resistance [69]. These are even considered for the development of new structural materials [70, 71].

Because of extensive use, several studies are conducted in this system concentrating mainly on the growth kinetics of the phases [72-73]. Most of the studies report the growth of $MoSi_2$ and $Mo_5Si_3$ in an interdiffusion zone of the Mo/Si diffusion couple. However, another phase *i.e.* $Mo_3Si$ is also found to exist as a thin layer [75]. Moreover, the thickness of $MoSi_2$ is found to be much higher compared to a thin layer of $Mo_5Si_3$. Therefore, logically $MoSi_2$/Mo incremental diffusion couples were prepared to study the growth of the $Mo_5Si_3$ with reasonable thickness. Following a similar logic, $Mo_5Si_3$/Mo diffusion couples were prepared to study the growth of $Mo_3Si$. Tracer diffusion coefficients are the ideal for understanding the atomic mechanism of diffusion. Therefore, Salmon et al. [84] conducted these experiments in the $MoSi_2$ phase. Their analysis indicates that Si has much higher diffusion rate (by few orders of magnitude) compared to Mo. Subsequent analysis indicates that vacancies are present mainly on the Si sublattice explaining the higher rate of diffusion of Si. The Kirkendall marker experiments were conducted by Yoon et al. [76] for the knowledge on relative mobilities of the components in $Mo_5Si_3$ and $Mo_3Si$. However, they did not estimate the diffusivities quantitatively, which could explain the types of defects present in these compounds. Prasad and Paul [10] conducted extensive analysis based on their own experiments and the results available from others, which will be described in detail here.

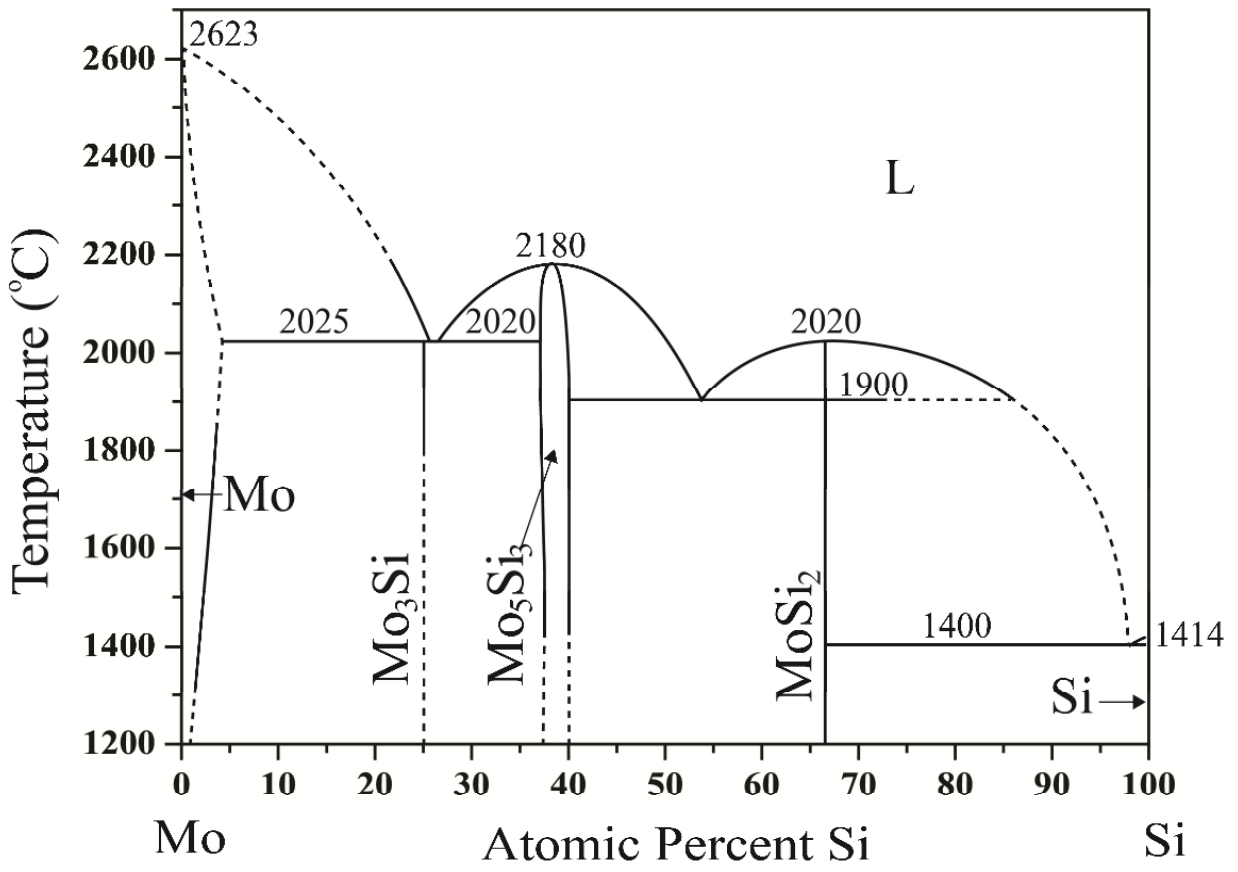

Figure 55 The Mo-Si phase diagram [10].

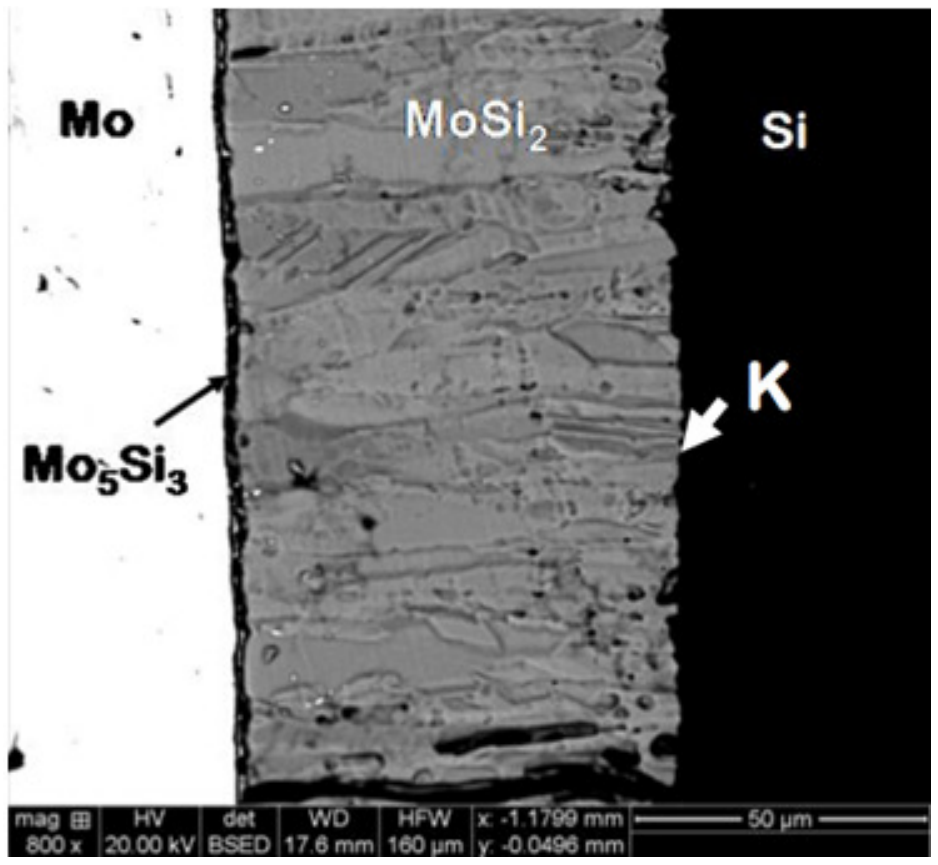

Figure 56 The Mo/Si interdiffusion zone developed at 1200 °C after annealing for 16 hrs [10].

Figure 55 shows the Mo-Si phase diagram. The interdiffusion zone developed in the Mo/Si diffusion couple at 1200 °C after annealing for 16 hrs is shown in Figure 56. Only two-phase layers, $MoSi_2$ and $Mo_5Si_3$ could be detected. The thickness of the $MoSi_2$ layer is found to be much higher than that of the $Mo_5Si_3$ phase. $MoSi_2$ phase layer has grown with columnar grains indicating the location of the Kirkendall marker plane at the $MoSi_2$/Mo interface. This means the diffusion rate of Si must be much higher compared to Mo. Previous studies conducted by Tortorici and Dayananda [83, 84] reported similar morphological features in this phase. Salmon et al. [31] have reported a similar difference in diffusion rates by the tracer diffusion experiments utilizing $^{99}$Mo , $^{31}$Si and $^{71}$Ge radio isotopes. The parabolic growth constant, $k_p$ $(\Delta x^2 = 2k_p t)$ for the $MoSi_2$ phase that is grown in Mo/Si diffusion couple, is shown in Figure 57. The data are compared between Prasad and Paul [10] and Toritorici and Dayananda [83, 84]. It can be seen that both the studies report similar results with respect to the growth constants and activation energy for diffusion.

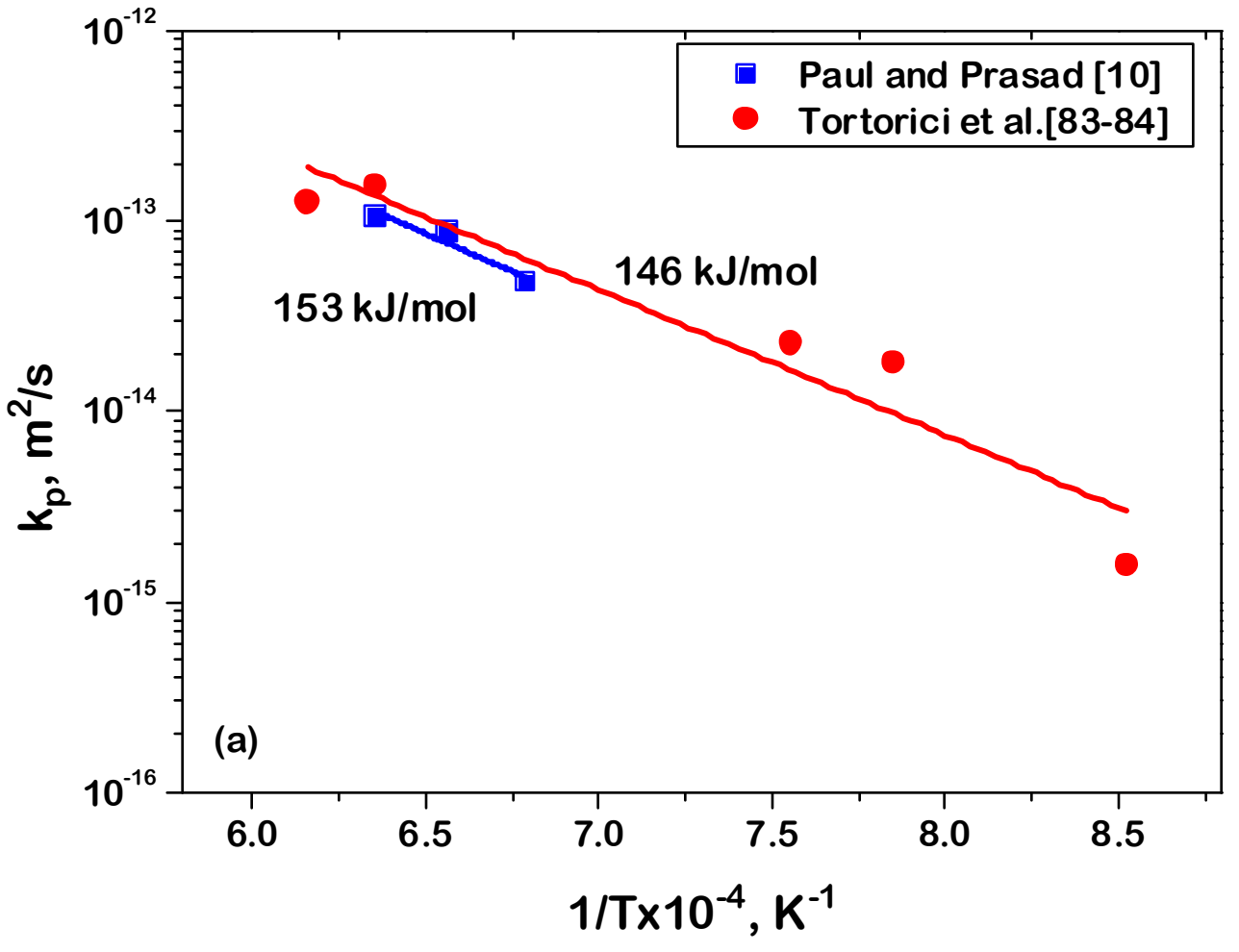

Figure 57 Parabolic growth constant of $MoSi_2$ in Mo/Si diffusion couple [10, 83-84].

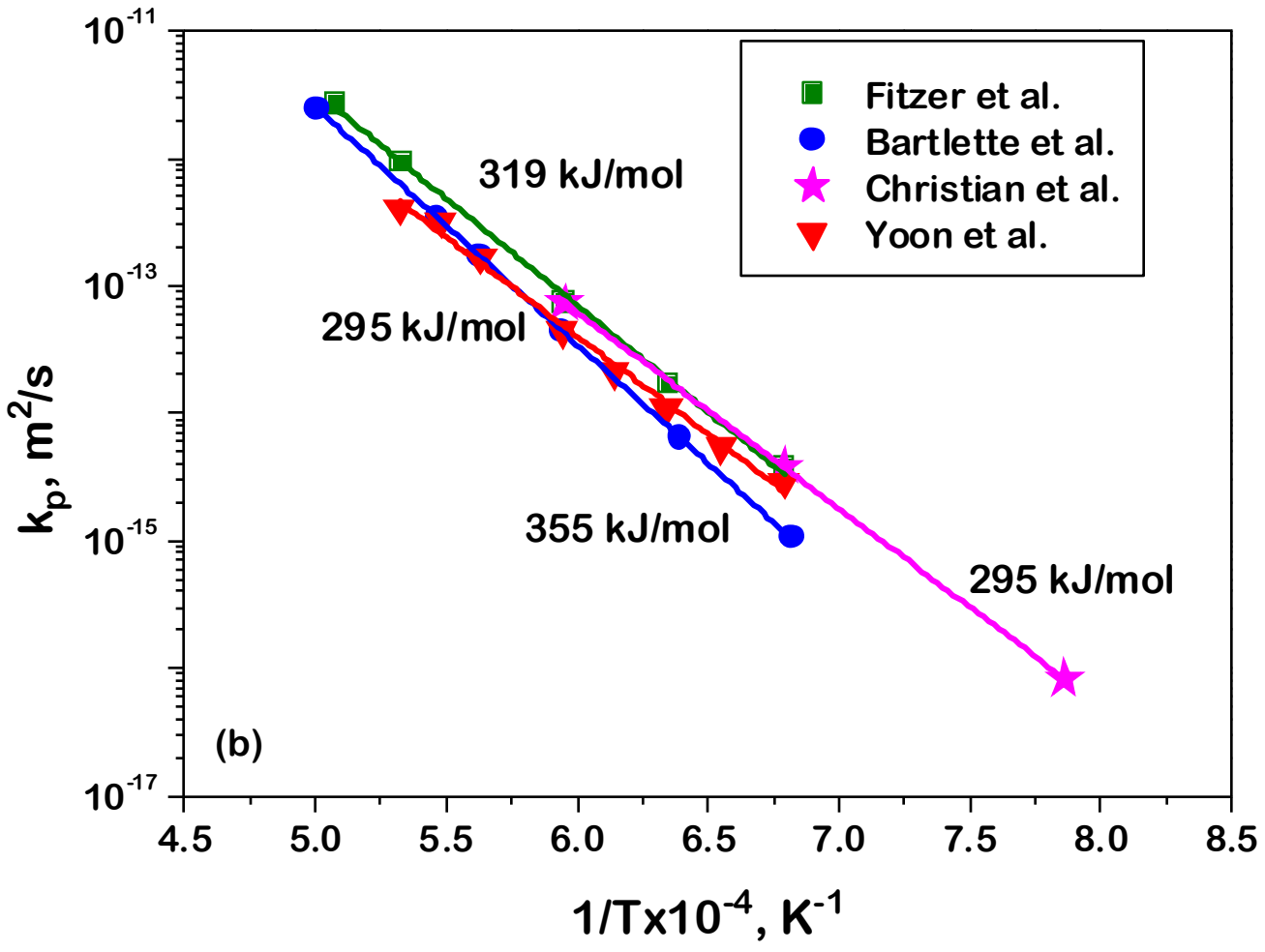

Figure 58 Parabolic growth constant of $Mo_5Si_3$ in $MoSi_2$/Mo diffusion couple [10]

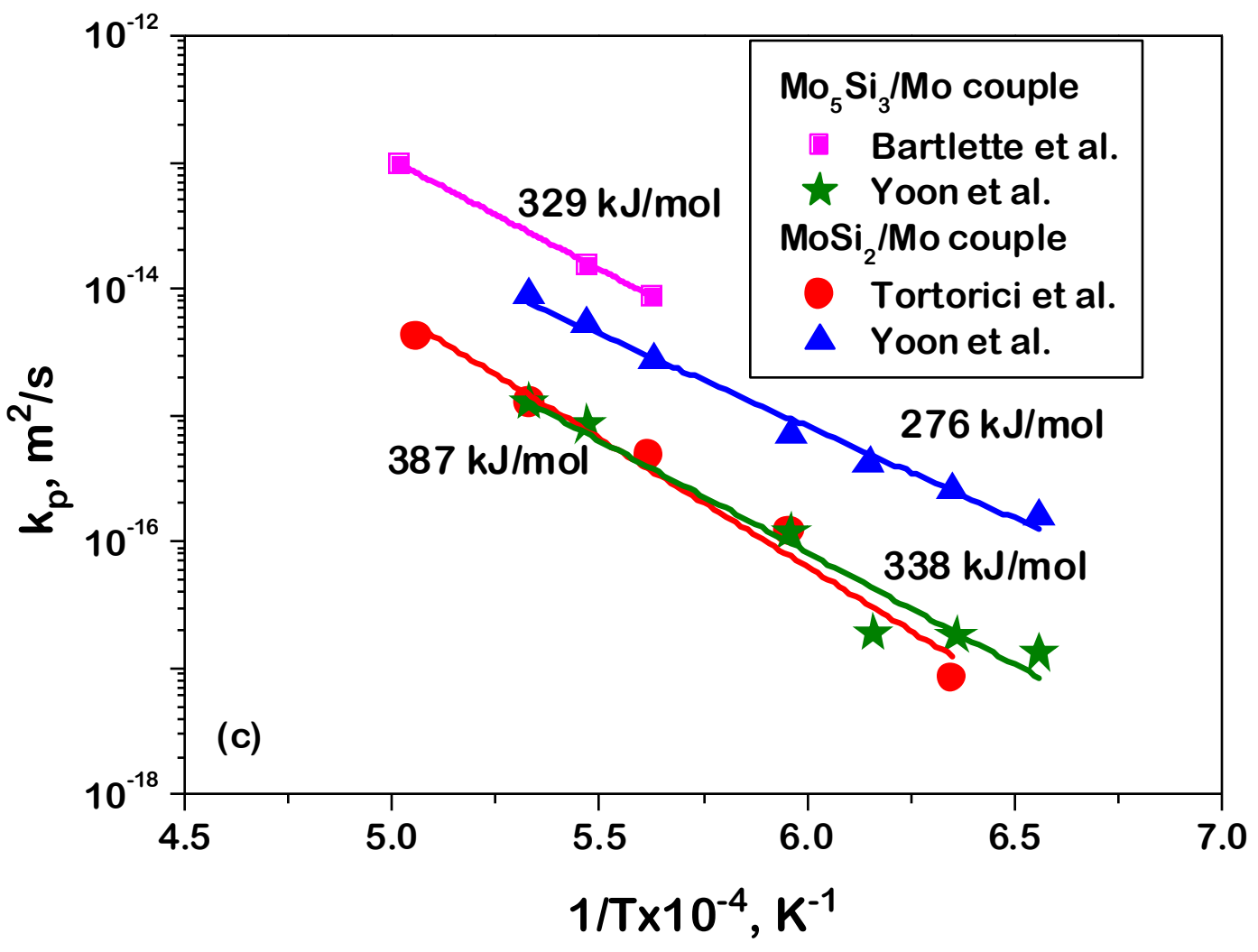

Figure 59 Parabolic growth constants of $Mo_3Si$ in different diffusion couples [10].

Figure 56 shows that the thickness of $Mo_5Si_3$ is very small in a Mo/Si diffusion couple. $Mo_3Si$ phase could not be even detected in SEM. Therefore, the parabolic growth constant values for the $Mo_5Si_3$ are estimated in the $MoSi_2$/Mo diffusion couple as reported in different studies. Yoon et al. [80-82] conducted these incremental diffusion couple experiments for the estimation of the growth kinetics of $Mo_5Si_3$ in the $MoSi_2$/Mo diffusion couple and the growth kinetics of $Mo_3Si$ in the $Mo_5Si_3$/Mo diffusion couple. It should be noted here that in the $MoSi_2$/Mo diffusion couple, both $Mo_5Si_3$ and $Mo_3Si$ grow in the interdiffusion zone, although $Mo_3Si$ grows with a very small thickness. Therefore, we consider the growth of $Mo_3Si$ only in $Mo_5Si_3$/Mo for our further analysis. They studied the growth of the phases with increase in annealing time at different temperatures to calculate the parabolic growth constant and the activation energy. The parabolic growth constants of $Mo_5Si_3$ and $MoSi_3$ are shown in Figure 58 and 59.

Parabolic growth constant values are not the materials constants and therefore the integrated diffusion coefficients should be estimated, which do not depend on the end member compositions. Molar volume of the phases used for these calculations is $V_m^{MoSi_2} = 8.1$ , $V_m^{Mo_5Si_3} = 8.6$ and $V_m^{Mo_3Si} = 8.8$ cm$^3$/mol. $\tilde{D}_{\text{int}}$ values estimated for the $MoSi_2$ phase by Prasad and Paul [10] and Toritorici and Dayananda [83-84] are compared as shown in Figure 60. $\tilde{D}_{\text{int}}$ calculated for $Mo_5Si_3$ phase by Toritorici and Dayananda [83-84] are shown. Yoon et al. [80-82] calculated only the parabolic growth constants. Utilizing these values, Prasad and Paul estimated the integrated diffusion coefficients of $Mo_5Si_3$ and $Mo_3Si$, which are shown in the same plot. It can be seen that the diffusion coefficients decrease and the activation energies increase in Mo-rich phases.

Since the Kirkendall marker plane in the Mo/Si diffusion couple is found to be located at the $MoSi_2$/Si interface, it is evident that Si has much higher diffusion rate in the $MoSi_2$ phase such that $D_{Si}^* / D_{Mo}^* = \infty$. This is in fact not actually infinitely higher diffusion rate but the diffusion rate of Si must be few orders of magnitude higher than Mo. Tracer diffusion studies indicate the same [31]. Yoon et al. used the inert markers to locate the Kirkendall marker plane; however, they did not estimate the intrinsic diffusion coefficients in the other two phases. In the $MoSi_2$/Mo diffusion couple, the marker plane was located in $Mo_5Si_3$. Utilizing the results published by them, Prasad and Paul [10] estimated the ratio of the tracer diffusion as $D_{Si}^* / D_{Mo}^* = 103 \pm 40$. Therefore, the Si diffusion rate is almost two orders of magnitude higher than Mo in the $Mo_5Si_3$ phase. Similarly, Yoon et al. prepared $Mo_5Si_3$/Mo diffusion couples at 1600 $^oC$ in which the Kirkendall marker plane was found in the $Mo_3Si$ phase. The ratio of the tracer diffusion was then estimated by Prasad and Paul [10] as $D^*_{Si} / D^*_{Mo} = 4.7$.

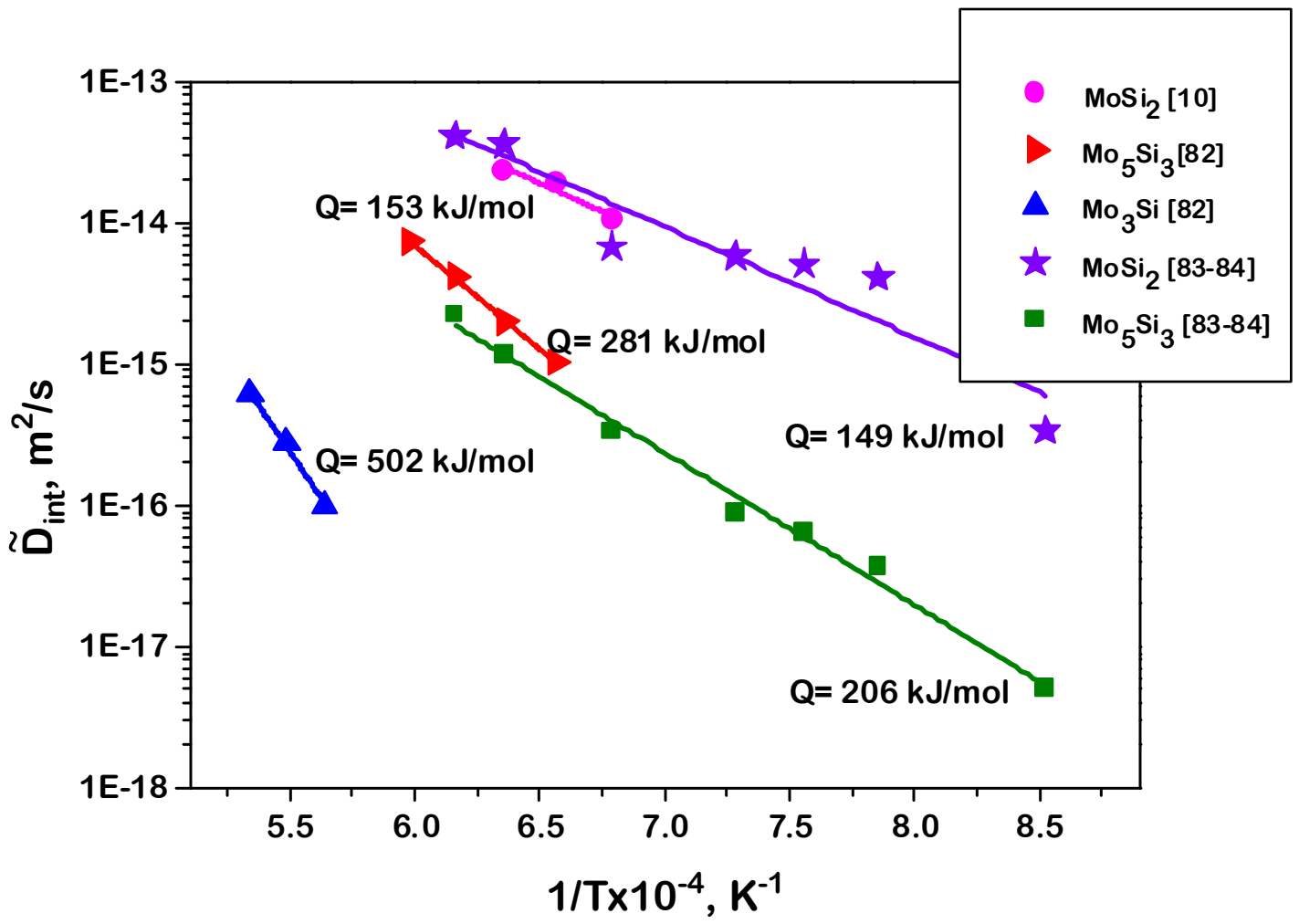

Figure 60 Integrated diffusion coefficients of the phases in the Mo-Si system [10, 82, 83, 84].

The growth mechanism of the phases is understood well following the physicochemical approach considering reaction-dissociation at the interfaces [1, 85]. This is phenomenological in nature and there is no need of understanding the complicated/unknown atomic mechanism of diffusion. As already explained, three phases $MoSi_2$, $Mo_5Si_3$ and $Mo_3Si$ should grow in the interdiffusion zone. Although $Mo_3Si$ phase does not grow with a measurable thickness in the interdiffusion zone and Mo has negligible diffusion rate in the $MoSi_2$ phase, we need to consider the growth of all the phases and diffusion of both the components in all the phases for utilization of the physicochemical approach.

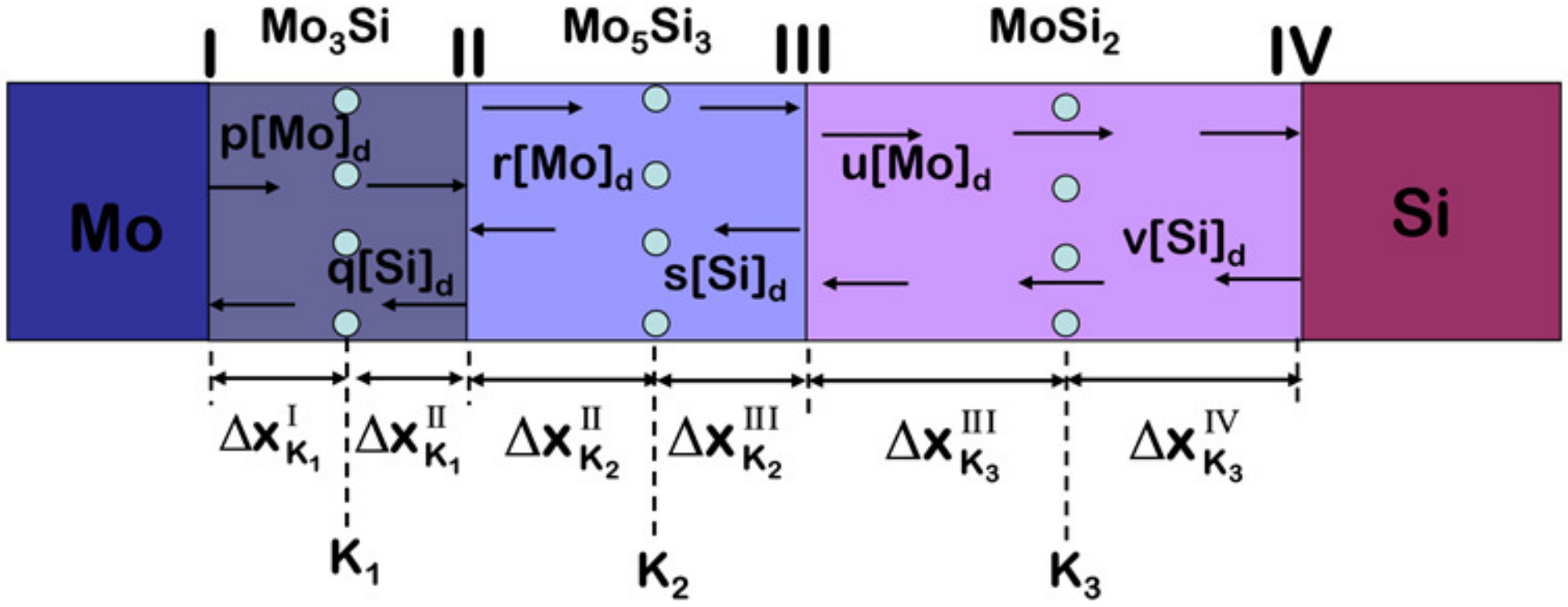

Figure 61 The phenomenological diffusion process in the Mo-Si diffusion [10].

The diffusion process is schematically explained in Figure 61. The reaction-dissociation equations at different interfaces can be expressed as

Interface I

$$3\,q\,Mo + q\,[Si]_d \Rightarrow 4\,q\,Mo_{3/4}Si_{1/4}$$

Interface II ($Mo_3Si$ side)

$$2\,p\,Mo_{5/8}Si_{3/8} + p\,[Mo]_d \Rightarrow 3\,p\,Mo_{3/4}Si_{1/4}$$

$$6\,q\,Mo_{5/8}Si_{3/8} \Rightarrow q\,[Si]_d + 5\,q\,Mo_{3/4}Si_{1/4}$$

Interface II ($Mo_5Si_3$ side)

$$3\,r\,Mo_{3/4}Si_{1/4} \Rightarrow r\,[Mo]_d + 2\,r\,Mo_{5/8}Si_{3/8}$$

$$5\,s\,Mo_{3/4}Si_{1/4} + s\,[Si]_d \Rightarrow 6\,s\,Mo_{5/8}Si_{3/8}$$

Interface III ($Mo_5Si_3$ side)

$$\frac{9}{7}\,r\,Mo_{1/3}Si_{2/3} + r\,[Mo]_d \Rightarrow \frac{16}{7}\,r\,Mo_{5/8}Si_{3/8}$$

$$\frac{15}{7}\,s\,Mo_{1/3}Si_{2/3} \Rightarrow s\,[Si]_d + \frac{8}{7}\,s\,Mo_{5/8}Si_{3/8}$$

Interface III ($MoSi_2$ side)

$$\frac{8}{7}\,v\,Mo_{5/8}Si_{3/8} + v\,[Si]_d \Rightarrow \frac{15}{7}\,v\,Mo_{1/3}Si_{2/3}$$

$$\frac{16}{7}\,u\,Mo_{5/8}Si_{3/8} \Rightarrow u\,[Mo]_d + \frac{9}{7}\,u\,Mo_{1/3}Si_{2/3}$$

Interface IV

$$2\,u\,Si + u\,[Mo]_d \Rightarrow 3\,u\,Mo_{1/3}Si_{2/3}$$

Diffusion of components in different phases in moles per unit area are expressed with $p$ and $q$ in the $Mo_3Si$, $r$ and $s$ in the $Mo_5Si_3$ phase and $u$ and $v$ in the $MoSi_2$ phase for Mo and Si after total annealing time $t$. As already discussed in other systems, at interfaces II and III, the growth process is very complicated at which one particular phase grows by consuming another phase and at the same time the same phase gets consumed by because of growth of the other phase. At the interface II, $(3p+5q)$ moles of $Mo_3Si$ are grown by consuming $Mo_5Si_3$. At the same time, $(3r+5s)$ moles were consumed by the same phase. On the other side of the same interface $(2r+6s)$ moles of $Mo_5Si_3$ are grown by consuming $Mo_3Si$. However, at the same time $(2p+6q)$ moles were consumed by the same phase. Following, a similar process at the interface III, $\left(\frac{16}{7}r+\frac{8}{7}s\right)$ moles of $Mo_5Si_3$ are grown and $\left(\frac{16}{7}u+\frac{8}{7}v\right)$ moles are consumed. $\left(\frac{9}{7}u+\frac{15}{7}v\right)$ moles of $MoSi_2$ are grown and $\left(\frac{9}{7}r+\frac{15}{7}s\right)$ moles are consumed. Therefore, by relating them with molar volumes and the thickness of the phase layers, we can write

$$4\,q\,.V_m^{Mo_3Si} = x_{K_1}^{I} \tag{12a}$$

$$\left(3p+5q-3r-5s\right).V_m^{Mo_3Si}=x_{K_1}^{II} \tag{12b}$$

$$\left(2r+6s-2p-6q\right).V_m^{Mo_5Si_3}=x_{K_2}^{II} \tag{12c}$$

$$\left(\frac{16}{7}r+\frac{8}{7}s-\frac{8}{7}v-\frac{16}{7}u\right).V_m^{Mo_5Si_3}=x_{K_2}^{III} \tag{12d}$$

$$\left(\frac{15}{7}v+\frac{9}{7}u-\frac{9}{7}r-\frac{15}{7}s\right).V_m^{MoSi_2}=x_{K_3}^{III} \tag{12e}$$

$$3u.V_m^{MoSi_2}=x_{K_3}^{IV} \tag{12f}$$

Following, $\tilde{D}_{\text{int}}$ of the phases, are related with the diffusing components are expressed as [1, 37-38]

$$\tilde{D}_{\text{int}}^{Mo_3Si}=\frac{V_m^{Mo_3Si}}{2t}\left(N_{Mo}^{Mo_3Si}q+N_{Si}^{Mo_3Si}p\right)\left(x_{K_1}^{I}+x_{K_1}^{II}\right) \tag{13a}$$

$$\tilde{D}_{\text{int}}^{Mo_5Si_3}=\frac{V_m^{Mo_5Si_3}}{2t}\left(N_{Mo}^{Mo_5Si_3}s+N_{Si}^{Mo_5Si_3}r\right)\left(x_{K_2}^{II}+x_{K_2}^{III}\right) \tag{13b}$$

$$\tilde{D}_{\text{int}}^{MoSi_2}=\frac{V_m^{MoSi_2}}{2t}\left(N_{Mo}^{MoSi_2}v+N_{Si}^{MoSi_2}u\right)\left(x_{K_3}^{III}+x_{K_3}^{IV}\right) \tag{13c}$$

The ratio of the tracer diffusion coefficients of the diffusing species in each of the phase can be expressed in terms of $p$, $q$, $r$, $s$, $u$ and $v$ as

$$\left.\frac{D_{Si}^*}{D_{Mo}^*}\right|_{Mo_3Si}=\frac{q}{p} \tag{13d}$$

$$\left.\frac{D_{Si}^*}{D_{Mo}^*}\right|_{Mo_5Si_3}=\frac{s}{r} \tag{13e}$$

$$\left.\frac{D_{Si}^*}{D_{Mo}^*}\right|_{MoSi_2}=\frac{v}{u} \tag{13f}$$

Let us consider the interdiffusion zone that is grown at 1200 °C. $\tilde{D}_{\text{int}}$ of the $MoSi_2$ phase was determined by Prasad and Paul [10] as $1.06\text{x}10^{-14}$ $m^2/s$. The same is estimated as $4.81\text{x}10^{-16}$ for $Mo_5Si_3$ and $1.0\text{x}10^{-19}$ $m^2/s$ for $Mo_3Si$ utilizing the growth kinetics details reported by Yoon et al. [80-82]. The average ratio of diffusivities in different phases is already mentioned previously, which are also used for the analysis. Considering the annealing time of 16 hrs, we have these details by solving the above equations

$p$ = 0.223 and $q$ = 1.046 mol/$m^2$

$r$ = 0.019 and $s$ = 2.006 mol/$m^2$

$u = 0.0$ and $v = 6.206$ mol/m$^2$

$x_{K_1}^{I} = 36.849$ μm and $x_{K_1}^{II} = -36.848$ μm

$x_{K_2}^{II} = 46.03$ μm and $x_{K_2}^{III} = -40.93$ μm

$x_{K_3}^{III} = 72.58$ μm and $x_{K_3}^{IV} \approx 0$ μm

Few interesting facts can be seen here. $x_{K_1}^{II}$ of $Mo_3Si$ is negative, which indicates the length of the phase that is consumed by the neighboring phase. Therefore, the addition of $x_{K_1}^{I} + x_{K_1}^{II}$ = 1 nm is the total thickness of this phase. This is indeed too thin to detect in SEM even if present at the $Mo/Mo_5Si_3$ interface. A TEM analyis indeed indicated the presence of this phase at this interface [73] as a very thin layer. Additionally, the Kirkendall plane cannot be present in this phase since one of the sublayers along with an addiional part of the other sublayer is consumed by the growth of the neighboring phase [1]. A similar discussion is applicable to the $Mo_5Si_3$ for the absence of the Kirkendall marker plane in this phase and the total thickness of the phase is estimated as $x_{K_2}^{II} + x_{K_2}^{III}$ = 8 μm. In the $MoSi_2$ phase, a sublayer thickness is almost zero and the other sublayer covers the whole thickness. Therefore, it is evident that the Kirkendall marker plane is present at the $MoSi_2/Si$ interface. The velocity of the Kirkendall marker planes in various phases can be estimated as [1]:

$$V_m^{Mo_3Si}(q-p) = 2t\,v_K^{Mo_3Si} = \left(x_K^{Mo_3Si} - x_o\right) \quad (14a)$$

$$V_m^{Mo_5Si_3}(s-r) = 2t\,v_K^{Mo_5Si_3} = \left(x_K^{Mo_5Si_3} - x_o\right) \quad (14b)$$

$$V_m^{MoSi_2}(v-u) = 2t\,v_K^{MoSi_2} = \left(x_K^{MoSi_2} - x_o\right) \quad (14c)$$

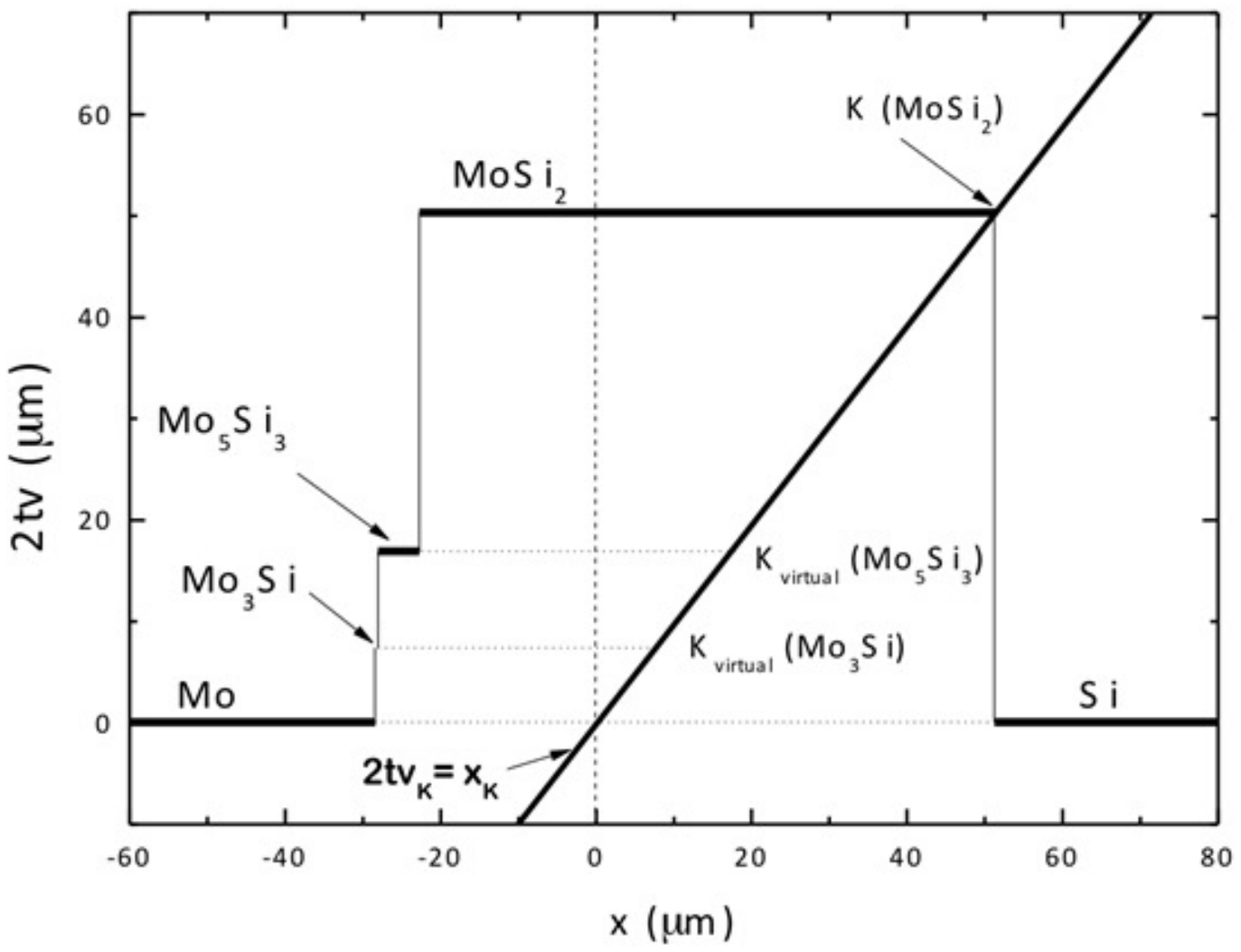

Figure 62 The velocity diagram for Mo/Si diffusion couple [10].

These are plotted in Figure 62 along with the line directly estimated by

$$v_K = \frac{dx}{dt} = \frac{x_K - x_0}{2t} = \frac{x_K}{2t}$$

$x_K$ is the position of the Kirkendall marker plane and $x_0 (= 0)$ is the location of the initial contact plane. The intersection of these lines indicates the location of the Kirkendall marker plane in the $MoSi_2$ phase (at the $MoSi_2$/Si interface) and the virtual Kirkendall marker planes of $K_{vir}^{Mo_5Si_3}$ and $K_{vir}^{Mo_3Si}$ in other phases.

### 4.3 Interdiffusion study in the W-Si system

Tungsten−silicon is another important system for integrated circuits because of low electrical resistivity and thermal stability of $WSi_2$ [86]. This phase is grown by reactive diffusion. This is also considered oxidation resistant coating [87-89]. This system is studied following both bulk [90-92] and thin film condition [88, 93-100]. However, most of these studies concentrated mainly on growth kinetics of the phases. Roy and Paul [7] estimated all the important diffusion parameters which will be discussed in detail here.

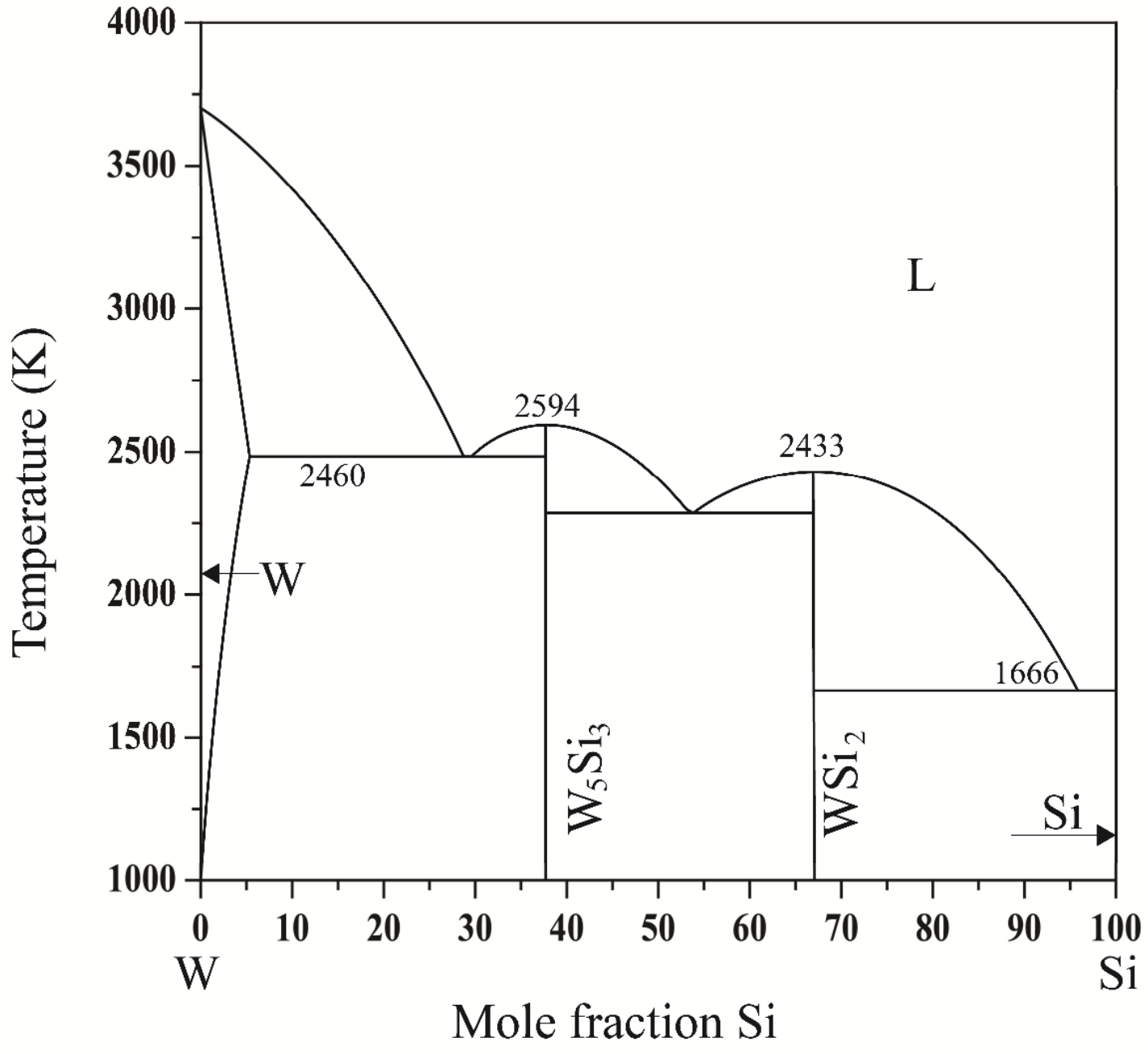

Figure 63 W-Si phase diagram [60].

Following the phase diagram, as shown in Figure 63, two phases are expected to grow in the interdiffusion zone of W/Si diffusion couple. As can be seen in Figure 64, $WSi_2$ could be detected after annealing at 1225 °C for 9 hrs with considerable thickness. $W_5Si_3$ was found to be present with very small thickness indicating the differences in growth rate between these two phases. The integrated diffusion coefficients are estimated based on experiments at different temperatures as shown in Figure 65. The activation energy is estimated following the Arrhenius equation as 152±7 kJ/mol. Other studies report this value in the range of 159-209 kJ/mol [88, 91, 96, 99]. The line of pores identifying the location of the Kirkendall marker plane is utilized to estimate the ratio of the tracer diffusion coefficients as $D_{Si}^* / D_W^* = 13.7$.

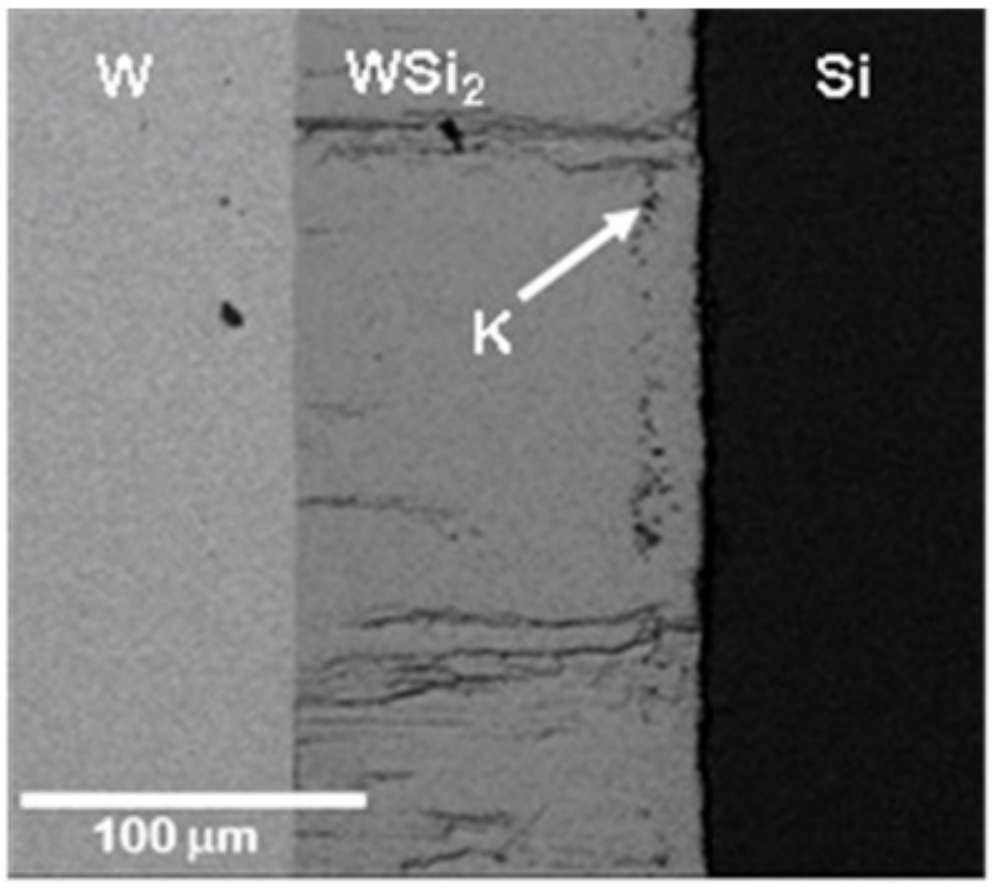

Figure 64 W/Si interdiffusion zone annealed 1225 °C for 9 hrs [7].

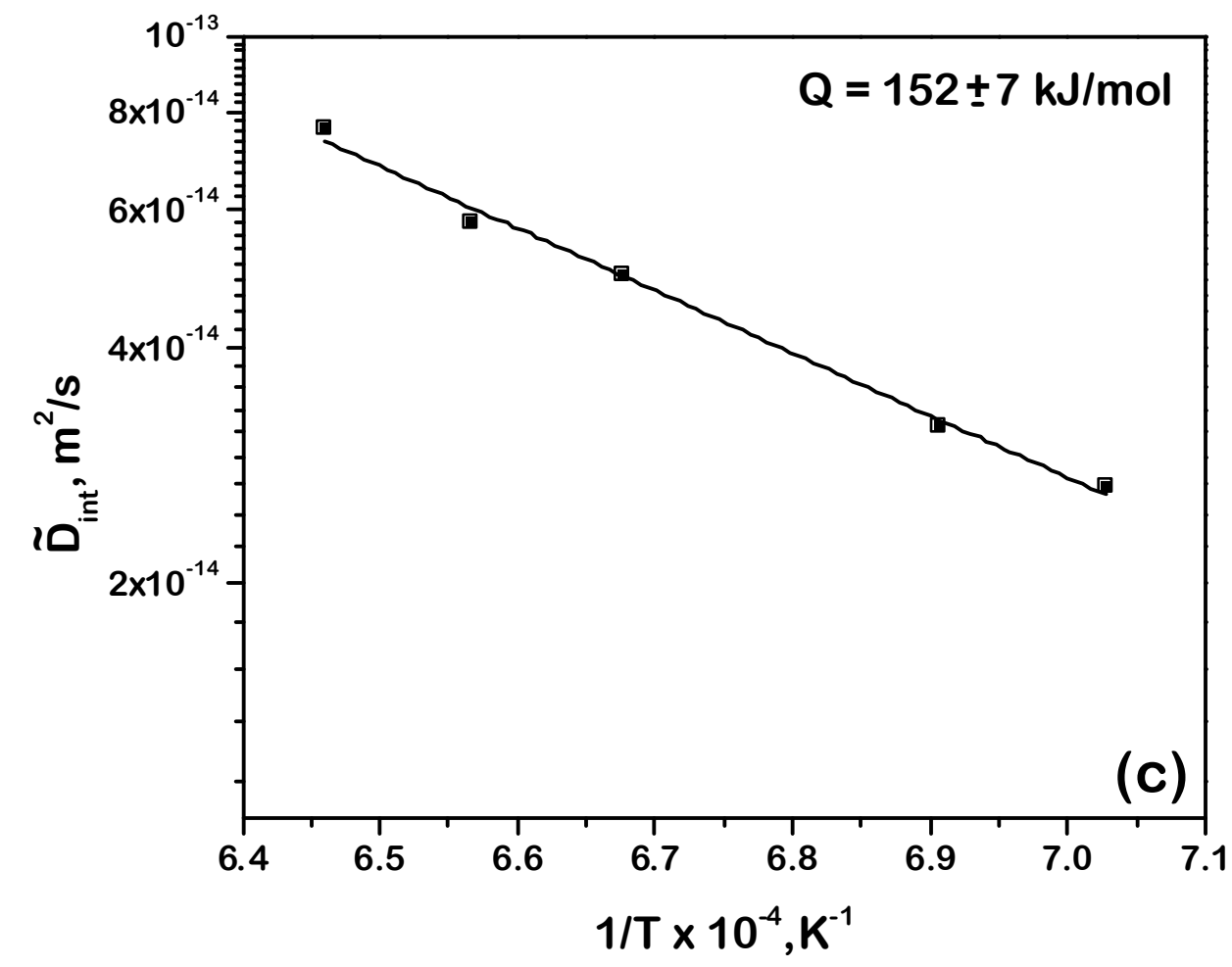

Figure 65 Integrated diffusion coefficients of $WSi_2$ [7].

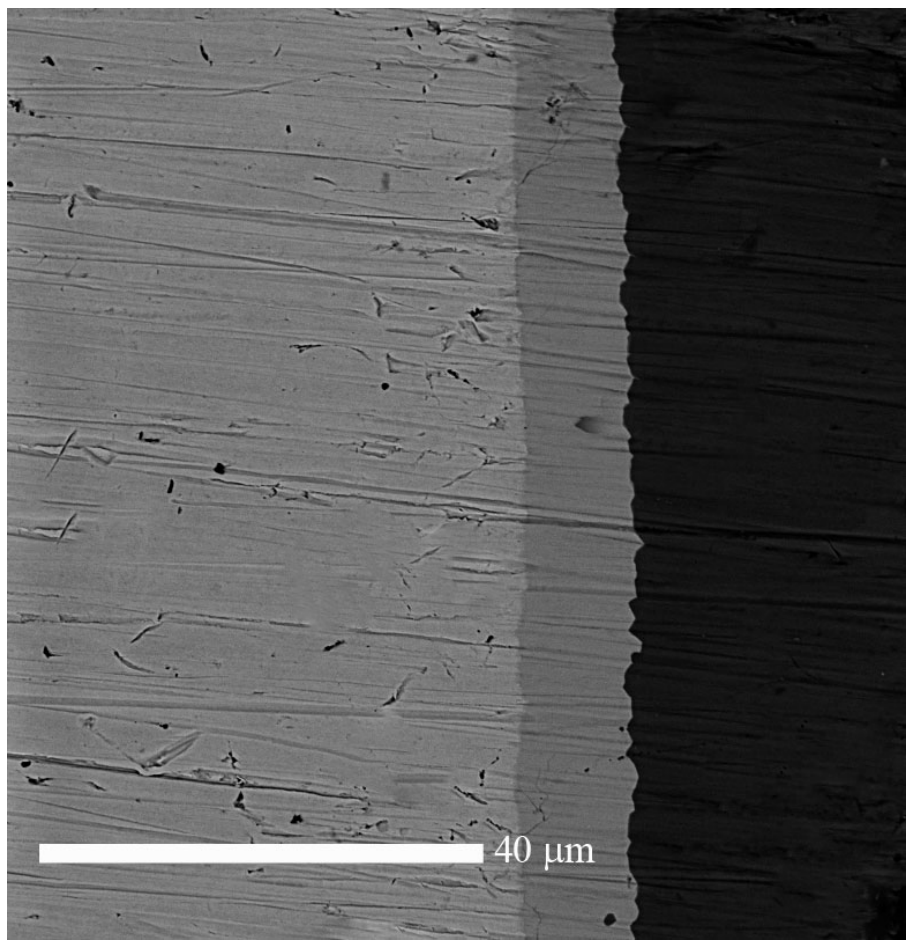

Figure 66 Diffusion couple of $WSi_2$/Si for the growth of $W_5Si_3$ at 1350 °C for 16 hrs [7].

$W_5Si_3$ phase did not grow with a reasonable thickness in the W/Si diffusion couple. Therefore, $WSi_2$/W incremental diffusion couples were prepared after removing Si from the W/Si diffusion couples such that only the $W_5Si_3$ phase could grow in the interdiffusion zone with measurable thickness. Such an interdiffusion zone is shown in Figure 66, which was annealed at 1350 °C for 16 hours. The estimated diffusion coefficients at different temperatures are shown in Figure 67. The activation energy was estimated at 301±40 kJ/mol. This is reported in the range of 289 – 360 kJ/mol in different other reports [90-92]. Lee et al. [90] conducted similar incremental diffusion couple experiment at 1400 °C identifying the location of the Kirkendall marker plane. The ratio of the tracer diffusivities is estimated as $D^*_{Si} / D^*_W = 11.9$.

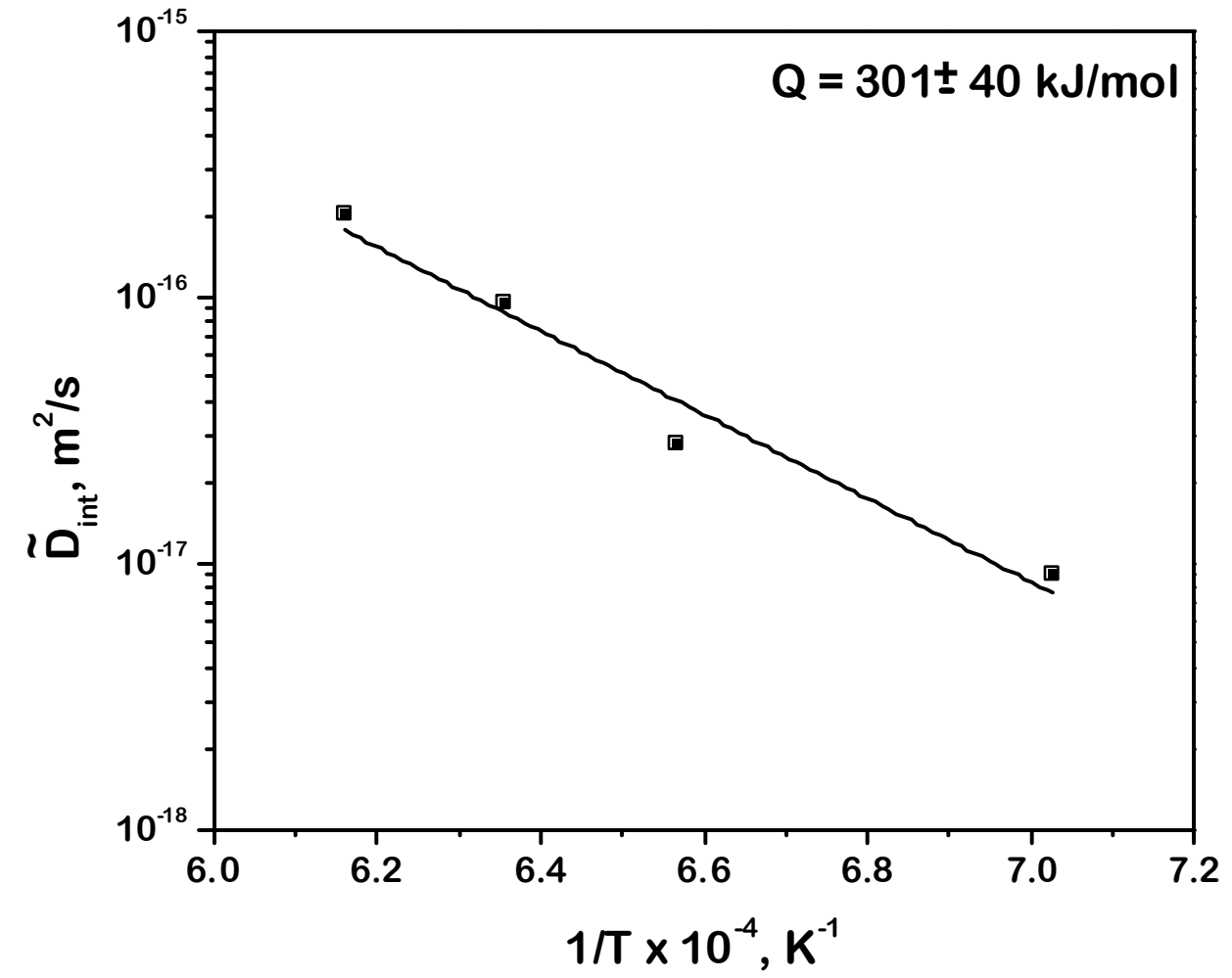

Figure 67 Integrated diffusion coefficient of $W_5Si_3$ at different temperatures [7].

### 4.4 Summary of interdiffusion studies in Group VIB Metal – Silicon systems

Now we compare the results in different systems of this group. Since only one experimental result is available in the Cr-Si system, the diffusion data in the Mo-Si and W-Si systems are compared. Figure 68 shows the comparison of the integrated diffusion coefficient of the disilicide phase. Similar to the other systems in the two other groups, it can be seen that the integrated diffusion coefficients are higher for the higher atomic number of the refractory component. This becomes even more pronounced in a normalized temperature plot, as shown in Figure 69. A similar nature of variation of the integrated diffusion coefficients is found even in the 5:3 silicide, as shown in Figure 70 and 71. It is evident that the integrated diffusion coefficient increases with the increase in an atomic number of the refractory component.

Based on the estimated ratio of the tracer diffusion coefficients, we can discuss the atomic mechanism of diffusion. In $CrSi_2$ phase, we have found that the diffusion rate of Si is infinitely (by few orders of magnitude) higher than Cr. In fact, it has the crystal structure of hP9, as shown in Figure 52. In this Cr is surrounded by 5 Si and Si is surrounded by 5 Cr atom and 5 Si. Therefore, Si can easily diffuse via its own sublattice and Cr cannot diffuse in the absence of antisites. Since the diffusion rate of Cr is found to be negligible, it is evident that the concentration of Cr antisites is negligible. As shown in Figure 72, both $MoSi_2$ and $WSi_2$ have the same crystal structure as tI6, $C11_b$. With respect to nearest neighbour surroundings, this is similar to the previously discussed crystal structure. The metal atom (Mo and W) are surrounded by 10 Si and Si atoms are surrounded by 5Si and 5 W. Therefore, metal atoms cannot diffuse unless antisites are present. Therefore, in MoSi2 phase indeed the concentration of antisites must be negligible; however, in the $WSi_2$ phase, W antisites must be present.

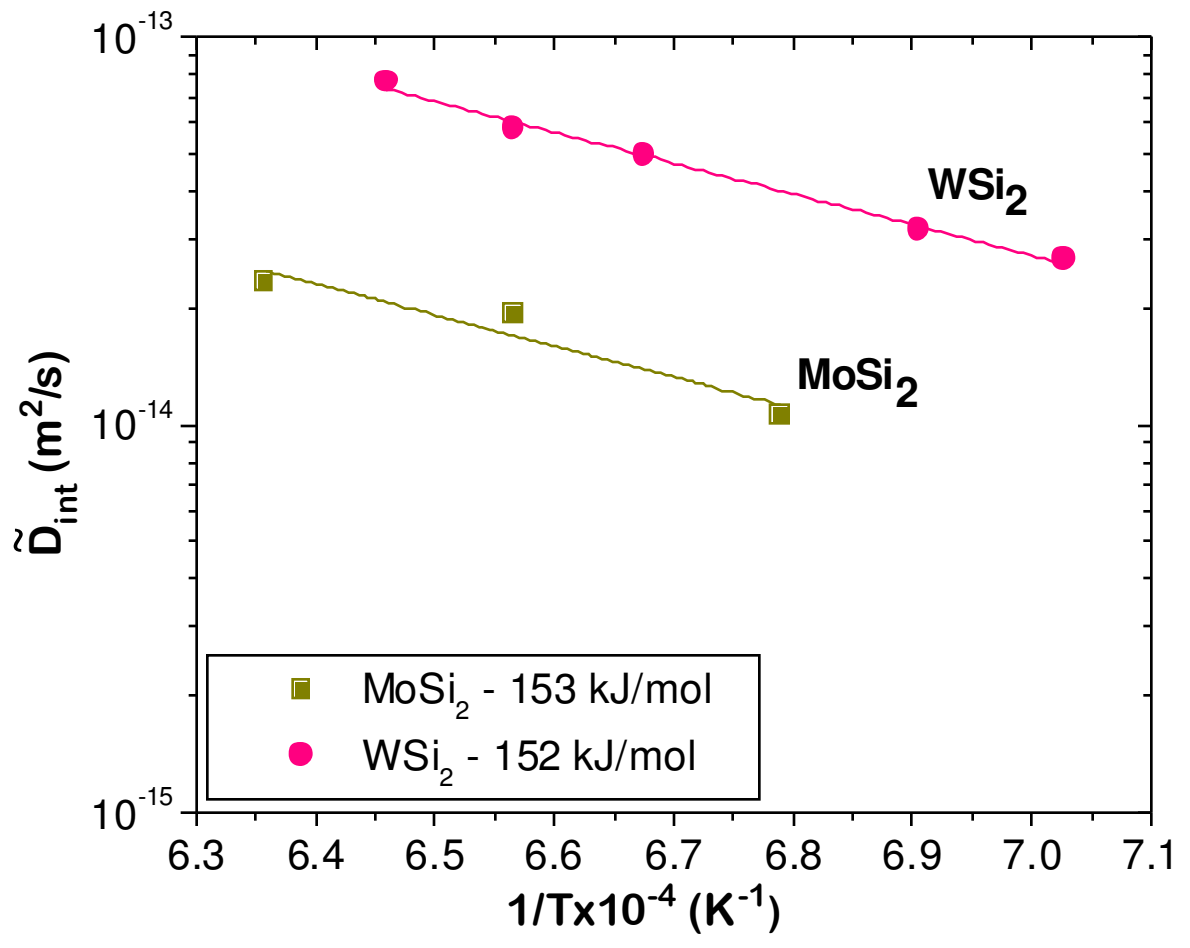

Figure 68 Comparison of integrated diffusion coefficients measured in disilicides of Mo-Si and W-Si systems [6, 60].

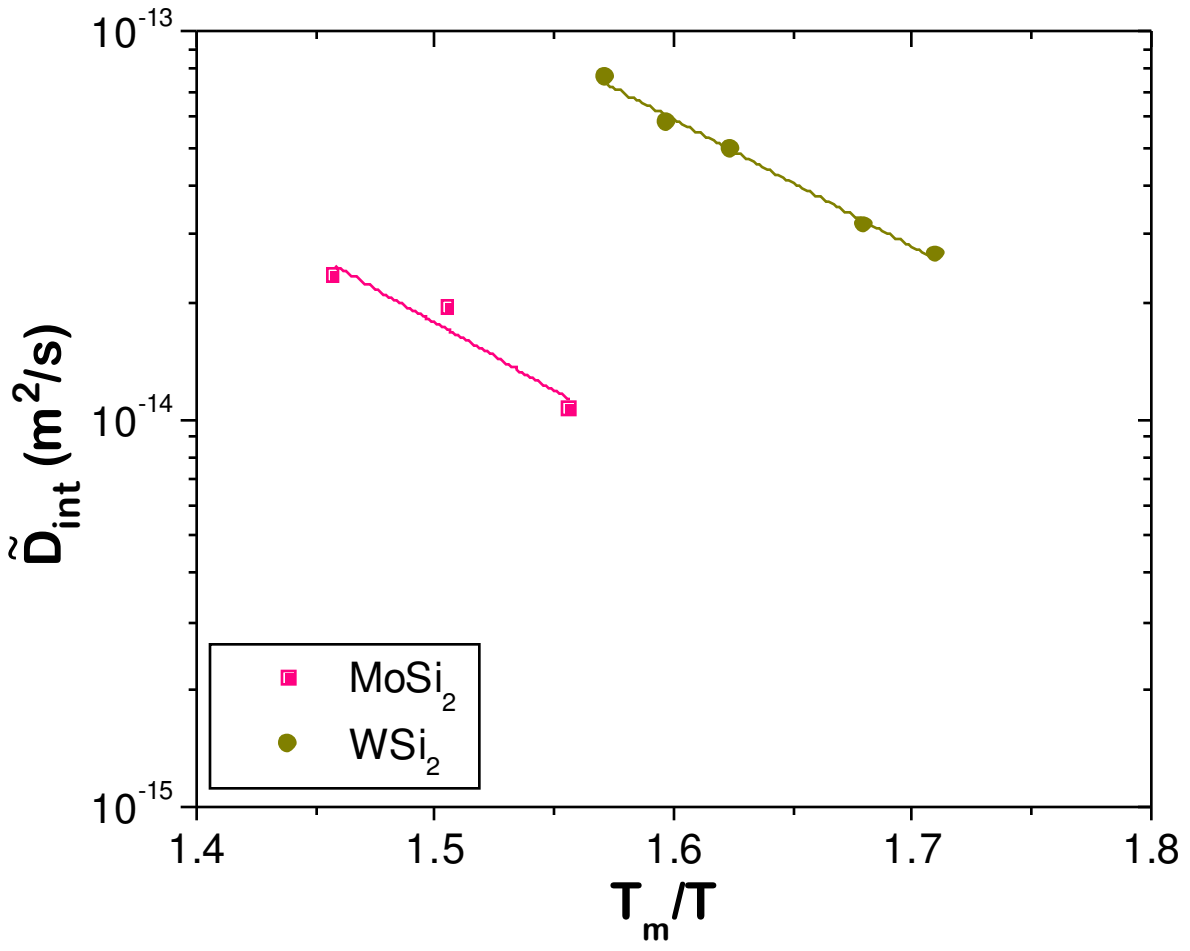

Figure 69 Normalized temperature dependent integrated diffusion coefficients measured in disilicides [6, 60].

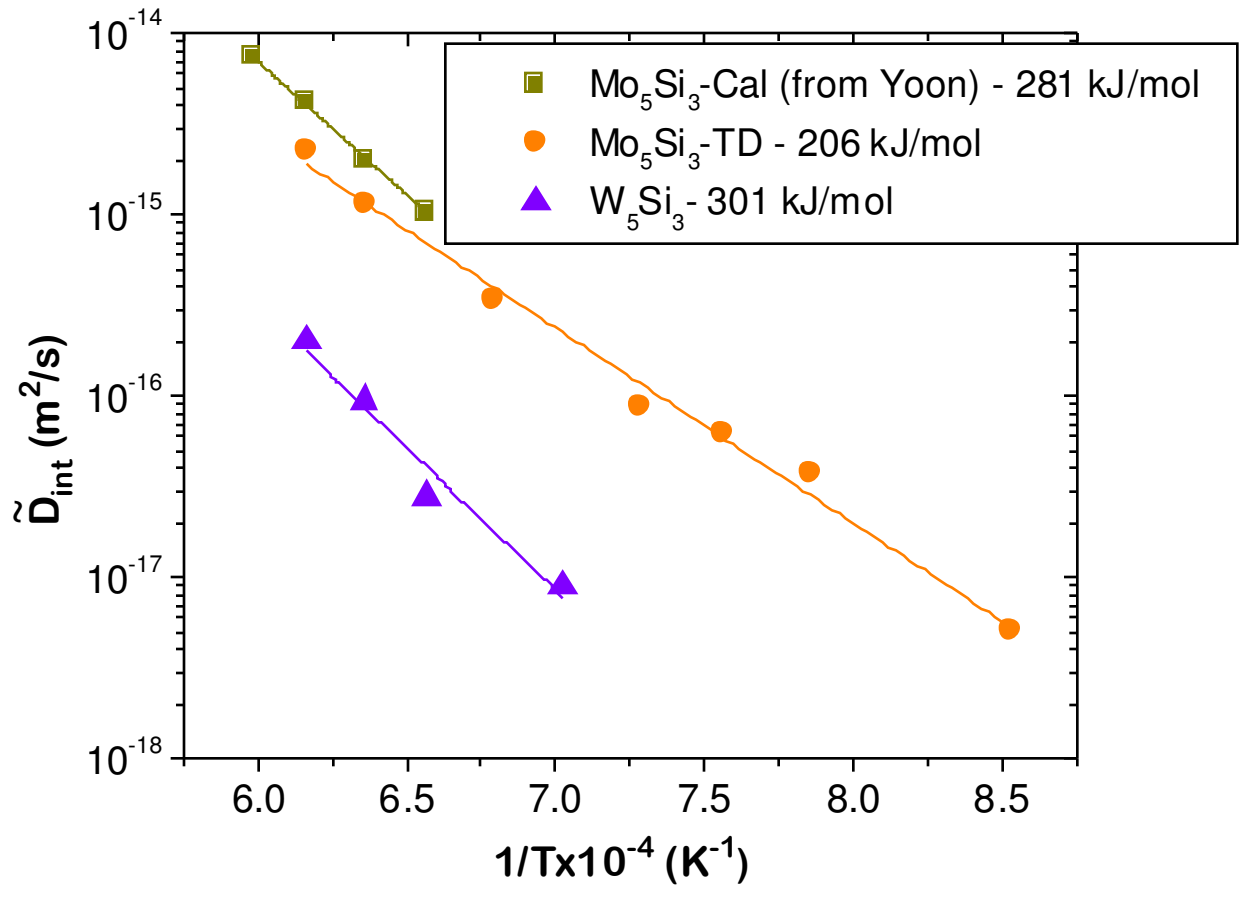

Figure 70 Comparison of integrated diffusion coefficients measured in 5:3 silicides of Mo-Si and W-Si systems [6, 60].

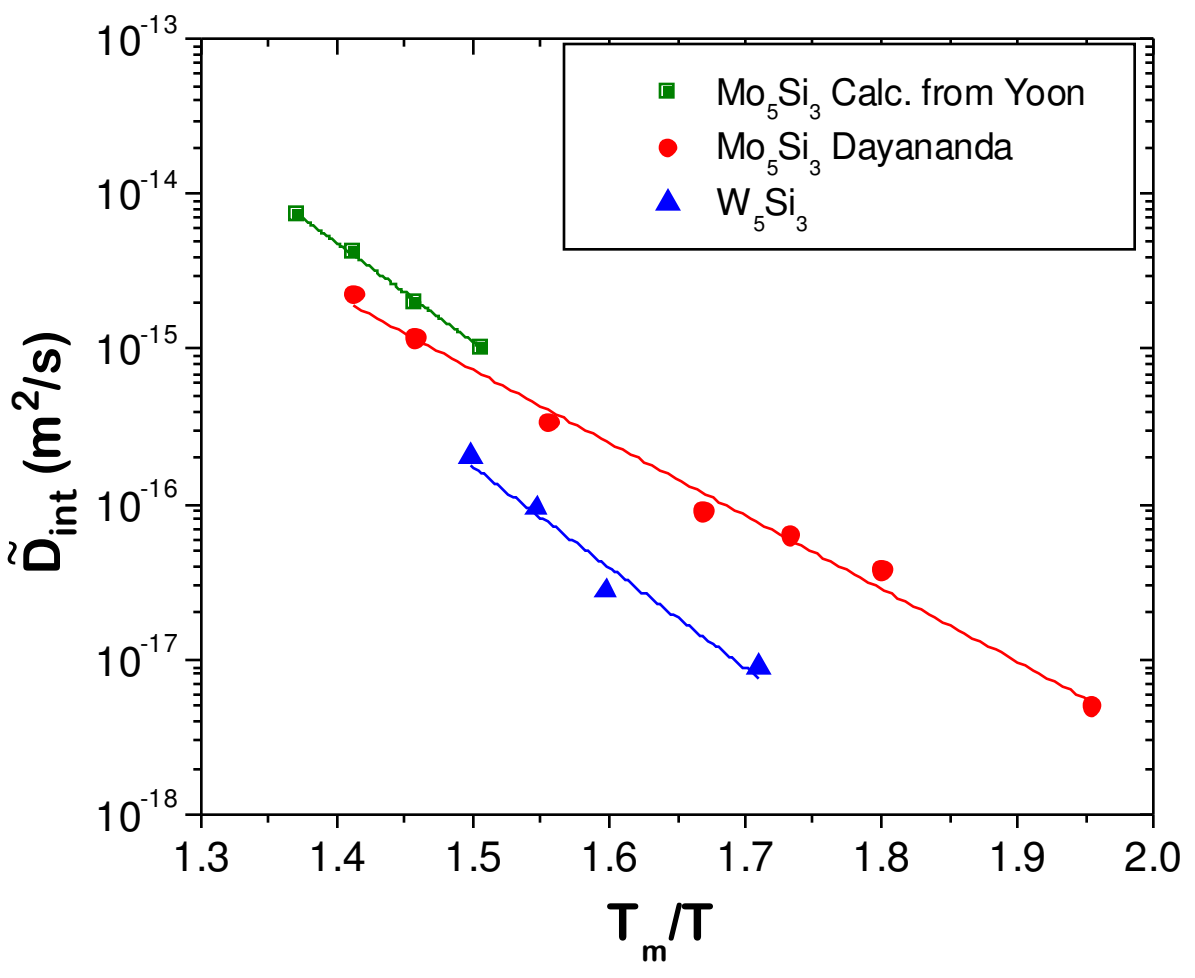

Figure 71 Homologous temperature normalized integrated diffusion coefficients measured in 5:3 silicides [6, 60].

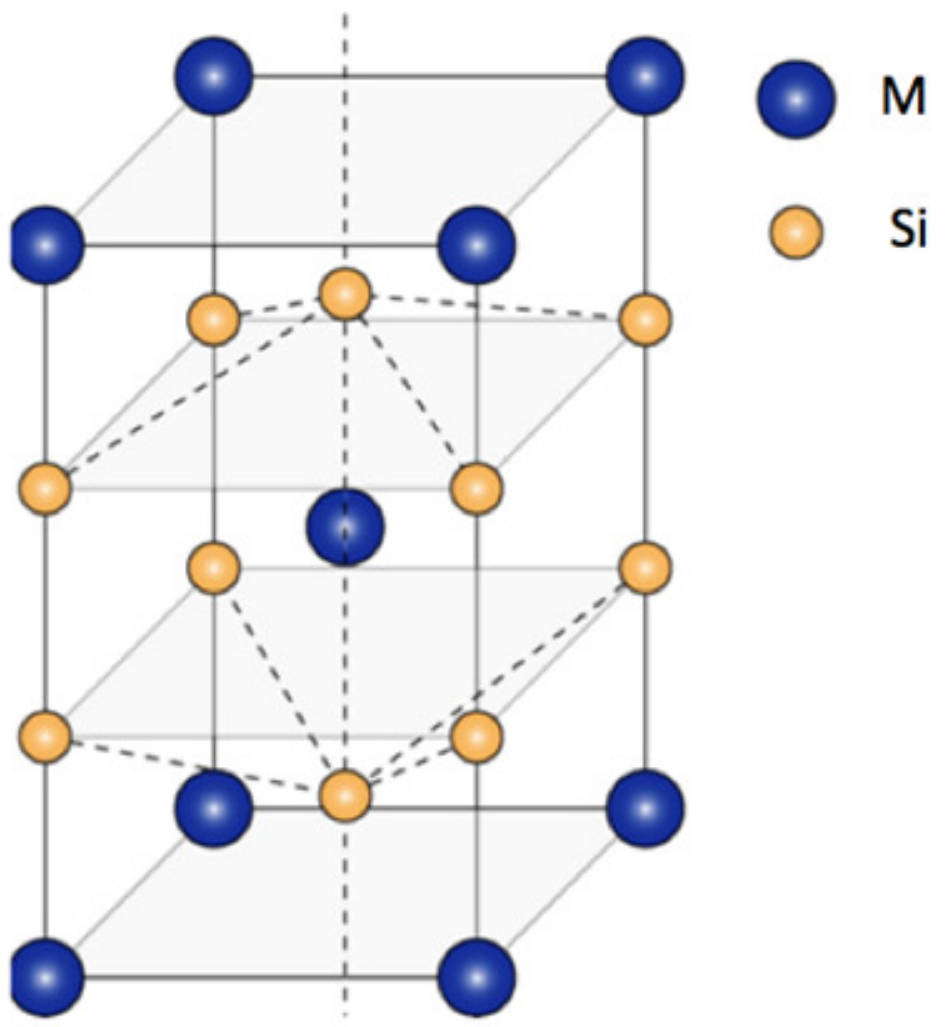

Figure 72 $C11_b$ crystal structure of $MoSi_2$ and $WSi_2$ with 6 atoms per unit cell (*tI6*) [6, 60].

The $Mo_5Si_3$ and $W_5Si_3$ compounds also have the same crystal structure of *tI32* as shown in Figure 73. There are only 2 Si-Si bonds compared to 8 Mo-Mo bonds in the structure. In the presence of an equal concentration of vacancies, one would expect a higher diffusion rate of Mo. However, in both the phases Si is found to have a higher diffusion rate compared to the metal component. Therefore, it is evident that the concentration of vacancies must be higher on Si sublattice. Additionally, Si has more than two orders of magnitude higher diffusion rate compared to Mo in the $Mo_5Si_3$ phase, whereas, Si has around one order of magnitude higher diffusion rate than W in the $W_5Si_3$ phase. Therefore, it is well possible that overall vacancy concentration in $W_5Si_3$ must be higher compared to the vacancy concentration in the $Mo_5Si_3$ phase. This argument finds support because of the higher integrated diffusion coefficient of $W_5Si_3$ compared to $Mo_5Si_3$.

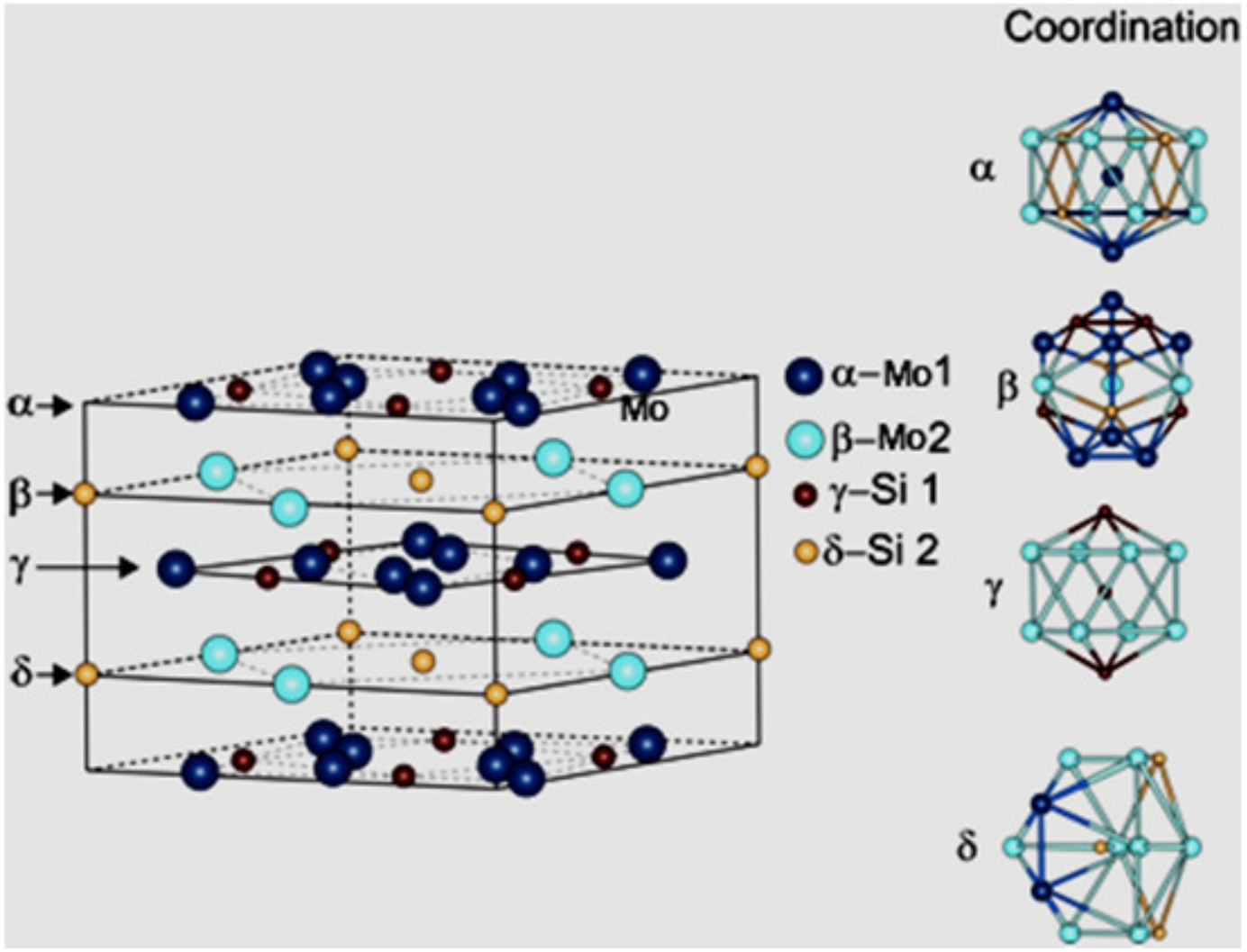

Figure 73 The *tI32* crystal structure of $Mo_5Si_3$ and $W_5Si_3$ [6, 60].

**5. Conclusions**

Diffusion-controlled the growth of refractory metal silicides is discussed in detail. Group IVB, VB and VIB refractory metals are considered for these discussions. In most of the systems, $MSi_2$ and $M_5Si_3$ silicides grow in an interdiffusion zone when diffusion couples are prepared by pure components or in incremental diffusion couples. Following the diffusion coefficients estimated are compared between systems in a particular group. Irrespective of the systems in different groups, the variation of the diffusion coefficients follow a certain pattern. In both $MSi_2$ and $M_5Si_3$ silicides, the integrated diffusion coefficient increases with the increase in an atomic number of the refractory component in a particular group. This trend follows when the data are compared at the same melting point normalized temperature ($T_m/T$). This indicates that there must be an increase in the overall concentration of defects with the increase in atomic number. Additionally, the measured ratio of diffusivities also indicates the change in the concentration of defects on different sublattices following a certain pattern. For example, if we consider the disilicide phases, Si has always much higher diffusion rate compared to the metal component in the group IVB-Si system. This indicates that antisite defects of metal components must be negligible. However, when we consider the disilicides of group VB M-Si systems, metal components also could diffuse, which increased with the increase in the atomic number of the metal component. A similar effect is seen even in the disilicides of group VIB M-Si systems. Therefore, although the defects concentrations in these phases are not known, which are even difficult to estimate, the diffusion studies indicate the kind of defects present in different intermetallic compounds.